FLORIDA STATE UNIVERSITY

COLLEGE OF ARTS & SCIENCES

# TRANSPORT PROPERTIES OF SEMIMETALLIC TRANSITION METAL DICHALCOGENIDES

By

QIONG ZHOU

A Dissertation submitted to the
Department of Physics
in partial fulfillment of the
requirements for the degree of
Doctor of Philosophy

2017

# TABLE OF CONTENTS

**Appendix**

# LIST OF TABLES

# LIST OF FIGURES

# ABSTRACT

The Weyl semimetal requires the breaking of either the time-reversal symmetry (TRS) or the lattice inversion symmetry. When the TRS and inversion symmetry coexist, a pair of degenerate Weyl points may exist, leading to the related Dirac semimetal phase. In other words, a Dirac semimetallic state can be regarded as two copies of Weyl semimetal states. In this dissertation, we study tellurium based compounds like the Weyl semimetal candidate $MoTe_2$ and the Dirac semimetal candidate $PtTe_2$ within the transition metal dichalcogenides family.

Firstly, we report a systematic study on the Hall-effect of the semi-metallic state of bulk $MoTe_2$, which was recently claimed to be a candidate for a novel type of Weyl semi-metallic state. The temperature ($T$) dependence of the carrier densities and of their mobilities, as estimated from a numerical analysis based on the isotropic two-carrier model, indicates that its exceedingly large and non-saturating magnetoresistance may be attributed to a near perfect compensation between the densities of electrons and holes at low temperatures. A sudden increase in hole density, with a concomitant rapid increase in the electron mobility below $T \sim 40$ K, leads to comparable densities of electrons and holes at low temperatures suggesting a possible electronic phase-transition around this temperature.

Secondly, the electronic structure of semi-metallic transition-metal dichalcogenides, such as $WTe_2$ and orthorhombic $\gamma-MoTe_2$, are claimed to contain pairs of Weyl points or linearly touching electron and hole pockets associated with a non-trivial Chern number. For this reason, these compounds were recently claimed to conform to a new class, deemed type-II, of Weyl semi-metallic systems. A series of angle resolved photoemission experiments (ARPES) claim a broad agreement with these predictions detecting, for example, topological Fermi arcs at the surface of these crystals. We synthesized single-crystals of semi-metallic $MoTe_2$ through a Te flux method to validate these predictions through measurements of its bulk Fermi surface (FS) *via* quantum oscillatory phenomena. We find that the superconducting transition temperature of $\gamma-MoTe_2$ depends on disorder as quantified by the ratio between the room- and low-temperature resistivities, suggesting the possibility of an unconventional superconducting pairing symmetry. Similarly to $WTe_2$, the magnetoresistivity of $\gamma-MoTe_2$ does not saturate at high magnetic fields and can easily surpass $10^6$ %. Remarkably, the analysis of the de Haas-van Alphen (dHvA) signal superimposed onto the magnetic torque, indicates that the geometry of its FS is markedly distinct from the calculated one. The dHvA signal also reveals that the FS is affected by the Zeeman-effect precluding the extraction of the Berry-phase. A direct comparison between the previous ARPES studies and density-functional-theory (DFT) calculations reveals a disagreement in the position of the valence bands relative to the Fermi level $\varepsilon_F$. Here, we show that a shift of the DFT valence bands relative to $\varepsilon_F$, in order to match the ARPES observations, and of the DFT electron bands to explain some of the observed dHvA frequencies, leads to a good agreement between the calculations and the angular dependence of the FS cross-sectional areas observed experimentally. However, this relative displacement between electron- and hole-bands eliminates their crossings and, therefore, the Weyl type-II points predicted for $\gamma-MoTe_2$.

Finally, we investigate the electronic structure and transport properties in single crystals of the semi-metallic platinum ditelluride ($PtTe_2$), recently claimed to be a novel type-II Dirac semimetal, via a methodology similar to that applied to $\gamma-MoTe_2$, i.e. the temperature and angular dependence of the SdH and dHvA effects. Our high-quality $PtTe_2$ crystal displays a large non-saturating magnetoresistance under magnetic field up to 61 T. The dHvA oscillation and SdH effect reveal several high and low frequencies suggesting a rather complex Fermi surface. We also find evidence for a non-trivial Berry phase. The crystal quality improved considerably under subsequent annealing at high-temperatures leading to the observation of linear in field magnetoresistivity. Combined with effective masses in the order of $\sim 0.1$ free electron mass, these results further suggest that $PtTe_2$ displays bulk Dirac-like bands.

# CHAPTER 1

# INTRODUCTION

## 1.1 Transition Metal Dichalcogenides

Transition metal dichalcogenides (TMDs) are layered materials characterized by strong intra-layer covalent bonding and weak inter-layer van-der-Waals coupling. These materials belong to a family of any compounds with the generalized formula $MX_2$, where M is typically an element from group IV to group VI ( for example Zr, Hf, V, Nb, Mo, W and so on) and X is a chalcogen (S, Se, Te). Each individual layer consists of 3 atomic planes with chalcogen atoms in two hexagonal planes segregated by a plane of metal atoms, as shown in Figure 1.1(a). In bulk form, adjacent layers are weakly coupled in a variety of polytypes, as shown in Figure 1.1(b). The most common polytypes are the $2H$ (with hexagonal symmetry with trigonal prismatic coordination , two layers per unit cell), the $3R$ (rhombohedral symmetry with trigonal prismatic coordination and 3 layers per unit cell) and the $1T$ (tetragonal symmetry with octahedral coordination and one layer per unit cell) phase.

TMDs compounds range from insulators (for instance $HfO_2$), semiconductors ( $MoSe_2$, $MoS_2$, $WSe_2$, etc.) to semimetals($WTe_2$ and $\beta-$ and $\gamma-$phase $MoTe_2$) or metals($NbSe_2$, etc.). Some TMDs ($1T$-$TaS_2$, etc.) [10] are reported to exhibit charge density wave and superconductivity. Typically in metallic TMDs ($1T$ phase), the metal atoms have octahedral coordination, while the metal atoms are trigonal prismatic coordinated in the semiconducting phase ($2H$ or $3R$).

Graphene, the monolayer counterpart of graphite has garnered tremendous attention among two-dimensional materials due to its remarkable electronic properties and its optical attributes [107]. Similarly to graphene, the weak interactions in TMDs allow the exfoliation of these materials into two-dimensional layers with just a few or a single layer thickness [11, 110]. Graphene displays an extraordinary high mobility exceeding $10^5$ $cm^2V^{-1}s^{-1}$ at room temperature [12] and exceeding $10^6$ $cm^2V^{-1}s^{-1}$ at 2 K in devices encapsulated in BN dielectric layers [108]. For this reason a lot of attempts have been made to use graphene to develop new ultra thin and high efficient electrical conductors [13]. However one issue regarding field effect transistors (FETs) made from graphene is that they can't be effectively switched off due to its lack of a band gap, which poses a problem in particular with digital electronics.

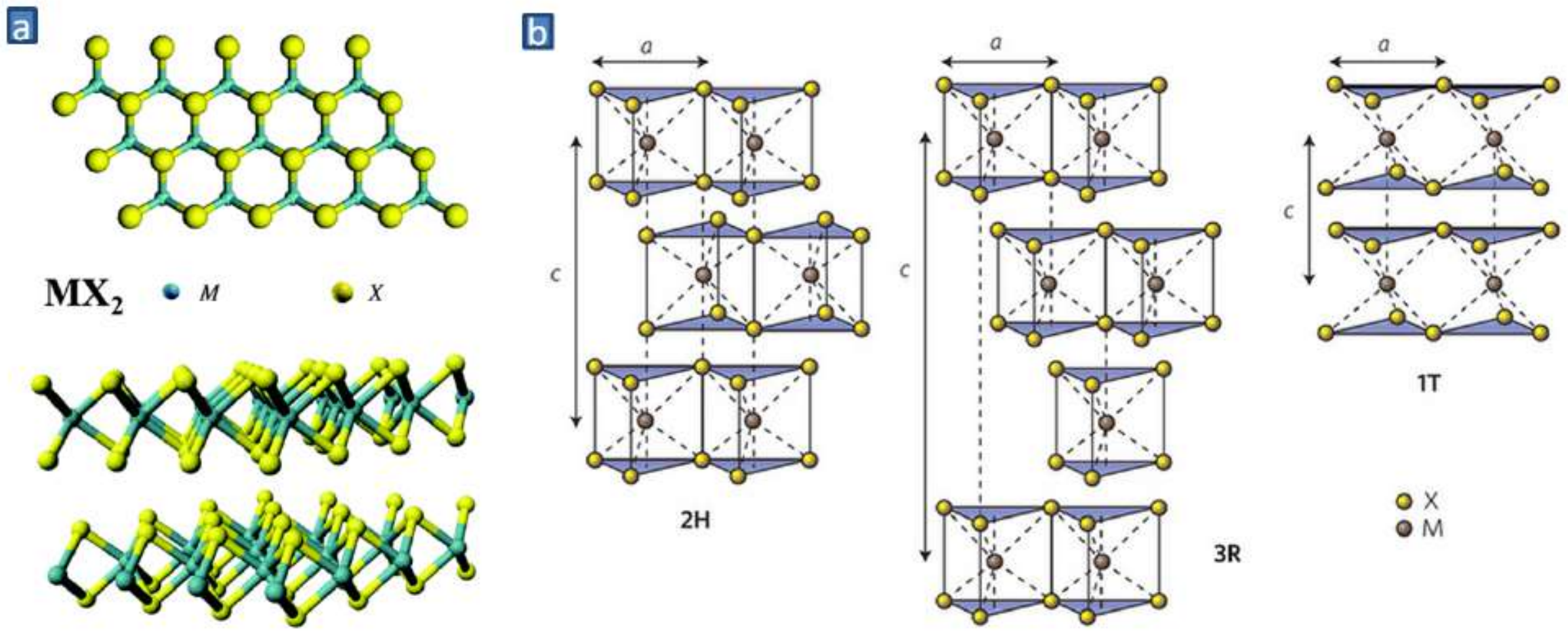

Figure 1.1: Structure of TMD materials. (a) Three-dimensional schematics of a typical layered $MX_2$ structure. (b) Schematics of TMD structures with various geometries: $2H$, $3R$, and $1T$. a is the lattice constant and c is the inter-layer lattice constant. Figure from Ref. [1].

Unlike graphene, many of the semiconducting TMDs have band gaps of approximately 1-2 eV [14] and they experience a transition from indirect band gap to direct band gap with decreasing number of layers. A typical example of this transition is 2H-$MoS_2$. The band gap in bulk compounds is around 1.2 eV while the band gap of monolayer $MoS_2$ is 1.9 eV, as shown in Figure 1.2 . Such direct band gap transition with the number of layers number, which is one of the key characteristics of 2D $MoS_2$ and other TMDs, is due to quantum confinement effect [15] attributed to the contributions of atomic orbitals to the electronic band states at different points of momentum space [3, 15]. As the number of layers decreases, the transition at the $\Gamma$ point shift significantly from an indirect one to a larger and direct one. Figure 1.3(a) and 1.3(b) show strong photoluminescence emerging when the $MoS_2$ crystal is thinned to one single layer, indicating an indirect to direct band gap transition. Other semiconducting TMDs such as $MoSe_2$, $2H$-$MoTe_2$ and $WSe_2$ and their heterostructures, have been widely investigated in the last few years because of sizable band gaps and layered-dependent properties which show a wide range of electronic, optical , mechanical, chemical and thermal properties which could be potentially useful in engineering photodetectors, solar cells,

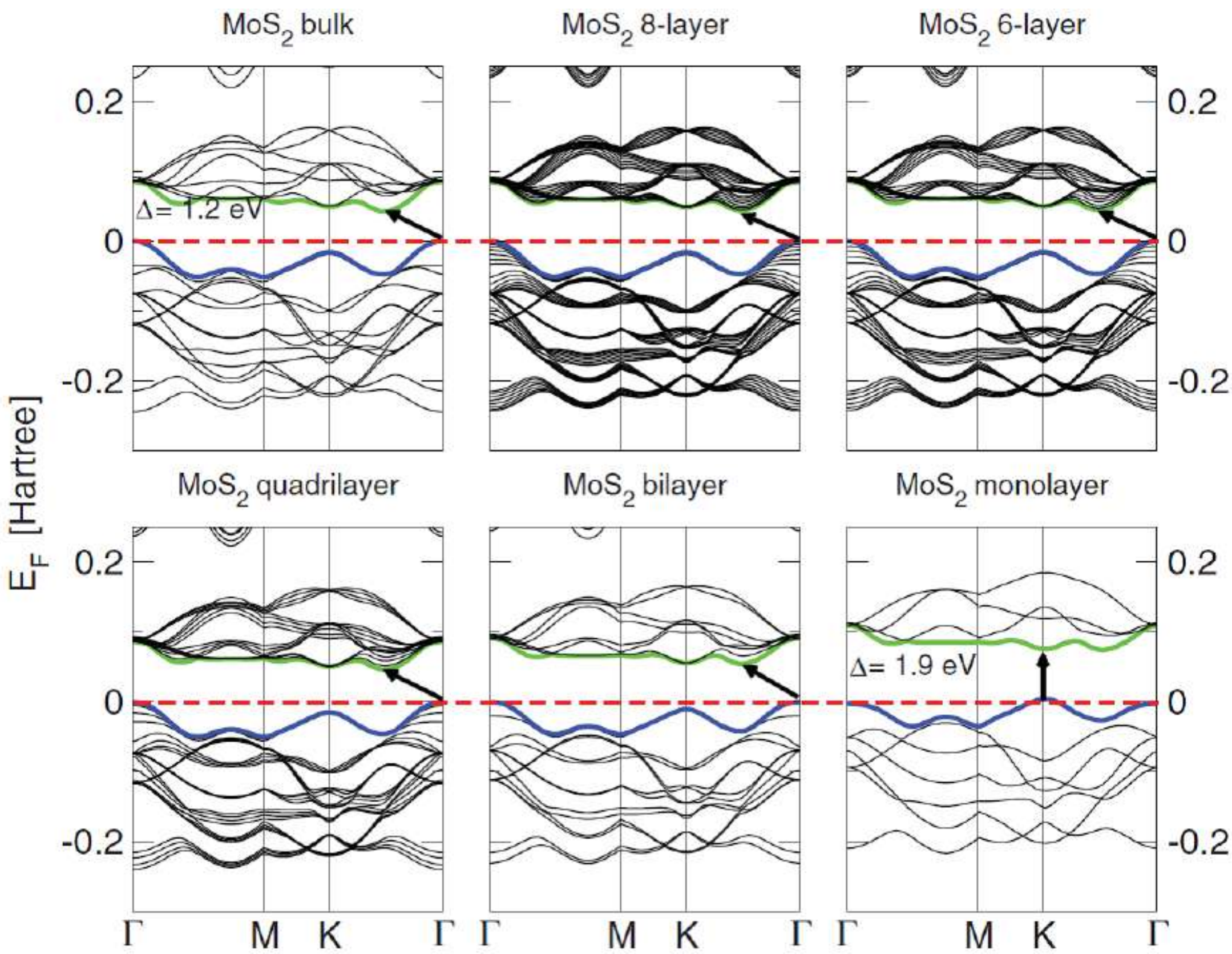

Figure 1.2: Calculated electronic band structures of bulk, multi-layer, down to single-layer MoS2 (from top left to bottom right) [2] via *ab initio* and first principles calculations using density functional theory , in which black arrows indicate the smallest indirect or direct band gap as the thickness was reduced to single layer.

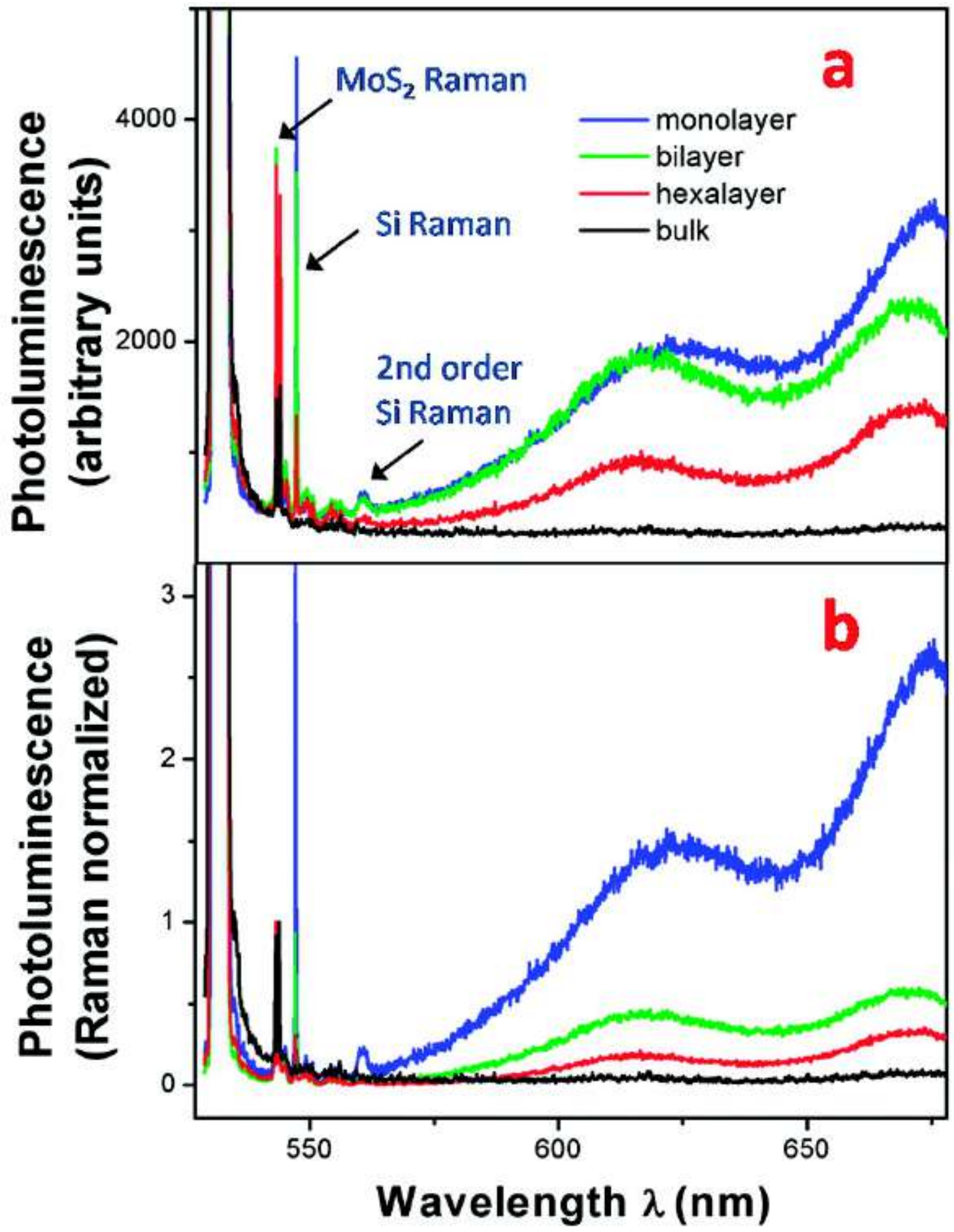

Figure 1.3: (a) Photoluminescence and Raman spectra for various layers of $MoS_2$. Three Raman modes can be identified for $MoS_2$, along with first and second order Raman peaks of Si substrate. (b) Photoluminescence spectra normalized by Raman intensity from Ref. [3] .

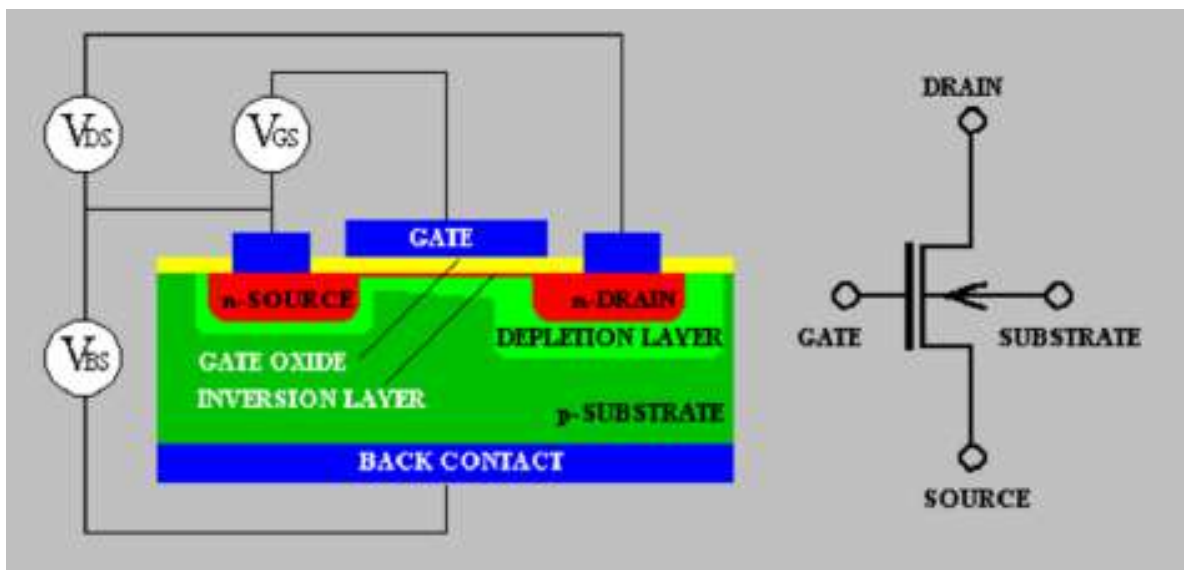

(a)

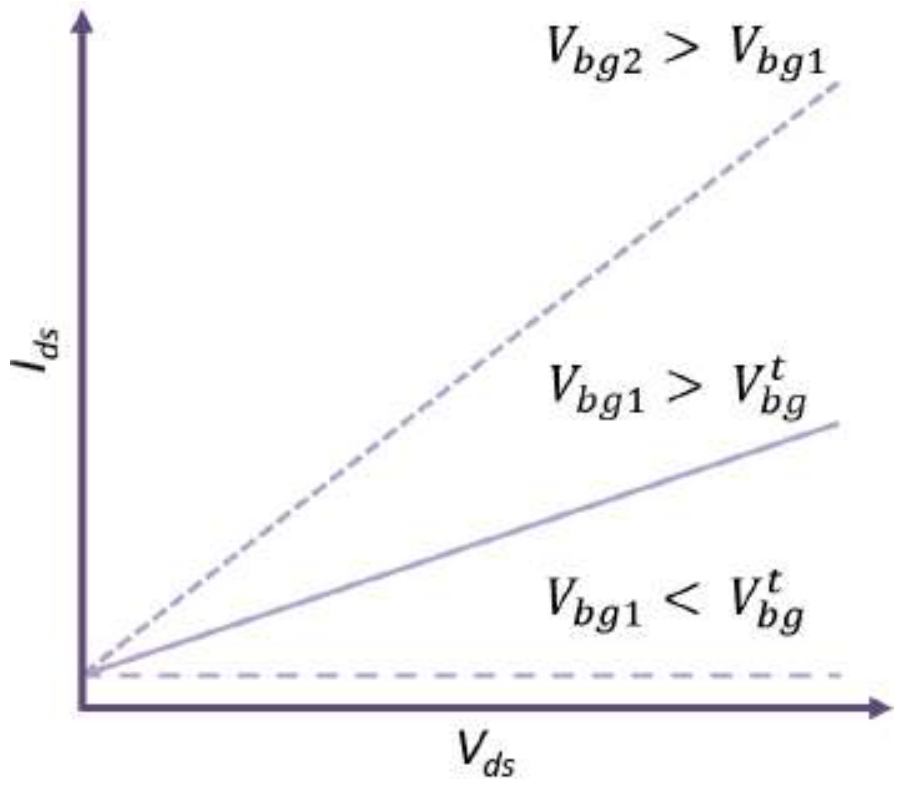

(b)

Figure 1.4: (a) Cross-section and circuit symbol of an n-type MOSFET. This figure is referenced from [4]; (b) The behavior of the device in the linear regime. When $V_{bg} < V_{bg}^t$, the device is turned off. When $V_{bg} > V_{bg}^t$ the transistor is switched into the on-state. The current maintains a linear regime and continues to increase until the saturation mode.

phototransistors and other electronics.

A typical n-type metal-oxide-semiconductor field-effect transistor (MOSFET) setup and the corresponding circuit symbol are displayed on Figure 1.4(a). The nMOSFET consists of a source and a drain, two highly conducting n-type semiconductor regions, which are isolated from the p-type substrate by reversed-biased p-n diodes. A metal or polycrystalline gate is separated from the semiconductor by the gate oxide. The electrons at the oxide-semiconductor interface are concentrated in a thin (~10 nm thick) "inversion" layer [4]. By now, most MOSFETs are "enhancement-mode" devices. While a minimum requirement for amplification of electrical signals is power gain, a device with both voltage and current gain is highly desirable in circuit electronics. The behavior of the device in the linear regime is displayed in Figure 1.4(b). $I_{ds}$ is weakly dependent upon $V_{ds}$ and controlled primarily by the gate voltage. When the back-gate voltage $V_{bg}$ is less than the threshold voltage $V_{bg}^t$, the device is not turned on and does not conduct. With sufficient drain-source voltage and a $V_{bg} > V_{bg}^t$ and $V_{ds} < V_{bg} - V_{bg}^t$, the device will be switched on and maintain an ohmic mode. When $V_{bg} > V_{bg}^t$ and $V_{ds} \geqq V_{bg} - V_{bg}^t$, due to the high excitation voltage, the electrons spread out and conduction is not through a narrow channel but through a broader, two or three dimensional current distribution extending away from the interface and deeper in substrate. As the drain-source voltage increases, this creates a pinch-off at the inversion layer near the drain and $I_{ds}$ reaches a maximum and saturates at the saturation mode.

Driven by scaling transistors to smaller dimensions, the leading-edge processors of semiconductor device fabrication for silicon-based MOSFETs is limited to 10-nm node technology. As the channel length becomes very short, new physical effects arise. For example, carrier transport in the active mode may be dominated by velocity saturation. The saturation drain-source current is more linear than quadratic in $V_{bg}$. At even shorter lengths, carriers transport in ballistic regime and the carries traveling at an injection velocity that may exceed the saturation velocity approaching the Fermi velocity at high inversion charge density. Meanwhile, reduction in dimension will soon approach limits due to statistical and quantum effects and create difficulties with heat dissipation [16]. New materials and concepts on devices are eminently investigated for industrial needs. In particular, 2D semiconducting TMD materials are appealing for processability and lack of short-channel effects which can impede device performance [17, 18].

Analogous to traditional heterostructures, layered transition metal dichalcogenide heterostructures can be designed and built by assembling individual single or multiple layers flakes into stacked structures. The performance of these heterostructures, especially as the film thickness is reduced towards a single atomic layer, gets compromised due to the interlayer coupling at the interface, composition variation and disorder localization. The use of TMDs in FETs was firstly reported by V. Podzorov group in 2004 indicating the low-field threshold electric field and high charge carrier mobility up to 500 $cm^2V^{-1}$ $m^{-1}$ for the p-type conductivity in $WSe_2$ based FETs at room temperature [19]. This result was soon followed by devices based on thin layers of $MoS_2$ and other TMDs with a back-gated or top-gated configuration. A typical back-gated field-effect transistor setup is depicted in Figure 1.5, a semiconducting channel region is connected to the source and drain electrodes, and separated by a dielectric layer from a gate electrode. The current flowing between the drain and source is tuned by modulating the gating voltage.

In 2D TMD materials, typically we are interested in the resistivity, on/off ratio, so-called subthreshold swing and carrier mobility of our samples. For digital logic transistors, desirable properties are high charge carrier mobilities for fast operation (a high current on/off ratio of $10^4 \sim 10^7$ are generally required for effective switching), high conductivity and low off-state conductance for low power consumption during operation [11]. Transport and scattering of the carriers are confined to the plane of the layers. The carrier mobility is mainly affected by scattering mechanisms [20, 21], which is summarized as: (1) acoustic and optical phonon scattering; (2) Coulomb scattering by charged impurities; (3) surface interface phonon scattering; and (4) roughness scattering. The carrier mobility is also impacted by layer thickness, carrier density, temperature, effective carrier mass, electronic band structure and phonon band structure. Additionally, the Schottky barriers formed at the interface between the metal contacts and the TMDs sample interface have huge influence on the mobility during the mobility measurement, as shown for a multilayered $MoS_2$ transistor with scandium contacts [22]. Phonon scattering plays an increasing role in the mobility with increasing temperature, particularly the acoustic phonon dominates at low temperatures (T $< 100$ K) while the optical component dominates at high temperatures ( T $> 100$ K) when the sample is clean [23]. However, if the sample is not clean and it contains charged impurities at the interface, like gas adsorbates, chemical residues, and dangling bonds on the surface, the random charged impurities' Coulomb scattering, surface phonon scattering or roughness scattering can dominate over the phonon scattering and critically impede the carrier mobility of van der Waals material as it is exfoliated down to a single atomic layer, as it is for graphene[24]. Engineering the dielectric environment can enhance mobilities, as has been demonstrated for graphene and $MoS_2$ [1].

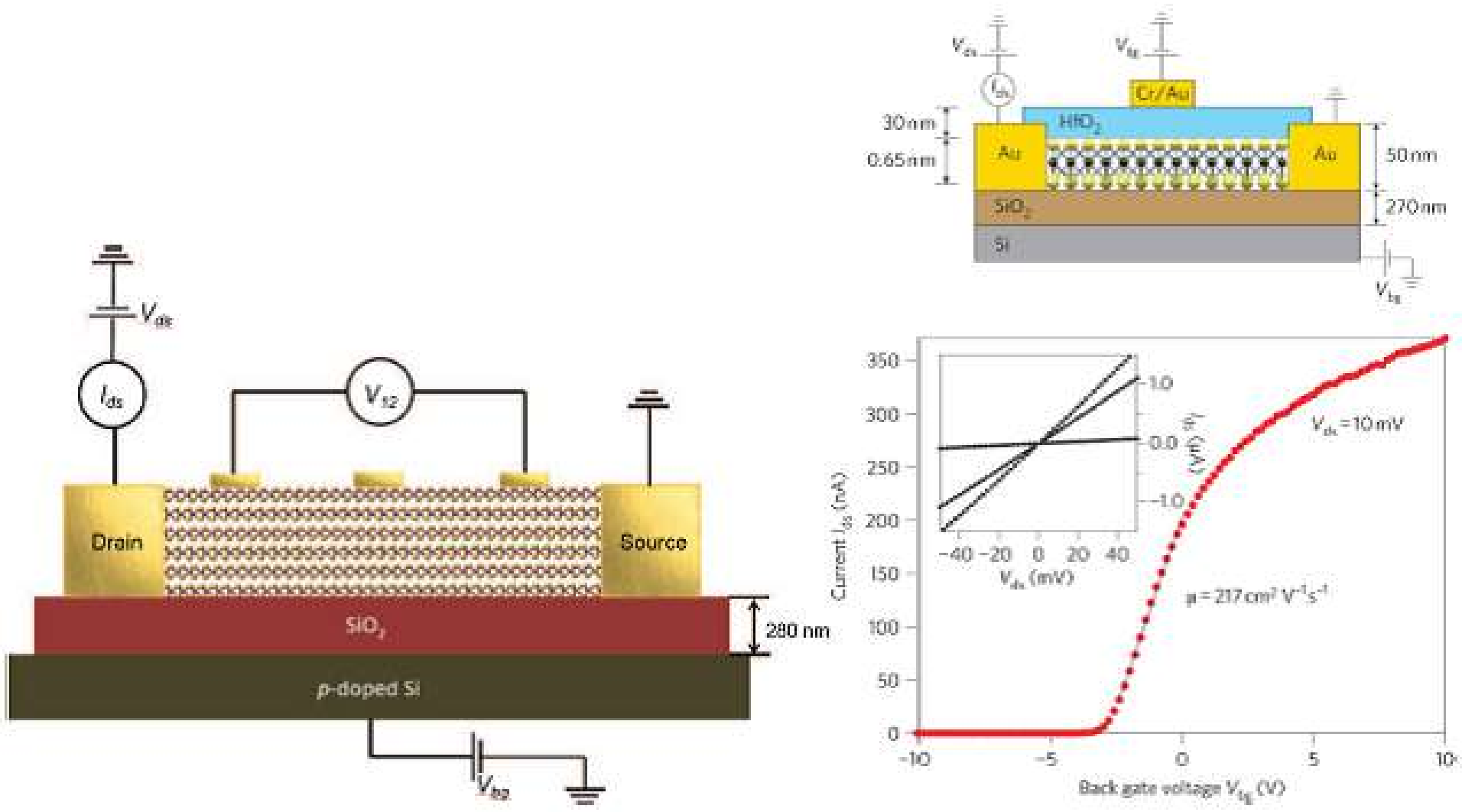

Figure 1.5: Skematic of a typical back-gated field-effect transistor. A few-layered TMDC flake is connected by source and drain electrodes sitting on top of 280-nm $SiO_2$ with voltage $V_{ds}$ and drain-source current $I_{ds}$. The three electrodes between those two are for possible electrical potential measurement ($V_{12}$) in a four-terminal measurement. The p-doped Si underneath the $SiO_2$ is the gate through which one applies a gate voltage $V_{bg}$.

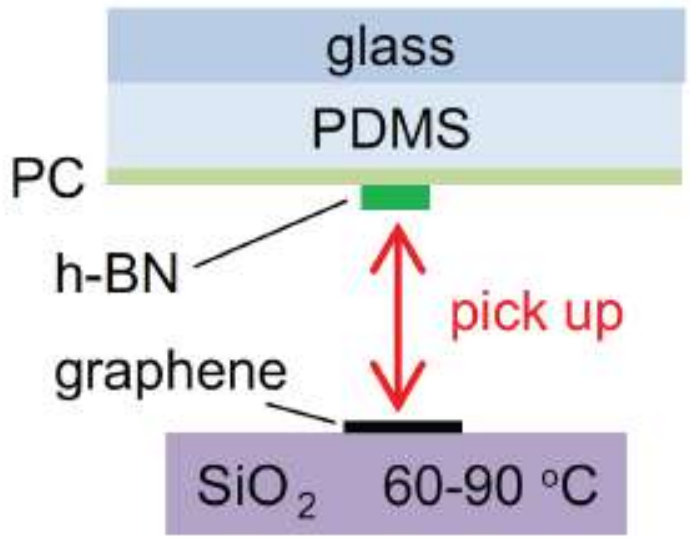

Figure 1.6: Side view of the fast pick up process using glass slide to pick up a graphene flake [5].

Furthermore, we can lower the Schottky barrier by changing the electrode contact material ( Ti, Au, Sc, Pd, Mo, Ni, etc.) or even by chemically changing the phase of TMDC where the contacts land on [25].

Here are a few steps of stacking few-layered TMD flakes together, as hexagonal boron nitride ($h-$BN) and graphene for example. TMDs flakes were mechanically exfoliated through the micromechanical cleavage technique onto 270-nm thick $SiO_2$ dielectric layers onto $p-$doped Si. The thickness of TMDs layer was later confirmed by Atomic Force Microspcopy (AFM). For two-layered heterostructures, one flake could be transferred onto the top of the other flake by PVA/PMMA transfer technique [26]. For multilayer heterostructure stacks like $h-$BN encapsulated devices, we will use the fast pick up techique from Ref. [5]. On top of a polycarbonate (PC) thin film we exfoliate $h-$BN crystals. By optical microscope we select an $h-$BN flake with a thickness ranging from 20 nm to 50 nm as determined by optical contrast. This $h-$BN layer would serve as the top layer of the stack. Then the PC film is attached onto the polydimethylsiloxane (PDMS) on a glass slide with the $h-$BN facing up. Flip the glass slide upside down to let the h-BN face down as displayed in Figure 1.6. With the help of micromanipulator, we can accurately align the $h-$BN and graphene ( or any other TMD flake) on a $SiO_2$ /Si substrate then bring them close to contact. As the temperature of the $SiQ_2$ wafer increases to values between 60 °C and 90 °C, the PC film expands until the contact of the flakes is made gradually. We switch off the heater in turn causing the PC film to slowly retract from substrate. In this way we can pick up another crystal to form a multilayer heterostructure as we desire. The whole stack can be released on to a substrate containing graphene or a TMD flake by heating the substrate up to ∼150 °C in order to melt the PC film. The stack is fabricated into an electronic device using standard electron beam lithography and electron beam evaporation techniques to couple the stacks to metal contacts (for instance, Ti/Au or Cr/Au). After the transfer, as well as the final deposition of metal, the samples are annealed at 200 $\sim$ 300°C for 2 $\sim$ 3 hours in forming gas ($H_2/N_2$ gas) then annealed at 120 °C for 24 hours to remove the moisture.

## 1.2 Weyl/Dirac Semimetals

Semimetallic TMDs including $WTe_2$, $MoTe_2$ and their mixture (W, Mo)$Te_2$ were introduced as candidates for a novel type of Weyl semi-metallic state called Weyl type-II [6, 27]. The Weyl semimetal (WSM) [28, 29, 30, 31, 32] is a topological semi-metallic state in which the electronic bands disperse linearly in the three dimensional momentum space through a node, called a Weyl point. Due to the lack of time-reversal or inversion symmetry these Weyl points act as topological charges that is, as sources and drains of Berry-phase. These materials with exotic transport properties and surface states that exhibit extremely large magnetoresistance (MR), even much larger than the traditional giant MR in thin film metals, Cr-based chalcogenide spinels and Mn-based pervoskites [33, 34, 35]. For example, they frequently exhibit extremely large magnetoresistance [30] (MR) which surpasses by far the one observed in thin metallic films displaying the giant magnetoresistive effect, or the magnitude of the MR observed in Cr-based chalcogenide spinels or in Mn-based pervoskites [33, 34, 35]. They have attracted attention due to the enormous potential for future applications in engineering and information technology like sensitive magnetic sensors and magnetic random access memories [36, 37]. Distinct from conventional type-I WSMs (e.g. TaAs, NbAs, NbP, TaP), which have linear dispersion along all momentum direction and the electron and hole pocket touching at the Fermi level as shown in left plane of Figure 1.7, the Weyl points in type-II WSMs result from linear touching points at the boundary between electron and hole Fermi surface pockets with strongly tilted Weyl cone , as depicted in the right plane.

Meanwhile, the exotic transport properties [38, 39, 40, 65, 66, 41, 58, 109] of these type-II WSM candidates have attracted considerable attention due to the observation of extremely large and non-saturating MR in $WTe_2$ [65, 66, 42]. The coexistence of multiple electron and hole pockets of approximately the same size were observed in $WTe_2$ by Angle-Resolved Photo-emission Spectroscopy (ARPES) [43, 44] measurements, suggesting near perfect carrier compensation at low temperatures, which is partially supported by the geometry of the Fermi surface extracted from Shubnikov de Haas(SdH) quantum oscillations [45] and also by a two-band analysis of the Hall-effect[93] . Nevertheless, the two band analysis of the Hall-effect indicates that the fraction of holes is $\sim 0.9$ times that of electrons, which contradicts the analysis and the claims in favor of perfect compensation in Ref. [65].

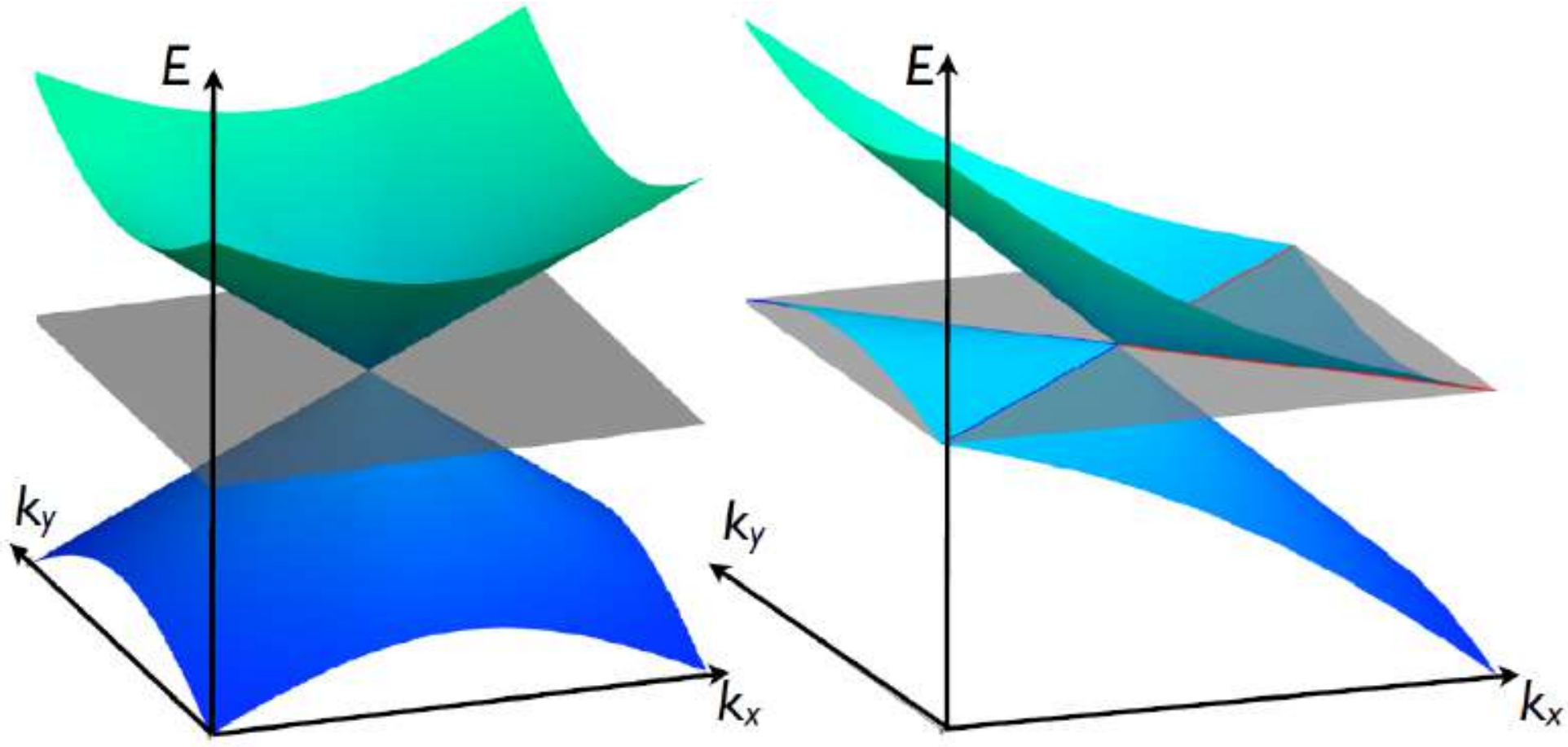

Figure 1.7: Two types of Weyl semimetals. Left panel: Type I Weyl point with point like Fermi surfaces. Right panel: Type II Weyl point linear touching at the boundry between electron and hole pockets. Grey plane refers as the position of Fermi level, and the blue(red) lines indicate the boundaries of hole (electron) pockets. This figure is taken from Ref.[6].

Our study of semimetallic TMDs focuses on the semimetallic compound $MoTe_2$ and $PtTe_2$. $MoTe_2$ is a layered transition metal dichalcogenide which can crystallize into two different structures: the $2H$ or the hexagonal $\alpha$-phase and the $1T'$ or the monoclinic $\beta$-phase which undergoes a structural phase-transition below $\sim 240$ K into an orthorhombic phase which is frequently called the $T_d$-phase. Throughout this chapter we will label the low temperature orthorhombic phase of $MoTe_2$ as the $\gamma$ phase or simply $\gamma$-$MoTe_2$. The $2H$ phase is semiconducting with the Mo atoms being trigonal-prismatically coordinated by Te atoms forming layers stacked upon one another with weak, or van der Waals (vdW) like, inter-planar coupling. The $\beta$ and $\gamma$ phases are semi-metallic with pseudo-hexagonal layers composed of zigzag metallic chains. The $\beta$ phase crystallizes in a monoclinic space group while the $\gamma$ phase results from the direct stacking of layers resulting in a higher-symmetry orthorhombic structure [86]. In contrast, $WTe_2$ is known to crystallize only in an orthorhombic phase which is crystallographically akin to $\gamma$-$MoTe_2$. Currently, $\gamma$-$MoTe_2$ is attracting a lot of interest due to its extremely large magnetoresistivity [66], the observation of pressure-driven superconductivity [86], and the aforementioned prediction of a novel type of WSM state [6, 71]. Recent ARPES measurements [32, 79, 78, 81, 46] claim to observe Fermi arcs at the surface of $MoTe_2$ and to reveal a bulk electronic electronic structure that is in broad agreement with band structure calculations and therefore with the prediction of the type-II WSM state in this system. Furthermore, impurity-dependent superconductivity accompanied by a non-trivial Berry phase was observed in gamma MoTe2 at ambient pressure [42].

The electronic structure of the TMDs belonging to the orthorhombic and non-centrosymmetric $Pmn2_1$ space group, e.g. $WTe_2$, were recently recognized as candidates for possible topologically non-trivial electronic states. For instance, their monolayer electronic bands were proposed to be characterized by a non-trivial $Z_2 = 1$ topological invariant based on the parity of their valence bands, making their monolayers good candidates for a quantum spin Hall insulating ground-state [61]. This state is characterized by helical edge states that are protected by time-reversal symmetry from both localization and elastic backscattering. Hence, these compounds could provide a platform for realizing low dissipation quantum electronics and spintronics [61, 62]. However, the majority of gapped TMDs, such as semiconducting $MoS_2$ or $WSe_2$, crystallize either in a trigonal prismatic coordination or in a triclinic structure with octahedral coordination [63, 64] as is the case of $ReS_2$. Those crystallizing in the aforementioned orthorhombic phase, e.g. $WTe_2$, are semi-metals albeit displaying remarkable transport properties such as an enormous, non-saturating magnetoresistivity [65]. Strain is predicted to open a band gap [61] in $WTe_2$, which might make it suitable for device development. In fact, simple exfoliation of its isostructural $\gamma-MoTe_2$ compound (where $\gamma$ refers to the orthorhombic semi-metallic phase) into thin atomic layers was claimed to induce a band gap [66] in the absence of strain. Such a transition would contrast with band structure calculations finding that $WTe_2$ should remain semi-metallic when exfoliated down to a single atomic layer [67]. The insulating behavior reported for a few atomic layers of $WTe_2$ was ascribed to an increase in disorder due to its chemical instability in the presence of humidity which would induce Anderson localization [68], although more recently it was claimed to be intrinsic from transport measurements on encapsulated few-layered samples [69]. Orthorhombic $\gamma-MoTe_2$ and its isostructural compound $WTe_2$ were also claimed, based on density functional theory calculations, to belong to a new class of Weyl semi-metals, called type-II, which is characterized by a linear touching between hole and electron Fermi surface pockets [70, 71, 72, 75]. As for conventional Weyl points [73, 74], these

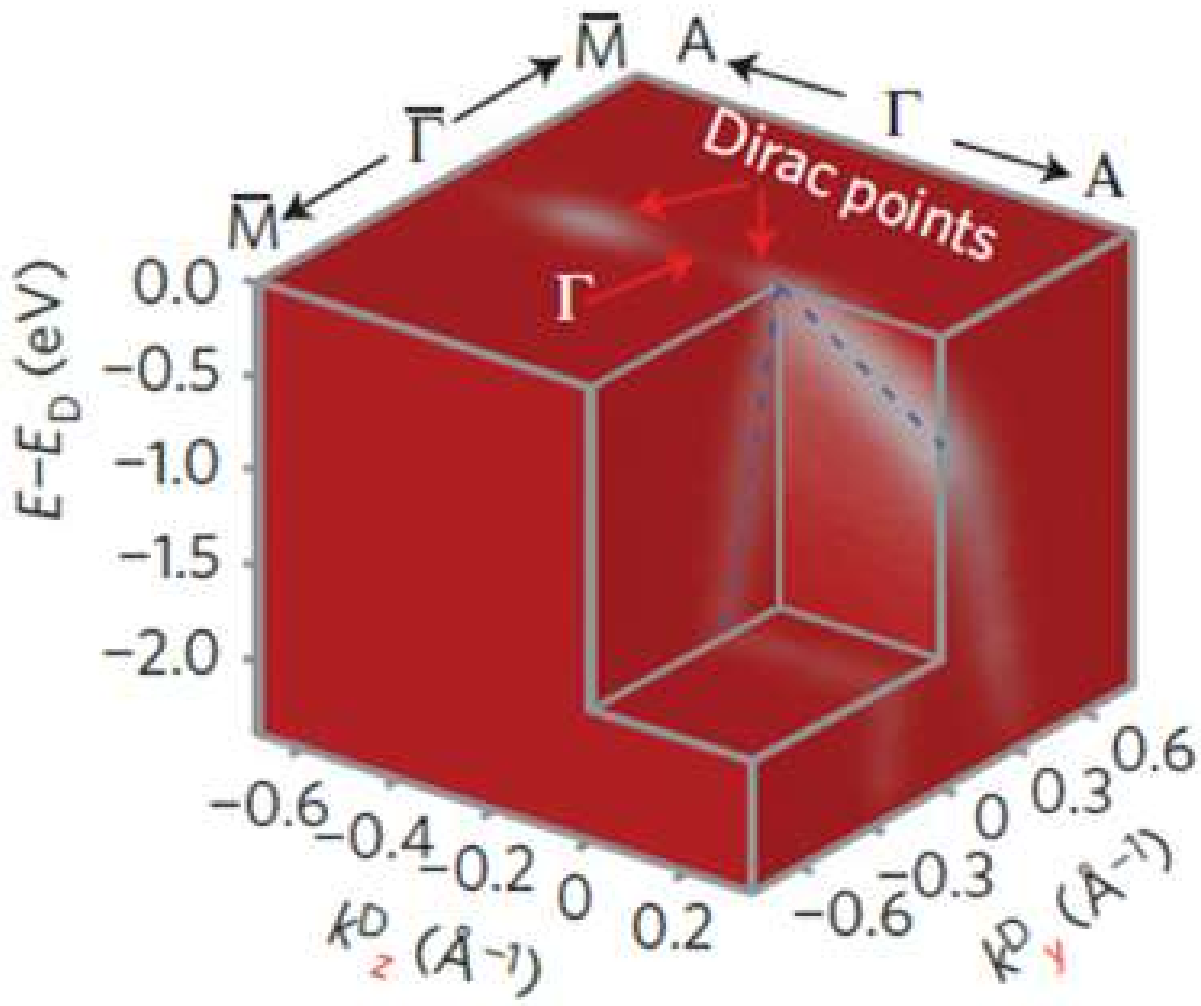

Figure 1.8: $Cd_3As_2$ ARPES projection of the 3D Dirac fermions into the $(k_y, k_z, E)$ space. 3-dimensional intensity plot of the photoemission spectra at the Dirac point in the $(k_y, k_z, E)$ space, clearly shows an enlongated Dirac cone along the $k_z$ direction [7].

Weyl type-II points would also act as topological charges associated with singularities, i.e., sources and sinks, of Berry-phase pseudospin [70, 71, 72, 75] which could lead to anomalous transport properties. A series of recent angle-resolved photoemission spectroscopy (ARPES) measurements [76, 77, 78, 79, 80, 81, 82, 83] claim to observe a good overall agreement with these predictions. These studies observe the band crossings predicted to produce the Weyl type-II points, which would be located slightly above the Fermi-level, as well as the Fermi arcs projected on the surface of this compound [76, 77, 78, 79, 80, 81, 82, 83].

Unlike 2D Dirac fermions in graphene, the Fermi surface of a Dirac semimetal consists entirely of such point like Dirac cones degeneracies. When time-reversal or inversion symmetry is broken, a spin-degenerate Dirac fermion splits into two Weyl fermions, and the topological surface states (TSS) evolve from a closed Fermi surface to open Fermi arcs [50]. Similar to Weyl fermions, type I Dirac fermions disperse linearly along all momentum directions, while type II Dirac fermions are characterized by strongly tilted Dirac cones. So far several Dirac semimetal materials such as $Cd_3As_2$ , $Na_3Bi$ are theoretically proposed and experimentally confirmed [51, 7, 52]. For example, bulk $Cd_3As_2$ was proposed to have dirac point and it is confirmed with the observation of Dirac fermions via Angle-Resolved Photoemission spectroscopy (ARPES) , as Figure 1.8 , which proves

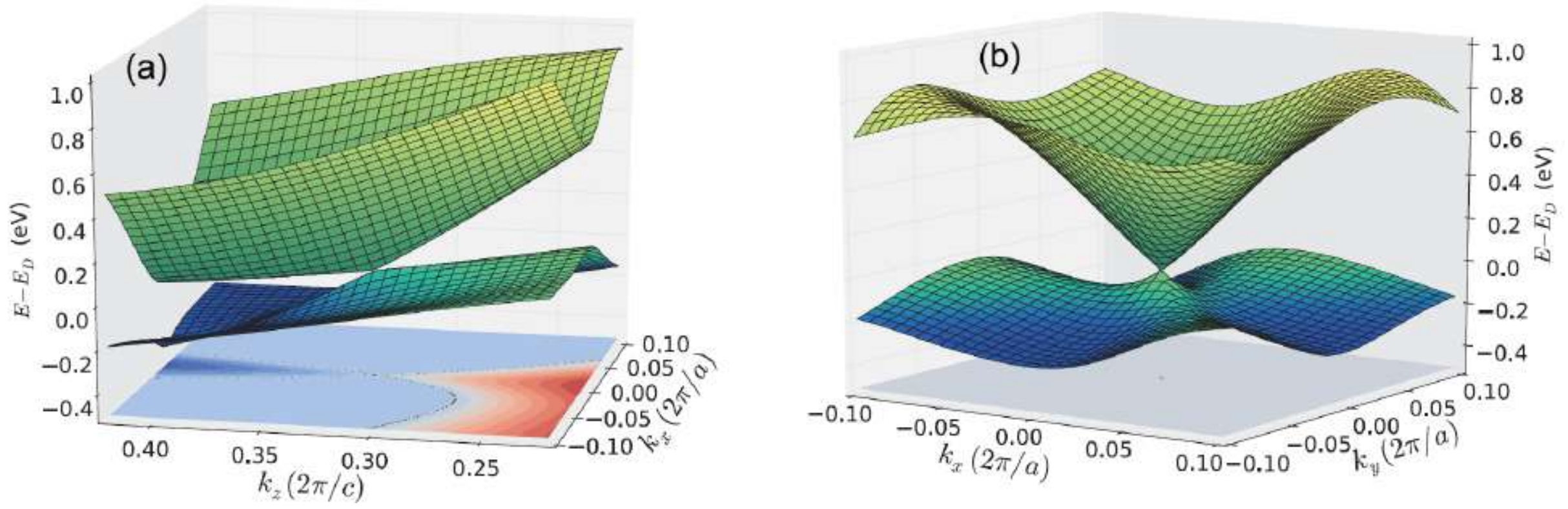

Figure 1.9: Theoretical calculations of type-II Dirac points in $PtSe_2$-type materials. (a) and (b) Three-dimentional band structure on the (a) $k_y = 0$ plane and (b) $k_z = k_z^c$ planes around the Dirac points [8].

that $Cd_3As_2$ is a model system of the 3D topological Dirac semimetal.

Similar to Weyl semimetals, Dirac semimetals give rise to interesting properties. Ultrahigh mobility and giant magnetoresistance have been observed in $Cd_3As_2$ and $Na_3Bi$ [7, 52]. Ref. [52] claims robust linear and large magnetoresistance What's more, it reveals that both the conductivity and resistivity tensors exhibit robust anormalies in magnetic fields [52] . However, only conventional type-I Dirac fermions exist in these materials.

When the time-reversal or inversion symmetry is not broken, some materials have unique tilted Dirac cone which is substantially different from the conventional type-I Dirac semimetals. Using first-principles calculations, one paper reports the existence of type-II Dirac fermions in $PtSe_2$-class TMD materials, like $PtSe_2$, $PtTe_2$, $PdTe_2$, and $PtBi_2$ [8]. Similar topological property is also expected in bulk $PtSe_2$-type materials. The calculation reveals its bulk type-II Dirac fermions as displayed in Figure 1.9, the two crossing bands exhibit linear dispersion along both the in-plane and out-of-pane directions with strongly tilted Dirac cones along the $k_z$ direction.

$PtTe_2$, similar to $PtSe_2$, crystalizes in the trigonal ($1T$) structure which is composed of edge-shared octahedra as shown in Figure 1.10 . A recent DFT calculations and ARPES study [50] on this compound discovered Lorentz-violating type-II Dirac fermions in single crystal $PtTe_2$. Their theoretical calculation exhibits the 3D plot of the electron and hole pockets at the Dirac point. And via ARPES, they observed a strongly tilted Dirac cone at the D point is revealed in the dispersion as a function of $k_z$ , which inplies Lorentz-violating type-II Dirac fermions in single crystal $PtTe_2$.

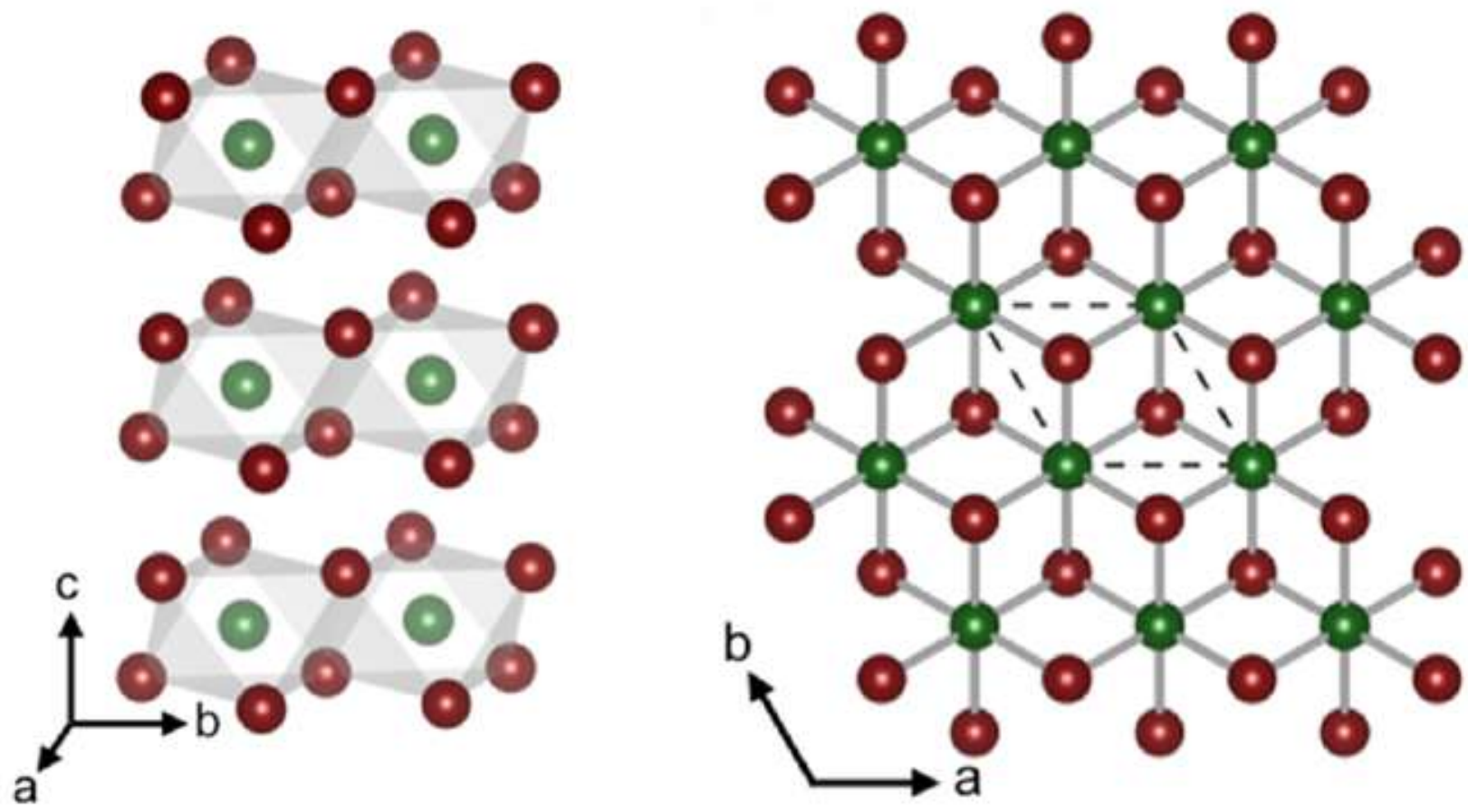

Figure 1.10: Side view (left panel) and top view (right panel) of the crystallographic structure of $PtTe_2$ [9]. Green symbols are Pt atoms while red are Te atoms. The dashed line indicates the unit cell.

We also investigated the electronic structure and transport properties via the temperature and angular dependence of the Shubnikov-de Haas (SdH) and de Haas-van Alphen (dHvA) effect in single crystals of the semi-metallic platinum ditelluride $PtTe_2$. Understanding of the bulk topological properties of this compound will help us predict the physical properties of heterostructures fabricated from single- or few atomic layers of this compound, which could provide huge impact on the quantum spin hall devices [53] engineering and technology.

# CHAPTER 2

# HALL EFFECT IN SEMIMETALLIC MOTE$_2$

Currently, $\gamma-$MoTe$_2$ is attracting a lot of interest due to its extremely large magnetoresistivity [66], the observation of pressure-driven superconductivit [86], and the aforementioned prediction for a Weyl type-II semi-metallic state [6, 71]. Recent ARPES measurements [80, 79, 78, 81, 46] claim to observe Fermi arcs at the surface of $\gamma-$MoTe$_2$ and to uncover a bulk electronic structure in broad agreement with the band structure calculations, and consequently with the predicted type-II WSM state in this system. The observation of a field-dependent Berry phase through quantum oscillatory phenomena indicates that its electronic structure is affected by Zeeman effect and a it reveals a Fermi surface topography markedly distinct from the calculated one [42]. Furthermore, the observation of superconductivity in $\gamma-$MoTe$_2$ at ambient pressure, with an impurity-dependent transition temperature, points towards an unconventional and probably topological superconducting state [42].

Therefore, understanding the electronic structure of $\gamma-$MoTe$_2$ is a critical step towards clarifying its topological nature, the properties of its superconducting state, and the mechanism leading to its enormous magnetoresistivity. Here, we investigated the electronic structure of $\gamma-$MoTe$_2$ through a systematic temperature dependent study of the Hall-effect in high-quality $\gamma-$MoTe$_2$ single-crystals. Our goal is to determine fundamental electrical transport variables such as the density of carriers and their mobilities as well as their evolution as a function of the temperature. Knowledge about these variables is crucial for understanding the electronic structure of this compound and for clarifying whether near perfect charge compensation, which is claimed to lead to the enormous magnetoresistivity [65] of WTe$_2$, is pertinent also to $\gamma-$MoTe$_2$.

## 2.1 Methods and Experimental Results

Single-crystals of MoTe$_2$ were grown at temperatures above 900 °C *via* a flux method using excess Te [42]. The magnetic field dependence of the longitudinal resistivity $\rho_{xx}$ and of the hall resistivity $\rho_{xy} = V_H t / I_{xx}$, where $V_H$, $t$ and $I_{xx}$ are the Hall voltage, sample thickness, and electrical current respectively, were measured under magnetic fields up to $H = 9$ T and under temperatures ranging from $T = 2$ K to 100 K. Measurements were performed in a Quantum Design Physical Properties Measurement System using a conventional four-terminal configuration when measuring the resistivity, or a six-wire one when studying the Hall-effect. The electrical current was applied along the crystallographic $a-$axis. Any mixture of $\rho_{xx}$ and $\rho_{xy}$ was corrected by reversing the direction of the applied magnetic field: $\rho_{xy}$ is determined by subtracting the negative field trace from the positive field one.

A careful fitting of the Hall conductivity $\sigma_{xy} = -\rho_{xy}/(\rho_{xy}^2 + \rho_{xx}^2)$ and of the longitudinal conductivity $\sigma_{xx} = \rho_{xx}/(\rho_{xy}^2 + \rho_{xx}^2)$ based on the two-carrier model [65], yields the densities and mobilities of both electrons and holes as a function of the temperature. As we discuss below, we observe an excellent agreement between our experimental data and the two-carrier model particularly at low temperatures confirming the coexistence of electrons and holes in $\gamma-$MoTe$_2$. A sharp change in carrier densities is observed at temperatures between 30 and 70 K indicating a possible electronic phase-transition. The density of holes increases dramatically below $\sim 40$ K, reaching parity with the electron density below $\sim 15$ K. Therefore, near perfect charge-carrier compensation is an important ingredient for the colossal magnetoresistivity observed in $\gamma-$MoTe$_2$ at low temperatures.

Figure 2.1 (a) displays $\rho_{xx}$ measured under zero magnetic field as a function of the temperature, with $T$ ranging from 2 to 300 K. For this particular crystal one obtains a residual resistivity ratio $RRR \equiv \rho(300\text{K})/\rho(2\text{K}) = 1064$, which is nearly one order of magnitude larger than previously reported values [38, 66, 86, 39]. The observed hysteresis around 250 K between the warm-up and the cool-down curves is associated with the structural phase-transition from the monoclinic $1T'$ to the orthorhombic $T_d$ structure. The upper inset displays a fit of $\rho_{xx}(T)$ to a combination of Fermi liquid and electron-phonon scattering mechanisms, or a fit to $\rho_{xx}(T) = \rho_0 + aT^2 + bT^5$ where $\rho_0 = \rho(T = 0$ K) with $a$ and $b$ being fitting parameters. Figures 2.1(b) and 2.1(c) display the magnetic field dependence of both $\rho_{xx}$ and $\rho_{xy}$, respectively. $\rho_{xx}$ follows a nearly quadratic dependence on magnetic field showing no signs of saturation. The resulting MR ratio $\Delta\text{MR} \equiv [\rho_{xx}(9\text{ T}) - \rho_{xx}(0\text{ T})]/\rho_{xx}(0\text{ T})$ reaches 74791 % at 2 K, which is the highest value reported so far for $\gamma-$MoTe$_2$ under $\mu_0 H = 9$ T [38, 66, 86]. In Ref. [42] , we include measurements up to $H = 65$ T in a lower quality crystal, which reveals a non-saturating $\Delta\text{MR} > 10^6$ %. As seen in Figs. 2.1(b) and 2.1(c), Shubnikov de Haas (SdH) oscillations are observed in both $\rho_{xx}(\mu_0 H)$ and $\rho_{xy}(\mu_0 H)$, respectively. Figure 2.1(d) shows the SdH oscillations superimposed onto the $\rho_{xx}$ trace taken at $T = 2$ K from a sample with a $RRR$ of approximately 1900 and a $\Delta$MR $\sim$ 265340 % at 9 T. $\rho_{xx}$ follows a near $(\mu_0 H)^2$ dependence as seen in the inset. Two frequencies $\alpha = 231$ T and $\beta = 242$ T are extracted from the Fast Fourier Transform of the oscillatory signal (bottom inset). A detailed analysis of the SdH effect can be found in Ref. [42]. The temperature dependence of $\rho_{xx}$ under several magnetic fields is shown in Figure 2.1(e). The observed behavior is similar to the one displayed by $WTe_2$. Namely, under a magnetic field $\rho_{xx}(T)$ essentially follows the zero-field curve until it reaches a minimum at a field dependent temperature $T^*$. Below $T^*$, $\rho_{xx}$ increases rapidly upon cooling. With increasing fields $T^*$ shifts to higher temperatures at a rate of 2.01 K/T.

Figure 2.2 displays $\rho_{xy}$ and $\rho_{xx}$ as functions of $\mu_0 H$ for selected temperatures ranging from $T = 2$ K to 90 K. Notice that $\rho_{xy}$ remains nearly unchanged as the temperature increases from 2 to 15 K. Its negative sign indicates that the electronic transport is dominated by electrons, or that the electrons have higher mobilities than the holes. Below $\sim$ 20 K, as well as above 70 K, $\rho_{xy}$ follows a linear dependence on $\mu_0 H$ for fields up to $\mu_0 H = 9$ T. Between 20 and 70 K, the Hall resistivity shows non-linear behavior, particularly at low fields, which is a clear indication for electrical conduction by both types of carriers. $\rho_{xx}$ on the other hand, shows a quadratic field dependence below $T \sim 45$ K which becomes linear at higher temperatures.

## 2.2 Isotropic Two-Carrier Model Analysis

Subsequently, we analyze the Hall response through the isotropic two-carrier model which was successfully used to describe the contributions of holes and electrons to the electrical transport properties of a number of compounds[54, 55, 93, 56, 57]. In this model, the conductivity tensor, in its complex representation, is given by: [65]

$$\boldsymbol{\sigma} = e\left[\frac{n_e\mu_e}{1+i\mu_e\mu_0 H} + \frac{n_h\mu_h}{1-i\mu_h\mu_0 H}\right], \tag{2.1}$$

Here, $n_e$ (or $n_h$) and $\mu_e$ (or $\mu_h$) are the densities of electrons (or holes) and the mobilities of electrons (or holes), respectively. To appropriately evaluate the carrier densities and their mobilities, we obtained the Hall conductivity $\sigma_{xy}$ and the longitudinal conductivity $\sigma_{xx}$ from the original experimental data, as previously defined, and as shown in Figure 2.3 (a) and 2.3 (b).

In the next step we fit $\sigma_{xy}$ and $\sigma_{xx}$ to the two-carrier model, where the field dependence of the conductivity tensor is given by [30, 56, 57]

$$\sigma_{xy} = \left[ n_e \mu_e^2 \frac{1}{1+(\mu_e \mu_0 H)^2} - n_h \mu_h^2 \frac{1}{1+(\mu_h \mu_0 H)^2} \right] e \mu_0 H, \tag{2.2}$$

$$\sigma_{xx} = \frac{n_e e \mu_e}{1+(\mu_e \mu_0 H)^2} + \frac{n_h e \mu_h}{1+(\mu_h \mu_0 H)^2}. \tag{2.3}$$

The fitting parameters from the Hall conductivity $\sigma_{xy}$ and longitudinal conductivity $\sigma_{xx}$ are listed in Table 2.1 and Table 2.2 , respectively.

Table 2.1: The fitting parameters for the Hall conductivity $\sigma_{xy}$ of $\gamma-MoTe_2$.

| Temperature (K) | $n_e (cm^{-3})$ | $\mu_e (cm^2/Vs)$ | $n_h (cm^{-3})$ | $\mu_h (cm^2/Vs)$ |
|---|---|---|---|---|
| 2 | $0.32229 \times 10^{20}$ | $5.82523 \times 10^4$ | $0.32244 \times 10^{20}$ | $1.58383 \times 10^4$ |
| 5 | $0.27281 \times 10^{20}$ | $5.45209 \times 10^4$ | $0.27401 \times 10^{20}$ | $1.37026 \times 10^4$ |
| 10 | $0.2759 \times 10^{20}$ | $2.76096 \times 10^4$ | $0.27382 \times 10^{20}$ | $0.95594 \times 10^4$ |
| 15 | $0.24258 \times 10^{20}$ | $1.26276 \times 10^4$ | $0.24072 \times 10^{20}$ | $0.4684 \times 10^4$ |
| 20 | $0.26401 \times 10^{20}$ | $0.56338 \times 10^4$ | $0.25757 \times 10^{20}$ | $1.26276 \times 10^4$ |
| 25 | $0.50865 \times 10^{20}$ | $0.25845 \times 10^4$ | $0.48507 \times 10^{20}$ | $0.17811 \times 10^4$ |
| 40 | $0.19302 \times 10^{20}$ | $0.10936 \times 10^4$ | $0.1761 \times 10^{19}$ | $0.2185 \times 10^4$ |
| 50 | $0.2221 \times 10^{20}$ | $0.05707 \times 10^4$ | $0.913 \times 10^{18}$ | $0.15397 \times 10^4$ |
| 70 | $0.9066 \times 10^{20}$ | $0.01282 \times 10^4$ | $0.214 \times 10^{17}$ | $0.20346 \times 10^4$ |
| 90 | $0.31685 \times 10^{20}$ | $0.01419 \times 10^4$ | $0.17 \times 10^{15}$ | $1.00427 \times 10^4$ |
| 100 | $0.13397 \times 10^{20}$ | $0.01882 \times 10^4$ | $0.33 \times 10^{15}$ | $0.37983 \times 10^4$ |

Table 2.2: The fitting parameters for the longitudinal conductivity $\sigma_{xx}$ of $\gamma-MoTe_2$.

| Temperature (K) | $n_e (cm^{-3})$ | $\mu_e (cm^2/Vs)$ | $n_h (cm^{-3})$ | $\mu_h (cm^2/Vs)$ |
|---|---|---|---|---|
| 2 | $0.57651 \times 10^{20}$ | $5.44709 \times 10^4$ | $0.81325 \times 10^{20}$ | $2.01521 \times 10^4$ |
| 5 | $0.50143 \times 10^{20}$ | $4.83227 \times 10^4$ | $0.78222 \times 10^{20}$ | $1.82115 \times 10^4$ |
| 10 | $0.59327 \times 10^{20}$ | $2.50465 \times 10^4$ | $0.47202 \times 10^{20}$ | $0.97992 \times 10^4$ |
| 15 | $0.46774 \times 10^{20}$ | $1.14613 \times 10^4$ | $0.59327 \times 10^{20}$ | $0.47519 \times 10^4$ |
| 20 | $0.47029 \times 10^{20}$ | $0.57413 \times 10^4$ | $0.52388 \times 10^{20}$ | $0.24776 \times 10^4$ |
| 25 | $0.39632 \times 10^{20}$ | $0.34612 \times 10^4$ | $0.65207 \times 10^{20}$ | $0.15702 \times 10^4$ |
| 40 | $1.09666 \times 10^{20}$ | $0.07001 \times 10^4$ | $0.5382 \times 10^{19}$ | $0.20386 \times 10^4$ |
| 50 | $1.15113 \times 10^{20}$ | $0.04697 \times 10^4$ | $0.183 \times 10^{19}$ | $0.18044 \times 10^4$ |
| 70 | $1.14942 \times 10^{20}$ | $0.02737 \times 10^4$ | $0.229 \times 10^{20}$ | $0.18848 \times 10^4$ |
| 90 | $1.14462 \times 10^{20}$ | $0.01884 \times 10^4$ | $0.619 \times 10^{16}$ | $0.20618 \times 10^4$ |
| 100 | $1.13695 \times 10^{20}$ | $0.01639 \times 10^4$ | $0.378 \times 10^{17}$ | $0.2255 \times 10^4$ |

The results of the fittings of $\sigma_{xy}$ and $\sigma_{xx}$ to Eqs.2.2 and 2.3 are displayed in Figs. 2.3 (a) and 2.3 (b), respectively. The excellent agreement between our experimental data and the fittings to the two-carrier model, over a broad range of temperatures (see Fig. 2.3), confirms the coexistence of electrons and holes in $\gamma-\text{MoTe}_2$.

The densities and mobilities of electrons and holes, as extracted from the fits of $\sigma_{xy}$ (as well as $\sigma_{xx}$) to the two-carrier model, are displayed in Figs. 2.4 (a) and 2.4 (b). The parameters extracted from $\sigma_{xy}$ and $\sigma_{xx}$ are close in value over a broad range of temperatures, particularly in what concerns the mobility, as illustrated by Figure 2.4 (b) and its inset. In Figure 2.4 (a) notice the drastic increase in $n_h$ as $T$ is reduced below $\sim 40$ K while $n_e$ remains nearly constant. At low-temperatures, the densities of electrons and holes become comparable while the mobility of the electrons becomes slightly higher than that of the holes. An equal concentration of electrons and holes is consistent with the observation of a linear Hall resistivity below $\sim 20$ K. Figure 2.4 (b) presents electron and hole mobilities as extracted from either $\sigma_{xy}$ or $\sigma_{xx}$ (inset). Both $\mu_e$ and $\mu_h$ increase dramatically, i.e. by one order of magnitude, as $T$ decreases below $\sim 40$ K. At $T = 2$ K, the extracted mobilities are $\mu_e \sim 5.8 \times 10^4$ cm$^2$/Vs and $\mu_h \sim 1.6 \times 10^4$ cm$^2$/Vs.

Assuming only a single band, the Hall mobility is determined by $\mu_{\text{H}}(T) = |R_{\text{H}}(T)|/\rho_{xx}(T)$ [56] while the Hall coefficient $R_{\text{H}} = \rho_{xy}/B$. The temperature dependence of the Hall mobility $\mu_{\text{H}}$ and the Hall coefficient $R_{\text{H}}$ are depicted in Figure 2.5 $R_{\text{H}}$(0 T) has been determined from the initial slope of $\rho_{xy}(B)$. The overlap of $R_{\text{H}}$(0 T) and $R_{\text{H}}$(9 T) especially at low temperatures indicates that below 30 K, $R_{\text{H}}$ remains nearly constant as the magnetic field is applied. For clarity, $\mu_{\text{H}}$(0 T) and $\mu_{\text{H}}$(9 T) are almost identical at various temperatures. It is worthy pointing out that we obtained relatively high Hall mobility of $\gamma$-MoT$_2$ at 2 K $\mu_{\text{H}} \sim 2.3 \times 10^4 cm^2V^{-1}s^{-1}$. The $\mu_{\text{H}}(T)$ high mobility remains approximately $10^4 cm^2V^{-1}s^{-1}$ below 10 K and exhibits a drastic downtrend as the temperatures increases above 20 K.

At $T = 2$ K the Hall mobility, or the ratio between the Hall constant as extracted from a linear fit of the Hall-effect and the resistivity, is $\mu_H \sim 2.3 \times 10^4$ cm$^2$/Vs, which is comparable to the values extracted from the two-carrier model. The extracted ratio between carrier densities $n_h/n_e$, and the ratio between carrier mobilities, or $\mu_h/\mu_e$ are displayed in Figs. 2.4 (c) and 2.4 (d), respectively. $n_h/n_e \simeq 1$ for temperatures below $T \simeq 40$ K but it decreases quickly as $T$ increases beyond this value. The mobility ratio indicates that the hole-mobility considerably exceeds the electron one at high temperatures. Finally, Fig. 2.4 (e) shows the Hall resistivity $\rho_{xy}$ as a function of $T$ under magnetic fields ranging from $\mu_0 H = 1$ to 9 T. Below a field-dependent temperature $T^{\mathrm{neg}}$, $\rho_{xy}$ shows a pronounced decrement towards negative values. It turns out that below $T^{\mathrm{neg}}$ one observes a sharp increase in the mobilities of both carriers and in the density ratio $n_h/n_e$. These observations, coupled to an anomaly seen in the heat capacity [42] around these temperatures, suggest a possible electronic phase-transition analogous to the $T$-dependent Lifshitz transition [106] reported for $WTe_2$.

## 2.3 Summary

In summary, our results indicate the coexistence of electrons and holes in $\gamma-MoTe_2$, with the density of holes increasing dramatically below $\sim 40$ K and finally reaching parity with the electron one below $\sim 15$ K. An equal density of electrons and holes is consistent with the observation of a linear Hall-resistivity below $T = 20$ K which implies that $\gamma-MoTe_2$ is in fact better compensated than $WTe_2$ [93, 41] whose Hall resistivity becomes non-linear at low temperatures. Hence, the extremely large and non-saturating magnetoresistance seen in $\gamma-MoTe_2$ should be primarily ascribed to the nearly perfect compensation between the densities of electrons and holes. The analysis of the Hall-effect through the two-carrier model yields high electron- and hole-mobilities, that is exceeding $10^4 \mathrm{cm}^2/\mathrm{Vs}$ at $T = 2$ K. Anomalies observed in the carrier densities, carrier mobilities, and in the heat capacity as a function of the temperature, suggest an electronic crossover or perhaps a phase-transition occurring between $\sim 30$ K and $\sim 70$ K, which is likely to lead to the electron-hole compensation and hence to the extremely large magnetoresistivity seen in $\gamma-MoTe_2$. It is important to clarify the existence and the role of this crossover, or phase-transition, since it is likely to affect the electronic structure at the Fermi level of semi-metallic $MoTe_2$ and its predicted topological properties [6, 27].

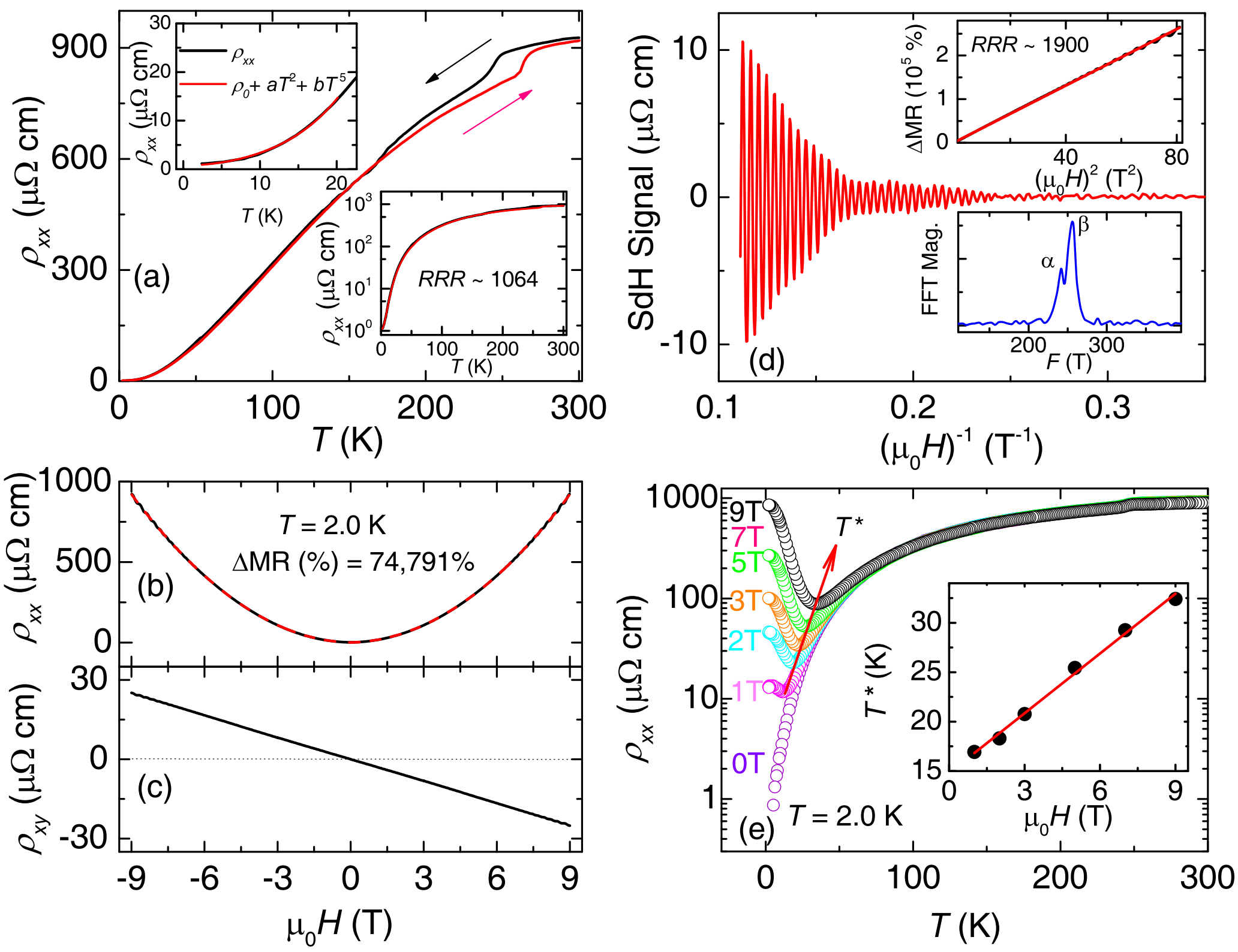

Figure 2.1: (a) Longitudinal resistivity $\rho_{xx}$ of semi-metallic $MoTe_2$ under zero magnetic field and as a function of the temperature. A hysteretic anomaly is observed around $\sim 240$ K, which is associated with the transition from the monoclinic $1T'$ to the orthorhombic $T_d$ structure. Upper inset: fit of $\rho_{xx}(T)$ to a combination of Fermi liquid and electron-phonon scattering mechanisms. Lower inset: $\rho_{xx}(T)$ in a logarithmic scale indicating a resistivity ratio $RRR = 1064$. (b) and (c) $\rho_{xx}$ and $\rho_{xy}$ as functions of $\mu_0 H$ at $T = 2$ K, respectively. (d) SdH oscillations as extracted from the $\rho_{xx}$ of a second sample as a function of $(\mu_0 H)^{-1}$ at $T = 2$ K. Bottom inset: Fourier transform of the oscillatory signal. Top inset: $\rho_{xx}$ as a function of $(\mu_0 H)^2$. Red-dashed line is a linear fit. (e) $\rho_{xx}$ in a logarithmic scale as a function of $T$ under various magnetic fields applied perpendicularly to the electrical current. $\rho_{xx}$ shows a minimum at $T^*$ as indicated by the red arrow. Inset: $T^*$ as a function of $\mu_0 H$ indicating a linear dependence in field.

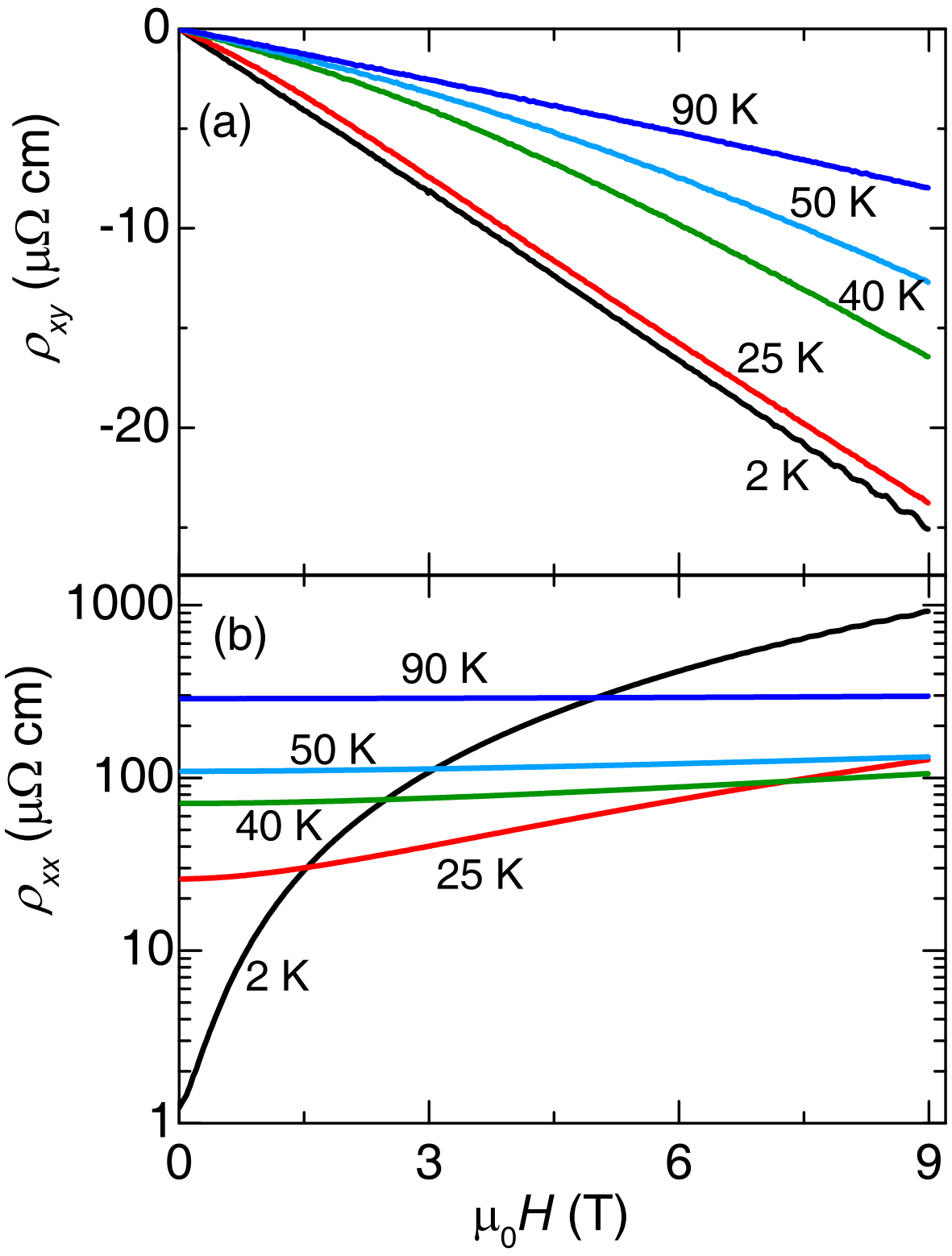

Figure 2.2: (a) Hall resistivity of a $\gamma-\mathrm{MoTe_2}$ single-crystal as a function of the field $\mu_0 H$ and for several temperatures ranging from $T = 2$ K to 90 K. A negative Hall resistivity indicates that electrons dominate the transport at all temperatures. (b) Longitudinal resistivity $\rho_{xx}$ as a function of $\mu_0 H$ and for the same temperatures.

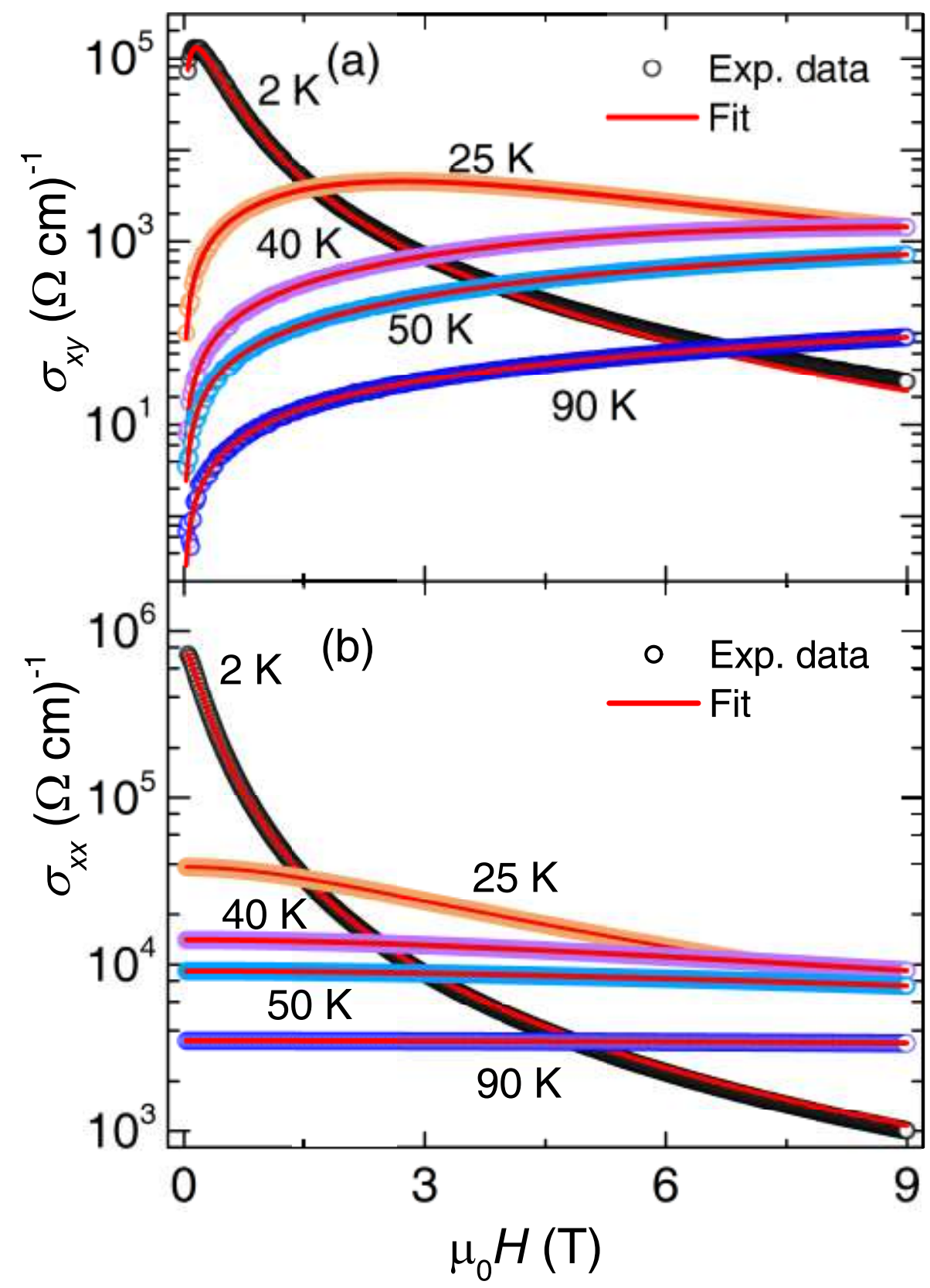

Figure 2.3: Components of the conductivity tensor, i.e $\sigma_{xy}$ and $\sigma_{xx}$ in panels (a) and (b) respectively, as functions of the the magnetic field for temperatures ranging from 2 to 90 K. Open circles represent experimental data and red solid lines the fitting curves based on the two-carrier model.

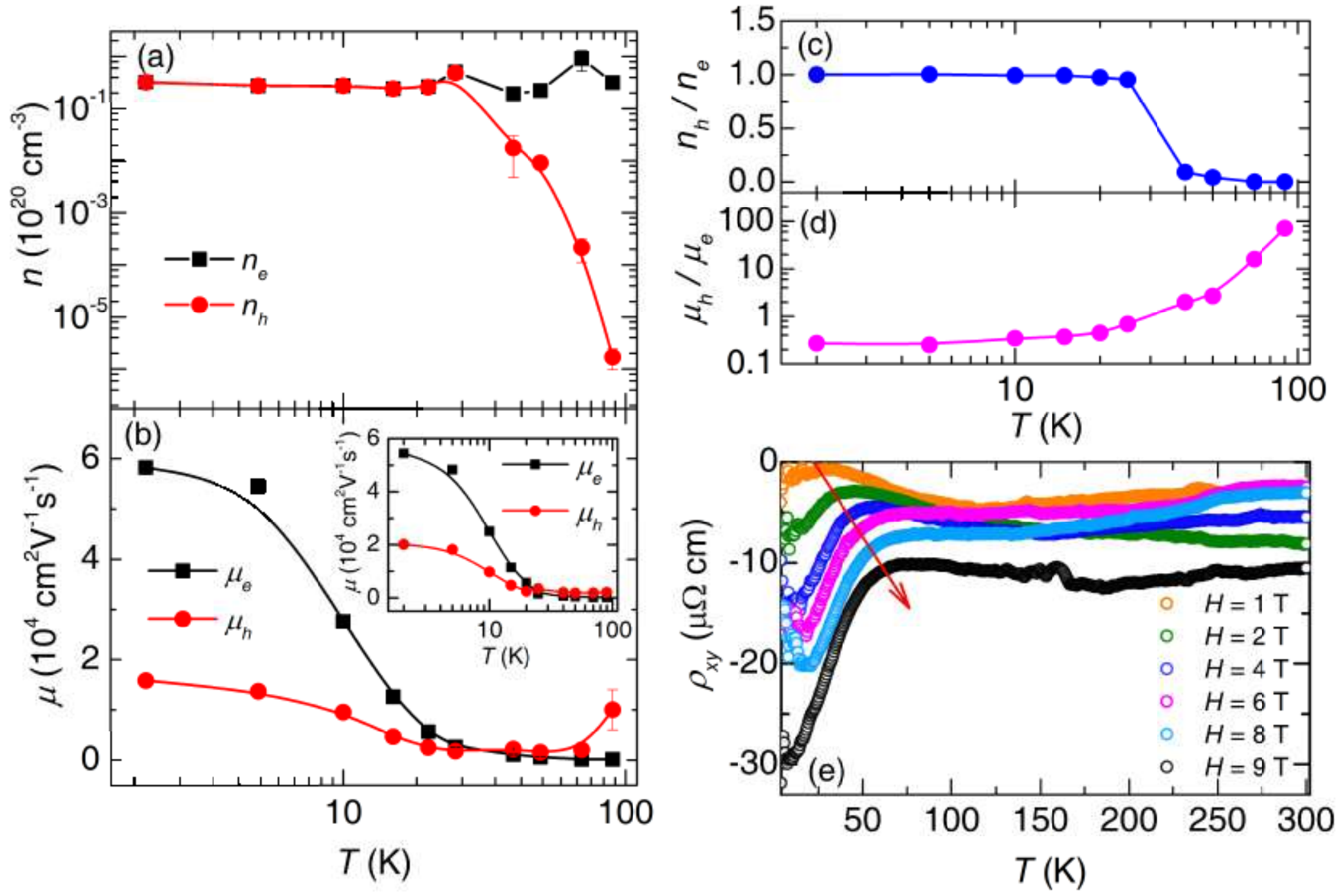

Figure 2.4: (a) Density of electrons $n_e$ and density of holes $n_h$ extracted from the two-carrier model analysis of $\sigma_{xy}$. (b) Carrier mobility $\mu_e$ and $\mu_h$ as a function of temperature deducted from $\sigma_{xy}$ (main panel) and from $\sigma_{xx}$(inset). (c) Density ratio and (d) mobility ratio between holes and electrons as a function of $T$. (e) Hall resistivity as a function of $T$ at various fields. A charge decrease in $\sigma_{xy}$ is indicated by the red arrow.

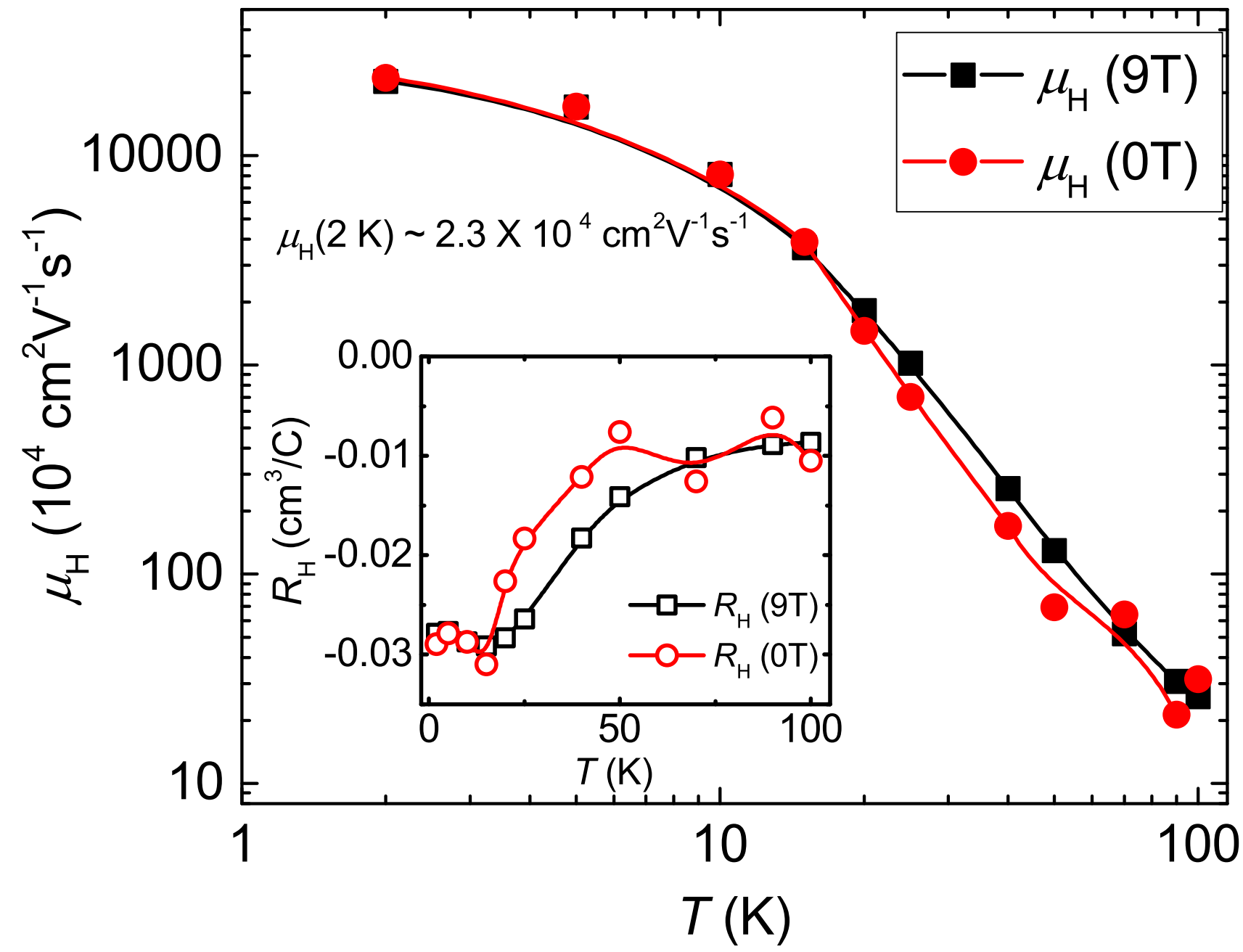

Figure 2.5: Hall mobility $\mu_{\mathrm{H}}$ as a function of $T$. Inset: Hall coefficient $R_{\mathrm{H}}$ as a function of $T$ in inset. Open red symbols represent $R_{\mathrm{H}}(9\mathrm{T})$ defined as $\rho_{xy}/B$ under $B = 9$ T, while black ones symbolize $R_{\mathrm{H}}(0\ \mathrm{T})$ as determined from the initial slope of the Hall resistivity $\rho_{xy}(B)$ as $B \rightarrow 0$.

# CHAPTER 3

# BULK FERMI-SURFACE OF THE WEYL TYPE-II SEMIMETAL CANDIDATE MOTE$_2$

Here, motivated by the scientific relevance and the possible technological implications of the aforementioned theoretical predictions [70, 71, 72, 75, 61], we evaluate, through electrical transport and torque magnetometry in bulk single-crystals, the electronic structure at the Fermi level and the topological character of orthorhombic $\gamma-$MoTe$_2$. Our goal is to contrast our experimental observations with the theoretical predictions and the reported ARPES results in order to validate their findings. This information could, for example, help us predict the electronic properties of heterostructures fabricated from single- or a few atomic layers of this compound. An agreement between the calculated geometry of the FS of $\gamma-$MoTe$_2$ with the one extracted from quantum oscillatory phenomena, would unambiguously support the existence of Weyl nodes in the bulk [73, 74] and, therefore, the existence of related non-trivial topological surface states or Fermi arcs [74, 75, 70, 71, 72, 75]. However, quantum oscillatory phenomena from $\gamma-$MoTe$_2$ single-crystals reveals a Fermi surface whose geometry is quite distinct from the one predicted by the DFT calculations based on its low temperature crystallographic structure. The extracted Berry-phase is found to be field-dependent. Still one does not obtain evidence for the topological character predicted for this compound when the Berry-phase is evaluated at low fields. Here, we show that shifts in the relative position of the electron and hole bands, implied by previous ARPES studies [76, 77, 78, 79, 80, 81, 82, 83], can replicate the angular dependence of the observed Fermi surface cross-sectional areas. However, these band shifts imply that the valence and electron bands would no longer cross and, therefore, that $\gamma-$MoTe$_2$ would not display the predicted Weyl type-II semi-metallic state.

## 3.1 Methods and Experimental Results

Very high quality single crystals of monoclinic $\beta-$MoTe$_2$ were synthesized through a Te flux method: Mo, 99.9999%, and Te 99.9999 % powders were placed in a quartz ampoule in a ratio of 1:25 heated up to 1050 °C and held for 1 day. Then, the ampoule was slowly cooled down to 900 °C and centrifuged. The “as harvested" single-crystals were subsequently annealed for a few days at a temperature gradient to remove the excess Te. Magneto-transport measurements as a function of temperature were performed in a Physical Property Measurement System using a standard four-terminal configuration. Measurements of the Shubnikov-de Haas (SdH) and the de Haas-van Alphen (dHvA) effects were performed in dilution refrigerator coupled to a resistive Bitter magnet, with the samples immersed in the $^3$He-$^4$He mixture. Measurements of the dHvA-effect were performed *via* a torque magnetometry technique, i.e. by measuring the deflection of a Cu-Be cantilever capacitively. Electrical transport measurements in pulsed magnetic fields were performed at the Dresden High Magnetic Field Laboratory using a 62 T magnet with a pulse duration of 150 ms. The sample temperature was controlled using a $^4$He bath cryostat (sample in He atmosphere) with an additional local heater for temperatures above 4.2 K. Synchrotron based X-ray measurements were performed in three single-crystals at the CHESS-A2 beam line using a combination of photon energies and cryogenic set-ups. The crystallographic data was reduced with XDS [84]. The structures were solved with direct methods using SHELXS [85]. Outlier rejection and absorption correction was done with SADABS. Least squares refinement on the intensities were performed with SHELXL [85]. For additional detailed information on the experimental set-ups used, see Appendix B.

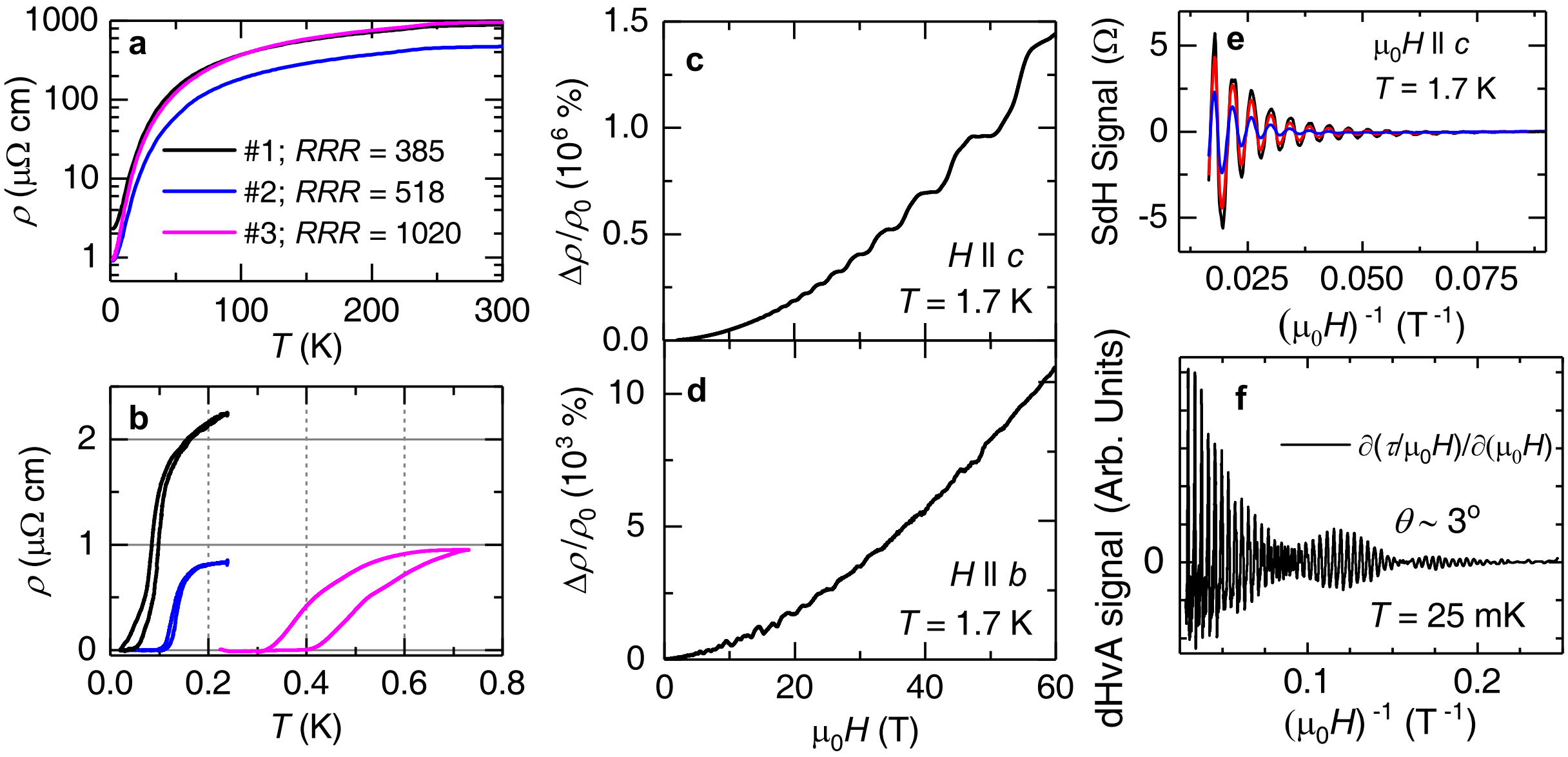

Figure 3.1: (a) Resistivity $\rho$, for currents flowing along the $a-$axis, as a function of the temperature $T$ for three representative single crystals displaying resistivity ratios $\rho(300\text{ K})/\rho(2)$ K between 380 and $\sim 1000$. (b) $\rho$ as a function of $T$ for each single-crystal indicating that $T_c$ depends on sample quality: it increases from an average transition middle point value of $\sim 130$ mK for the sample displaying the lowest ratio to $\sim 435$ mK for the sample displaying the highest one. The apparent hysteresis is due to a non-ideal thermal coupling between the single-crystals, the heater and the thermometer. (c) $\rho$ as a function of the field $H$ applied along the $c$-axis at a temperature $T = 1.7$ K for a fourth crystal characterized by a resistivity ratio of $\sim 207$. Notice i) the non-saturation of $\rho(H)$ and ii) that $\Delta\rho(\mu_0 H)/\rho_0 = (\rho(\mu_0 H) - \rho_0)/\rho_0$, where $\rho_0 = \rho(\mu_0 H = 0\text{ T}, T = 2\text{ K})$ surpasses $1.4 \times 10^6$ % at $\mu_0 H = 60$ T. (d) $\rho$ as a function of $\mu_0 H$ applied along the $b-$axis also at $T = 1.7$ K and for the same single-crystal. (e) Shubnikov-de Haas signal superimposed onto the magnetoresistivity for $\mu_0 H \| c-$axis and for three temperatures, $T = 8$ K (blue line), 4.2 K (red line) and 1.7 K (black line), respectively. (f) Oscillatory signal (black line) superimposed onto the magnetic susceptibility $\Delta\chi = \partial(\tau/\mu_0 H)/\partial\mu_0 H$, where $\tau$ is the magnetic torque.

As illustrated by Fig. 3.1(a), the as synthesized single-crystals display resistivity ratios $RRR$ $= \rho(T = 300 \text{ K})/\rho(T = 2 \text{ K})$ ranging from 380 to $> 2000$ which is one to two orders of magnitude higher than the $RRR$ values currently in the literature (see, for example, Ref. [86]) . Although not clearly visible in Fig. 3.1(a) due to its logarithmic scale, a hysteretic anomaly is observed in the resistivity around 240 K corresponding to the monoclinic to orthorhombic structural transition which stabilizes what we denominate as the orthorhombic $\gamma-MoTe_2$ phase as explained in Chapter 2. These single-crystals were subsequently measured at much lower temperatures allowing us to determine their superconducting transition temperature $T_c$. Remarkably, and as seen in Fig. 3.1(b), we find that $T_c$ depends on sample quality, increasing considerably as the $RRR$ increases, suggesting that structural disorder suppresses $T_c$. For these measurements, particular care was taken to suppress the remnant field of the superconducting magnet since the upper critical fields are rather small (see Appendix B Fig. A.1). The sample displaying the highest $RRR$ and concomitant $T_c$ was measured in absence of a remnant field. To verify that these differences in $T_c$ are not due to a poor thermal coupling between the sample and the thermometers, $T_c$ was measured twice by increasing and decreasing $T$ very slowly. The observed hysteresis is small relative to $T_c$ indicating that the measured $T_c$s are not an artifact. The values of the residual resistivities $\rho_0$ depend on a careful determination of the geometrical factors such as the size of the electrical contacts. Therefore, the $RRR$ provides a more accurate determination of the single-crystalline quality. In the past, the suppression of $T_c$ by impurities and structural defects was systematically taken as evidence for unconventional superconductivity [87, 88, 89], e.g. triplet superconductivity [90] in $Sr_2RuO_4$. Nevertheless, the fittings of the upper-critical fields $H_{c2}$ to a conventional Ginzburg-Landau expression, shown in Figure A.1 in Appendix B, points towards singlet pairing.

We have also evaluated the quality of our single crystals through Hall-effect [58] and heat capacity measurements (see, Appendix B Fig. A.2 ). Hall-effect reveals a sudden increase in the density of holes below $T = 40$ K, suggesting a possible temperature-induced Lifshitz-transition. While the heat capacity reveals a broad anomaly around $T = 66$ K, well-below its Debye temperature ($\Theta_D \simeq 120$ K), that would suggest that the structural degrees of freedom continue to evolve upon cooling below $T = 100$ K. Given that such structural evolution could affect the electronic band structure predicted for $\gamma-$MoTe$_2$ [71, 72, 75], we performed synchrotron X-ray scattering down to $\sim 12$ K (see, Appendix B Fig. A.3 ). We observe some variability in the lattice constants extracted among several single-crystals and a sizeable hysteresis in the range $125\text{K} \leq T \leq 250\text{K}$ associated with the structural transition observed at $T \simeq 250$ K, but no significant evolution in the crystallographic structure below 100 K. As we discuss below, there are negligible differences between the electronic bands calculated with the crystal structures collected at 100 K and at 12 K, respectively.

Figures 3.1(c) and 3.1(d) display the change in magnetoresistivity $\Delta\rho(\mu_0 H)/\rho_0 = (\rho(\mu_0 H) - \rho_0)/\rho_0$ as a function of the field $\mu_0 H$ for a crystal characterized by $RRR \sim 450$ when the electrical current flows along the crystalline $a$-axis and the field is applied either along the $c$- or the $b$-axes, respectively. Similarly to $WTe_2$, for both orientations $\Delta\rho/\rho_0$ shows no sign of saturation under fields all the way up to 60 T while surpassing $1 \times 10^6$ % for $\mu_0 H \| c$-axis [65]. For $WTe_2$ such anomalous magnetoresistivity was attributed to compensation between the density of electrons and holes [65, 91, 92]. Nevertheless, there are a number of subsequent observations [41] contradicting this simple scenario, such as i) a non-linear Hall response [93], ii) the suppression of the magnetoresistivity at a pressure where the Hall response vanishes [94] (i.e. at perfect compensation), and iii) the observation

of a pronounced magnetoresistivity in electrolyte gated samples with a considerably higher density of electrons with respect to that of holes [95]. It remains unclear if the proposed unconventional electronic structure [70, 71, 72] would play a role on the giant magnetoresistivity of $WTe_2$, while its measured FS differs from the calculated one [41]. In contrast, we have previously shown that $\gamma$-$MoTe_2$ indeed is a well compensated semi-metal [58]. The slightly smaller magnetoresistivity of $\gamma-MoTe_2$ relative to $WTe_2$ is attributable to heavier effective effective masses, according to de Haas-van Alphen-effect discussed below, or concomitantly lower mobilities.

The best $\gamma-MoTe_2$ samples, i.e. those with $RRR \geq 2000$, display even more pronounced $\Delta\rho/\rho_0$ under just $\mu_0 H \simeq 10$ T. The oscillatory component superimposed on the magnetoresistivity corresponds to the Shubnikov-de Haas (SdH) effect resulting from the Landau quantization of the electronic orbits. Figure 3.1(e) shows the oscillatory, or the SdH signal as a function of inverse field $(\mu_0 H)^{-1}$ for three temperatures. The SdH signal was obtained by fitting the background signal to a polynomial and then by subtracting it. Notice how for this sample and for this orientation, that is for $\mu_0 H \|$ $c$-axis, the SdH signal is dominated by a single frequency. However for all subsequent measurements performed under continuous fields (discussed below) one observes the presence of two main frequencies very close in value, each associated to a extremal cross-sectional area $A$ of the FS through the Onsager relation $F = A(\hbar/2\pi e)$ where $\hbar$ is the Planck constant and $e$ is the electron charge. To illustrate this point, we show in Fig. 3.1(f) the oscillatory signal extracted from the magnetic torque, i.e. $\tau = \mathbf{M} \times \mu_0 \mathbf{H}$, or the de Haas-van Alphen effect (dHvA) collected from a $\gamma-MoTe_2$ single-crystal for fields aligned nearly along its $c$-axis. Here $M = \chi \mu_0 H$ is the magnetization and $\chi(\mu_0 H, T)$ is its magnetic susceptibility. Figure 3.1(f) shows the oscillatory component of the magnetic susceptibility $\Delta\chi = \partial(\tau/\mu_0 H)/\partial(\mu_0 H)$. The envelope of the oscillatory signal displays the characteristic “beating" pattern between two close frequencies. This becomes clearer in the fast Fourier transform of the oscillatory signal shown below. According to The Lifshitz-Onsager quantization condition [96, 97], the oscillatory component superimposed onto the susceptibility is given by:

$$\Delta\chi[(B)^{-1}] \propto \frac{T}{B^{5/2}} \sum_{l=1}^{\infty} \frac{\exp^{-l\alpha\mu T_D/B} \cos(lg\mu\pi/2)}{l^{3/2} \sinh(\alpha\mu T/B)}$$

$$\times \cos\left\{2\pi\left[\left(\frac{F}{B} - \frac{1}{2} + \phi_B\right) l + \delta\right]\right\} \qquad (3.1)$$

where $F$ is the dHvA frequency, $l$ is the harmonic index, $\omega_c$ the cyclotron frequency, $g$ the Landé $g$-factor, $\mu$ the effective mass in units of the free electron mass $m_0$, and $\alpha$ is a constant. $\delta$ is a phase shift determined by the dimensionality of the FS which acquires a value of either $\delta = 0$ or $\pm 1/8$ for two- and three-dimensional FSs [96, 97, 98], respectively. $\phi_B$ is the Berry phase which, for Dirac and Weyl systems, is predicted to acquire a value $\phi_B = \pi$ [96, 97, 98]. Finally, $T_D = \hbar/(2\pi k_B \tau)$ is the so-called Dingle temperature from which one extracts $\tau$ or the characteristic quasiparticle scattering time.

In Appendix B Figs. A.4 and A.5 , we discuss the extraction of the Berry-phase of $\gamma-$MoTe$_2$ via fits to Eq. (1) of the oscillatory signal shown in Fig. 3.1(f). It turns out that the geometry of the FS of $\gamma-$MoTe$_2$ evolves slightly as the field increases due to the Zeeman-effect, which precludes the extraction of its Berry phase. More importantly, one cannot consistently extract a value $\phi_B \simeq \pi$ when one limits the range in magnetic fields to smaller values in order to minimize the role of the Zeeman-effect. In other words, the dHvA-effect does not provide evidence for the topological character of $\gamma-$MoTe$_2$. Nevertheless, it does indicate that the Dingle temperature decreases as the field increases implying a field-induced increase in the quasiparticle lifetime. This effect should contribute to its large and non-saturating magnetoresistivity. We reported a similar effect for WTe$_2$ [41].

Since the Berry-phase extracted from the dHvA-effect does not provide support for a topological semi-metallic state in $\gamma-$MoTe$_2$, it is pertinent to ask if the DFT calculations predict the correct electronic band-structure and related FS geometry for this compound , since both are the departing point for the predictions of Refs. [70, 71, 72] . To address this issue, we studied the dHvA-effect as a function of the orientation of the field with respect to the main crystallographic axes. Here, our goal is to compare the angular dependence of the cross-sectional areas determined experimentally, with those predicted by DFT.

Figures 3.2(a) and 3.22(b) display both the dHvA (red traces) and the SdH signals (black traces) measured in two distinct single crystals and for two field orientations, respectively along the $c-$ and the $a-$axes. As previously indicated, the dHvA and SdH signals were obtained after fitting a polynomial and subtracting it from the background magnetic torque and magnetoresistivity traces, respectively. The SdH signal was collected from a crystal displaying a $RRR \gtrsim 1000$ at $T \simeq 25$ mK under fields up to 18 T, while the dHvA one was obtained from a crystal displaying $RRR \gtrsim 2000$

at $T \simeq 35$ mK under fields up to 35 T. Both panels also display the Fast Fourier transform (FFT) of the oscillatory signal. For fields along the $c-$axis, one observes two main peaks at $F_\alpha = 231$ T and at $F_\beta = 242$ T, as well as their first- and second harmonics and perhaps some rather small frequencies which could result from imperfect background subtraction. We obtain the same two dominant frequencies regardless of the interval in $H^{-1}$ used to extract the FFTs. For the sake of completeness, Appendix B Fig. A.6 displays the dHvA signal for $H$ aligned nearly along the $b-$axis along with the corresponding FFT spectra which is again dominated by two prominent peaks. The observation of just two main frequencies for $\mu_0 H \| c-$axis is rather surprising since, as we show below, DFT calculations, including the effect of the spin-orbit interaction, predict several pairs of electron-like corrugated cylindrical FS along with pairs of smaller three-dimensional electron-like sheets in the First-Brillouin zone. Around the $\Gamma-$point, DFT predicts at least a pair of four-fold symmetric helix-like large hole sheets. This complex FS should lead to a quite rich oscillatory signal, contrary to what is seen. One might argue that the non-observation of all of the predicted FS sheets would be attributable to an experimental lack of sensitivity or to poor sample quality which would lead to low carrier mobility. Nevertheless, our analysis of the Hall-effect within a two-carrier model [58], yields electron- and hole-mobilities ranging between $10^4$ and $10^5$ cm$^2$/Vs at low $T$s which is consistent with both the small residual resistivities and the large resistivity ratios of our measured crystals. Given that the magnetic torque is particularly sensitive to the anisotropy of the FS, such high mobilities should have allowed us to detect most of the predicted FSs, particularly at the very low $T$s and very high fields used for our measurements. Hence, we conclude that the geometry of the FS ought to differ considerably from the one predicted by DFT.

In Figs. 3.2(c) and 3.2(d) we plot the amplitude of the main peaks observed in the FFT spectra for fields along the $c-$axis as a function of the temperature. Red lines are fits to the Lifshitz-Kosevich (LK) temperature damping factor, i.e. $x/\sinh x$ with $x = 14.69\mu T/H$ and with $\mu$ being the effective mass in units of the free electron mass, from which we extract the masses associated with each frequency. As seen, for $H\|c-$axis one obtains $\mu_\alpha = 0.85\ m_0$ and $\mu_\beta = 0.8\ m_0$, which contrasts with the respective values obtained for $H\|a-$axis, namely $\mu_{\alpha,\beta} \simeq 1.5\ m_0$, see Figs. 3.2(e) and 3.2(f). As previously mentioned for $\mu_0 H\|b-$axis, we observe two main frequencies, but by reducing the $H^{-1}$ window to focus on the higher field region, we detect additional frequencies (See, Fig. S6) which are characterized by heavier effective masses, i.e. in the order of $2.5 - 2.9\ m_0$. This indicates that $\gamma-$MoTe$_2$ displays a higher anisotropy in effective masses when compared to WTe$_2$ [41], although these masses are consistent with its sizeable $\gamma_e$ coefficient. In Appendix B Fig. A.7 displays several traces of the dHvA signal as functions of the inverse field for several angles between all three main crystallographic axes. These traces are used here to plot the angular dependence of the FS cross-sectional areas in order to compare these with the DFT calculated ones.

## 3.2 Comparison Between Experiments and the DFT Calculations

Several recent angle-resolved photoemission spectroscopy (ARPES) studies [76, 77, 78, 79, 80, 81, 82, 83] claim to find a broad agreement between the band structure calculations, the predicted geometry of the Fermi surface, the concomitant existence of Weyl type-II points [70, 71, 72], and the related Fermi arcs on the surface states of $\gamma-$MoTe$_2$. Several of these experimental and theoretical studies claim that the electronic structure of this compound is particularly sensitive to its precise crystallographic structure. Inter-growth of the $2H-$phase or the temperature used to collect to X-ray diffraction data, typically around 100 to 230 K, are claimed to have a considerable effect on the calculations [72, 81]. Given the few frequencies observed by us, it is pertinent to ask if the mild evolution of the crystallographic structure as a function of the temperature shown in Fig. S3 in Appendix B would affect the geometry of the FS of $\gamma-$MoTe$_2$. To address this question we performed a detailed angular-dependent study of the frequencies extracted from both the SdH and the dHvA effects in $\gamma-$MoTe$_2$ in order to compare these with the angular dependence of the FS cross-sectional areas predicted by the calculations.

In the subsequent discussion we compare the angular dependence of our dHvA frequencies with calculations performed with the Quantum Espresso [99] implementation of the density functional theory in the GGA framework including spin-orbit coupling (SOC). The Perdew-Burke-Ernzerhof (PBE) exchange correlation functional [100] was used with fully relativistic norm-conserving pseudopotentials generated using the optimized norm-conserving Vanderbilt pseudopotentials as described in Ref. [101] . The $4s$, $4p$, $4d$ and $5s$ electrons of Mo and the $4d$, $5s$ and $5p$ electrons of Te were treated as valence electrons. After careful convergence tests, the plane-wave energy cutoff was taken to be 50 Ry and a $k-$point mesh of $20 \times 12 \times 6$ was used to sample the reducible Brillouin Zone (BZ) used for the self-consistent calculation. The Fermi surfaces were generated using a more refined $k-$point mesh of $45 \times 25 \times 14$. FS sheets were visualized using the XCrysden software [102]. The related angular dependence of the quantum oscillation frequencies was calculated using the skeaf code [103]. As shown in Fig. 3 the results are very close to those obtained in an earlier version of this manuscript using VASP and the Wien2K implementations of DFT, and also to those reported by Refs. [71, 72, 76, 83].

Figure 3.3(a) displays the electronic band structure of $\gamma-$MoTe$_2$, based on its structure determined at $T = 100$ K, with and without the inclusion of SOC. As previously reported [71, 72], electron- and hole-bands intersect along the $\Gamma - X$ direction at energies slightly above $\varepsilon_F$ creating a pair of Weyl type-II points. Figure 3.3(b) shows a comparison between band structures based on the crystallographic lattices determined at 12 K and at 100 K, respectively. Both sets of electronic bands are nearly identical and display the aforementioned crossings between hole- and electron-bands thus indicating that the electronic structure remains nearly constant below 100 K. Figures 3.3(c) and 3.3(d) provide a side perspective and a top view of the overall resulting FS within BZ, respectively. The main features of the DFT calculations are the presence of two-dimensional electron pockets, labeled $e_1$ and $e_2$ in Figs. 3.3(e) and 3.3(f) and of large "star-shaped" hole-pockets near the $\Gamma-$point, labeled as the $h_2$ and the $h_3$ sheets in Figs. 3.3(k) and 3.3(l). These electron and hole pockets nearly "touch". Due to the broken inversion symmetry, these bands are not Kramer's degenerate, and hence the spin-orbit split partners of the corresponding electron and hole pockets are located inside the corresponding bigger sheets. The $h_1$ hole pocket and the $e_3$ and $e_4$ electron pockets are very sensitive to the position of $\varepsilon_F$ disappearing when $\varepsilon_F$ is moved by only $\pm 15$ meV.

Figures 3.4(a) and 3.4(b) present the angular dependence of the calculated and of the measured FFT spectra of the oscillatory signal (raw data in Fig. S7 Appendix B), respectively. In this plot the Onsager relation was used to convert the theoretical FS cross-sectional areas into oscillatory frequencies. In Fig. 3.4(b) $\theta$ refers to angles between the $c-$ and the $a-$axis, where $\theta = 0^\circ$ corresponds to $H \parallel c-$axis, while $\phi$ corresponds to angles between the $c-$ and the $b-$axis, again relative to the $c-$axis. As seen, there are striking differences between both data sets with the calculations predicting far more frequencies than the measured ones. More importantly, for fields along the $c$-axis, one observes the complete absence of experimental frequencies around $\sim 1$ kT which, according to the calculations, would correspond to the cross-sectional areas of the hole-pockets $h_2$ and $h_3$. In addition, while many of the predicted electron orbits show a marked two-dimensional character, diverging as the field is oriented towards the $a-$ or the $b-$axis, the experimentally observed frequencies show

finite values for fields along either axis. This indicates that these orbits are three-dimensional in character, despite displaying frequencies close to those predicted for the $e_1$ and the $e_2$ pockets for fields along the $c-$axis. These observations, coupled to the non-detection of all of the predicted orbits, in particular the large hole $h_1$ and $h_2$ Fermi surfaces, indicate unambiguously that the actual geometry of the FS of $\gamma-\mathrm{MoTe_2}$ is different from the calculated one. Notice that frequencies inferior to $F = 100$ T, which correspond to the smaller electron- and hole-pockets and which are particularly sensitive to the position of $\varepsilon_F$ as previously mentioned, were not included in Fig. 3.4(a) for the sake of clarity.

The calculation shows a significant difference between the SOC-split theoretical bands, which is highlighted by the absence of a frequency around 0.5 kT associated with $e_2$ pocket, along with its presence in association with the $e_1$ pocket. This contrast between both orbits is due to the presence of a "handle-like" structure (see Fig. 3.3(e)) in $e_1$ which gives a maximum cross-section close to the BZ edge. However, at this position there is no maximum cross-section within the BZ for the $e_2$ pocket since its "handle" is missing (see Fig. 3.3(f)). This marked difference in topology between the FSs of both spin-orbit split partners indicates that the strength of the SOC provided by the DFT calculations tends to be considerably larger than the one implied by our experiments. In fact, from the twin peaks observed in the experimental FFT spectra having frequencies around 250 T for fields along the $c$-axis, which are likely to correspond to SOC-split bands due to their similar angular dependence, we can infer that the actual SO-splitting is far less significant than the value predicted by the calculations. We have investigated the possibility of an overestimation of the strength of the SOC within our calculations which is the mechanism driving the DFT prediction of a large number of dHvA frequencies displaying remarkably different angular dependencies. For instance, we calculated the angular dependence of the FSs without the inclusion of SOC. This leads to just one, instead of a pair of distinct SOC-split bands, which in fact display angular dependencies very similar to those of orbits $h_3$, $e_1$ and $e_3$ in Fig. 4(a). Notice that part of the discrepancy is attributable to the inter-planar coupling which is not well captured by the DFT calculations [104]. DFT suggests that this compound is van der Waals like by predicting several two-dimensional (i.e. cylindrical like) FS sheets, when the experiments indicate that the overall FS displays a marked three-dimensional character. This indicates that the inter-planar coupling is stronger than implied by DFT. In any case, from Figs. 3.4(a) and 3.4(b) and the above discussion, it is clear that there are significant discrepancies between the calculated and the measured FSs. Given that the proposed Weyl type-II scenario [70, 71, 72] hinges on a possible touching between electron- and hole-pockets, it is critical to understand their exact geometry, or the reason for the disagreement between predictions and experiments, before one can make any assertion on the existence of the Weyl type-II points in $\gamma-\text{MoTe}_2$.

To understand the source of the disagreement between calculations and our measurements, we now focus on a detailed comparison between our DFT calculations and a couple of ARPES studies. Figure 3.5(a) corresponds to data from Ref. [83] depicting an ARPES energy distribution map (EDM) along $k_y$ while keeping $k_x = 0$. Figure 3.5(b) plots its derivative. In both figures the $\Gamma-$point corresponds to $k_y = 0$. According to the calculations, this EDM should reveal two valence bands intersecting $\varepsilon_F$ around the $\Gamma-$point; the first leading to two small hole-pockets, or the $h_1$ sheets at either side of $\Gamma$, with the second SOC-split band producing the larger $h_2$ and $h_3$ sheets. Instead, ARPES observes just one band intersecting $\varepsilon_F$ which leads to a single FS sheet of cross-sectional area $S_{\text{FS}} \sim \pi(0.1 \text{ Å}^{-1})^2$ as indicated by the vertical blue lines in Fig. 3.5(b). This observation by ARPES questions the existence of the band (or of its intersection with $\varepsilon_F$) responsible for the large $h_2$ and $h_3$ hole-pockets with this band being the one previously reported to touch the electron band that produces the $e_1$ pocket creating in this way the Weyl type-II points [76, 77, 78, 81, 82]. Notice that our dHvA measurements do not reveal any evidence for the original $h_2$ and $h_3$ sheets, thus, being in agreement with this ARPES observation. Furthermore, the Onsager relation $F = S(\hbar/2\pi e)$ yields a frequency of $\sim 330$ T for $S_{\text{FS}}$ which, in contrast, is close to the frequencies observed by us for $\mu_0 H \| c-$axis. Figure 3.5(c) displays the band structure calculated "ribbons" obtained by projecting the $k_z$ dependence of the bands onto the $k_x - k_y$ plane. This representation of the band-structure provides a better comparison with the ARPES EDMs. As seen, there is a good overall agreement between the ARPES and the DFT bands, as previously claimed [76, 77, 78, 81, 82], except for the exact position of $\varepsilon_F$. The purple line depicts the position of $\varepsilon_F$ according to the DFT calculations while the white line depicts the position of $\varepsilon_F$ according to ARPES. A nearly perfect agreement between DFT and ARPES is achievable by shifting the DFT valence bands by $\sim -50$ meV, which is what ends suppressing the $h_2$ and $h_3$ FS hole sheets from the measured ARPES EDMs. As shown through Figs. 3.5(d) and 3.5(e), this disagreement between the ARPES and the DFT bands is observed in the different ARPES studies[81]. Figure 3.5(d) corresponds to an EDM along the $k_x$ ($k_y = 0$) direction of the BZ. As shown in Fig. 3.5(e), DFT reproduces this EDM quite well. Nevertheless, as indicated by the yellow lines in both figures, which are positioned at the top of the deepest valence band observed by ARPES, the ARPES bands are displaced by $\sim -45$ meV with respect to the DFT ones. Red dotted lines in Fig. 3.5(d) indicate the cross-sections of the observed electron pockets or $\sim \pi(0.1\ \text{Å}^{-1})^2$. Therefore, to match our main dHvA frequencies, the electron bands would have to be independently and slightly displaced towards higher energies to decrease their cross-sectional area. To summarize, DFT and ARPES agree well on the overall dispersion of the bands of $\gamma-$MoTe$_2$, but not on their relative position with respect to $\varepsilon_F$.

Therefore, guided by ARPES, we shifted the overall valence bands of $\gamma-$MoTe$_2$, shown in Figs. 3.3(a) and 3.3(b), by -50 meV and the electron ones by +35 meV to recalculate the FS cross-sectional areas as a function of field orientation relative to the main crystallographic axes. The comparison between the measured dHvA cross-sectional areas and those resulting from the shifted DFT bands are shown in Fig. 3.6. Figure 6(a) displays the Fourier spectra, previously shown in Fig. 3.4(b), with superimposed colored lines identifying shifted electron (magenta) and hole (blue) orbits according to Fig. 3.6(b) which displays these frequencies as a function of field orientation for shifted non-SOC-split DFT bands. As seen, the qualitative and quantitative agreement is good, but not perfect. In contrast, Fig. 3.6(c) displays these orbits/frequencies as a function of field orientation for SOC-split DFT bands. Clearly, and as previously discussed, the approach used to evaluate the effect of the SOC in $\gamma-$MoTe$_2$ overestimates it for reasons that remain to be clarified.

Concerning the Weyl physics in $\gamma-$MoTe$_2$, the displacement of the bands, introduced here to explain our observations based on the guidance provided by previous ARPES studies, would eliminate the crossings between the electron- and the hole-bands as shown in Fig. 3.7(a). The very weak SOC implied by our study further undermines the argument in favor of a Weyl type-II semi-metallic state in this compound. Finally, Figs. 3.7(b) to 3.7(g) display the geometry of the Fermi surface resulting from the shifted bands. Overall, the FS displays a distinctly more marked three-dimensional character, with the electrons and the-hole pockets remaining well-separated in $k$-space.

## 3.3 Conclusions

In conclusion, quantum oscillatory phenomena reveal that the geometry of the Fermi surface of $\gamma-MoTe_2$ is quite distinct from the one predicted by previous electronic band-structure calculations. Our low-temperature structural analysis *via* synchrotron X-ray diffraction measurements indicates the absence of an additional structural transition below the monoclinic to orthorhombic one that would explain this disagreement, while heat-capacity measurements provide no evidence for an electronic phase-transition upon cooling. In contrast, a direct comparison between DFT calculations and the band-structure reported by angle resolved photoemission spectroscopy reveals a disagreement on the position of the valence bands relative to the Fermi-level, with the experimental valence bands shifted by $\sim -50$ meV relative to the DFT ones. Therefore, one should be careful concerning the claims of a broad agreement between the calculations and the electronic bands revealed by ARPES measurements [76, 77, 78, 79, 80, 81].

Here, we show that it is possible to describe the angular-dependence of the observed de Haas-van Alphen Fermi surface cross-sectional areas by shifting the position of the DFT bands relative to the Fermi level as indicated by ARPES. However, with this adjustment, the Weyl points, which result from band-crossings that are particularly sensitive to small changes in the lattice constants, are no longer present in the band-structure of $\gamma-MoTe_2$. Although our approach of modifying the band structure in order to obtain an agreement with both ARPES and de Haas-van Alphen experiments is not based on *ab-initio* calculations and has only a phenomenological basis, our findings do shed a significant doubt on the existence of the Weyl points in the electronic band structure of $\gamma-MoTe_2$. In addition, our measurements indicate that the DFT calculations overestimate the role of spin-orbit coupling in this compound. Hence, our work which combines DFT calculations with measurements of quantum oscillatory phenomena, strongly suggests that the approach of including the role of spin-orbit coupling in the DFT implementations needs to be significantly revised, particularly in what concerns layered compounds.

Finally, this study combined with the ARPES results in Ref. [105] , indicate that there ought to be a Lifshitz-transition [106] upon W doping in the $\gamma-Mo_{1-x}W_xTe_2$ series, leading to the disappearance of the central hole pockets in $\gamma-MoTe_2$ in favor of the emergence of hole-pockets at either side of the $\Gamma-$point in $\gamma-Mo_{1-x}W_xTe_2$.

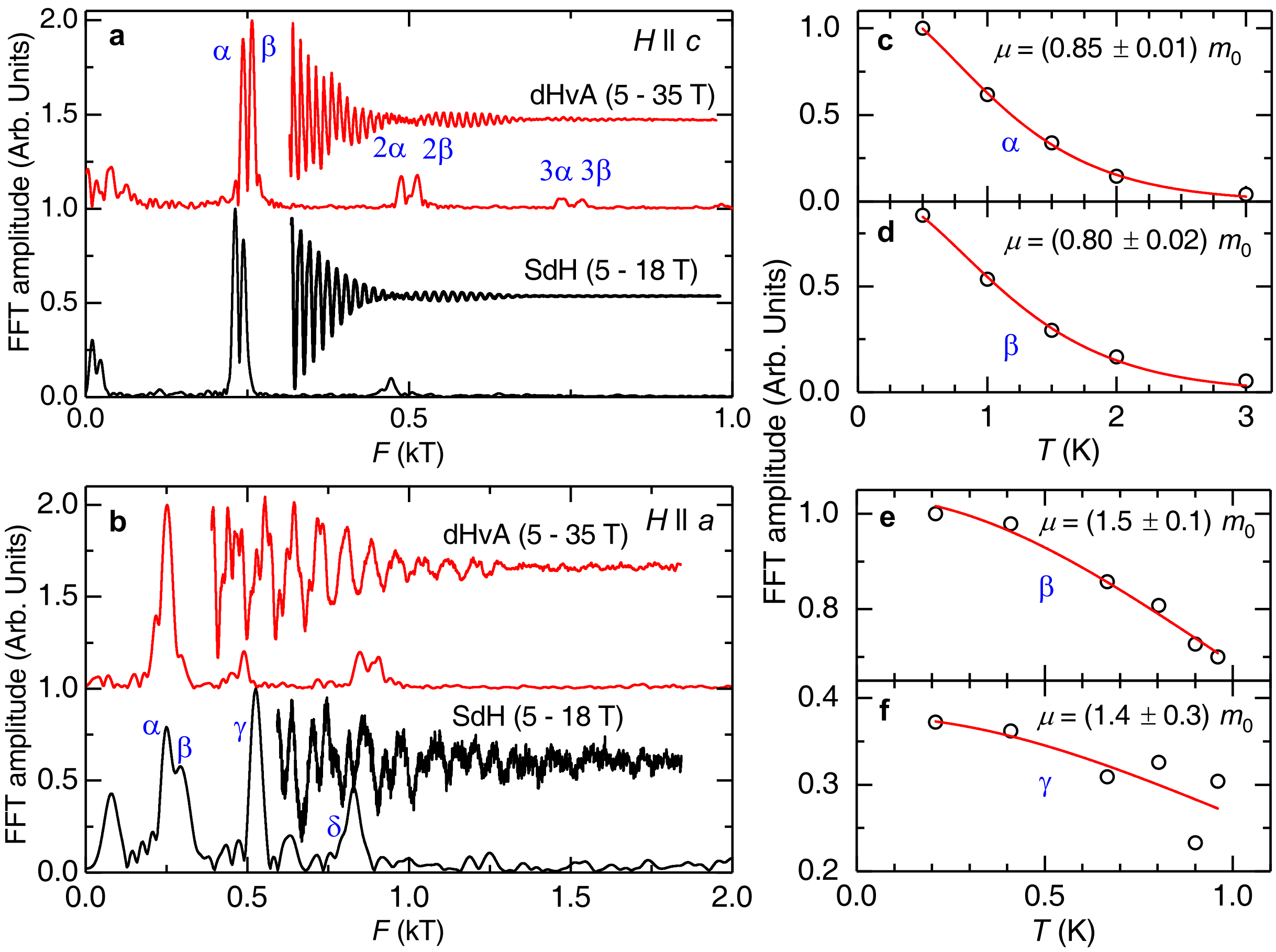

Figure 3.2: (a) Typical de Haas-van Alphen (red trance) and Shubnikov-de Haas (black trace) signals superimposed onto the magnetic torque and the magnetoresistivity respectively, for fields aligned along the $c$-axis of $\gamma-MoTe_2$ single-crystals at $T \simeq 30$ mK. The same panel shows the FFT spectra for each signal revealing just two main frequencies or FS cross-sectional areas. (b) dHvA and SdH signals and corresponding FFT spectra obtained from the same single-crystals but for $H$ aligned nearly along the $a$-axis of each single-crystal. (c) and (d), Amplitude of the peaks observed in the FFT spectra for $H \| c$-axis and as a function of $T$ including the corresponding fits to the LK formula from which one extracts the effective masses. (e) and (f) Amplitude of two representative peaks observed in the FFT spectra for $H \| a$-axis and as a function of $T$ with the corresponding fits to the LK formula to extract their effective masses.

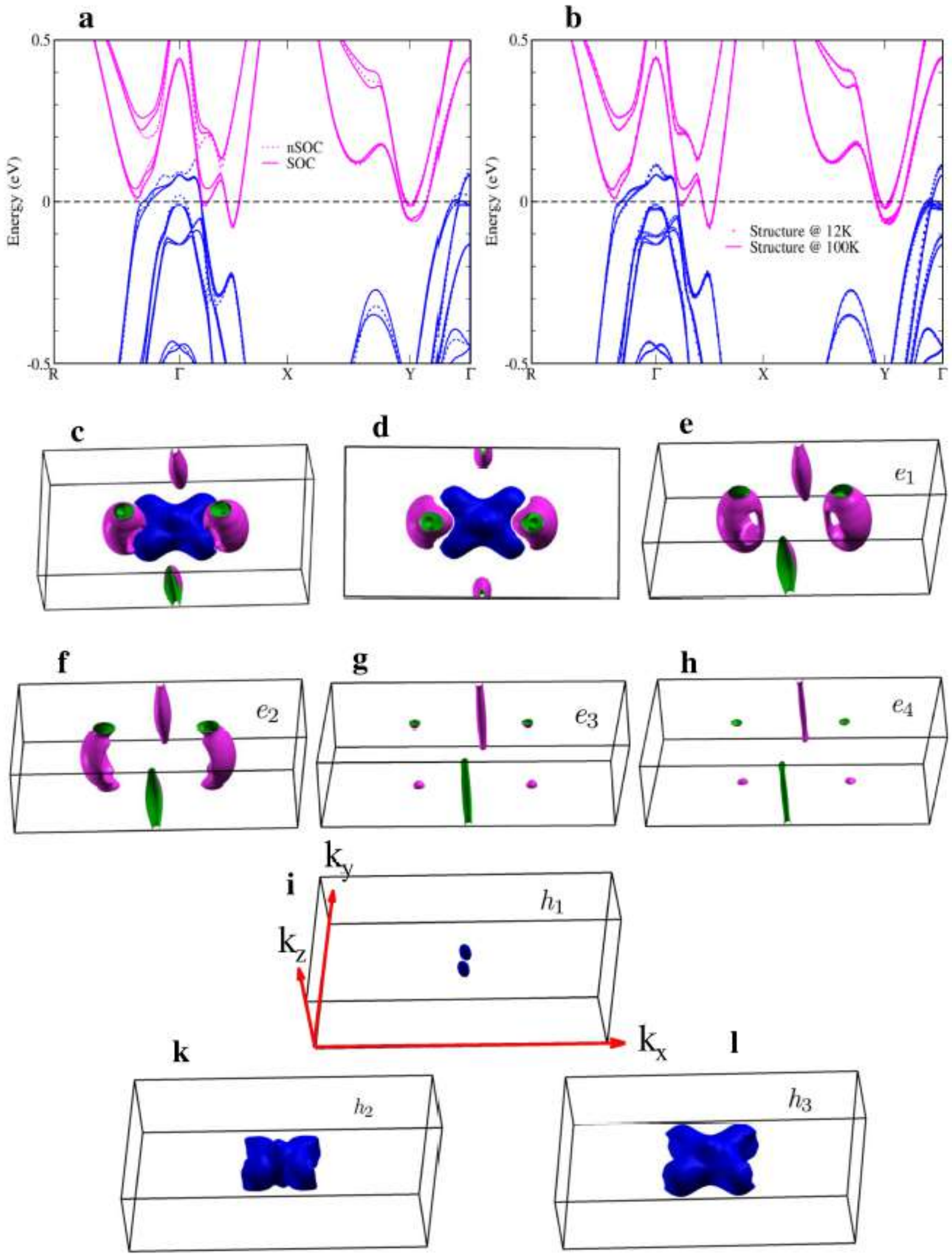

Figure 3.3: (a) Electronic band-structures of $\gamma-\mathrm{MoTe_2}$ calculated with and without the inclusion of spin-orbit coupling (SOC). These calculations are based on the crystallographic structure measured at $T = 100$ K. (b) Comparison between the calculated electronic bands based upon the crystallographic structures measured at $T = 100$ and $T = 12$ K, respectively. (c) and (d) Respectively, side and top views of the calculated FS. (e), (f), (g) and (h) Fermi surface sheets resulting from electron bands. Notice the marked two-dimensional character of several of the electron-like FS sheets. (i), (k) and (l), Hole-like sheets around the $\Gamma-$point.

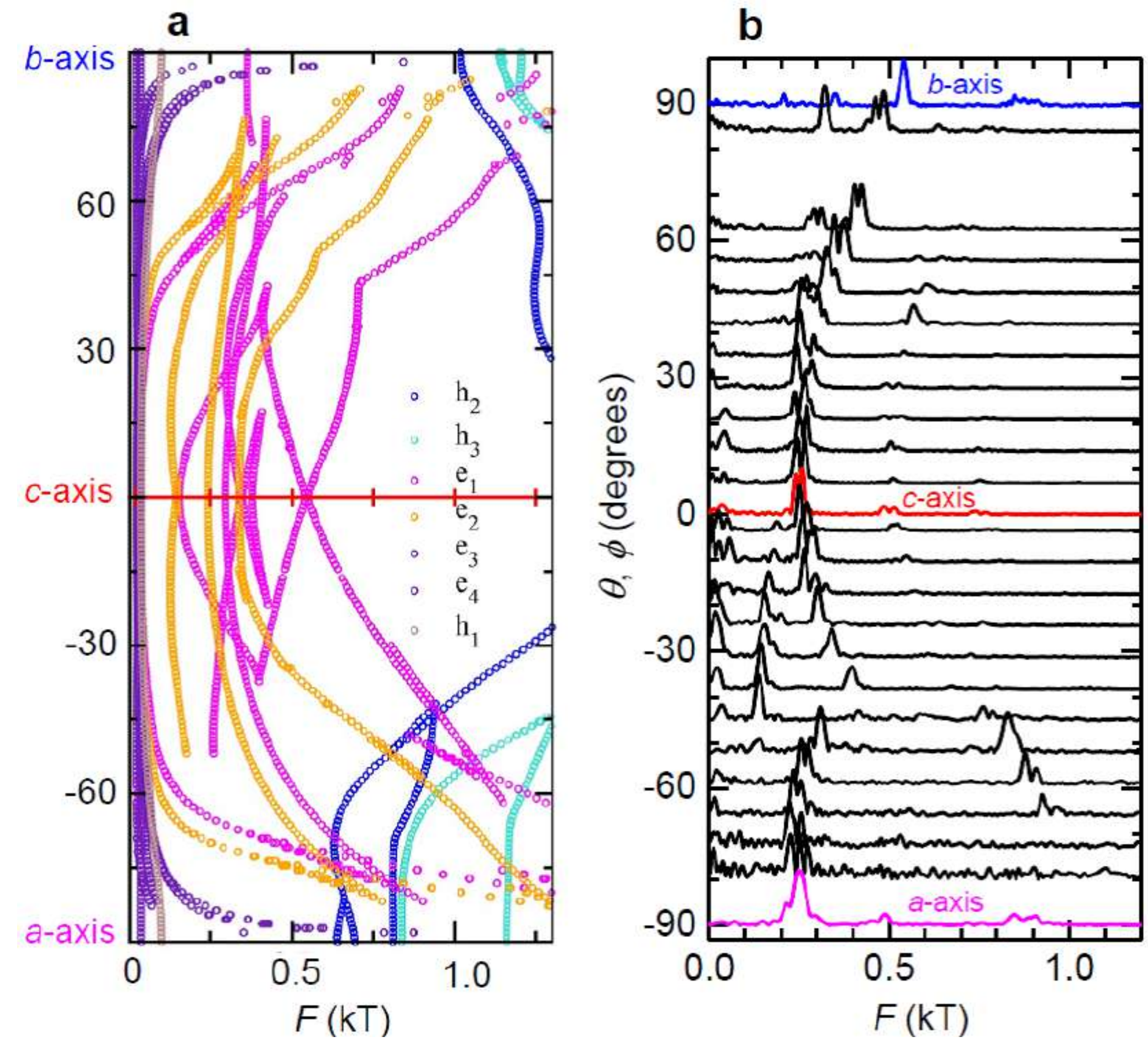

Figure 3.4: (a) Angular dependence of the FS cross-sectional areas or, through the Onsager relation, de Haas-van Alphen frequencies as with respect to the main crystallographic axes. (b) Experimentally observed dHvA spectra as a function of the frequency $F$ for several angles $\theta$ between the $c-$ and and the $b-$axis and for several values of the angle $-\phi$ between the $c-$ and the $a-$axes.

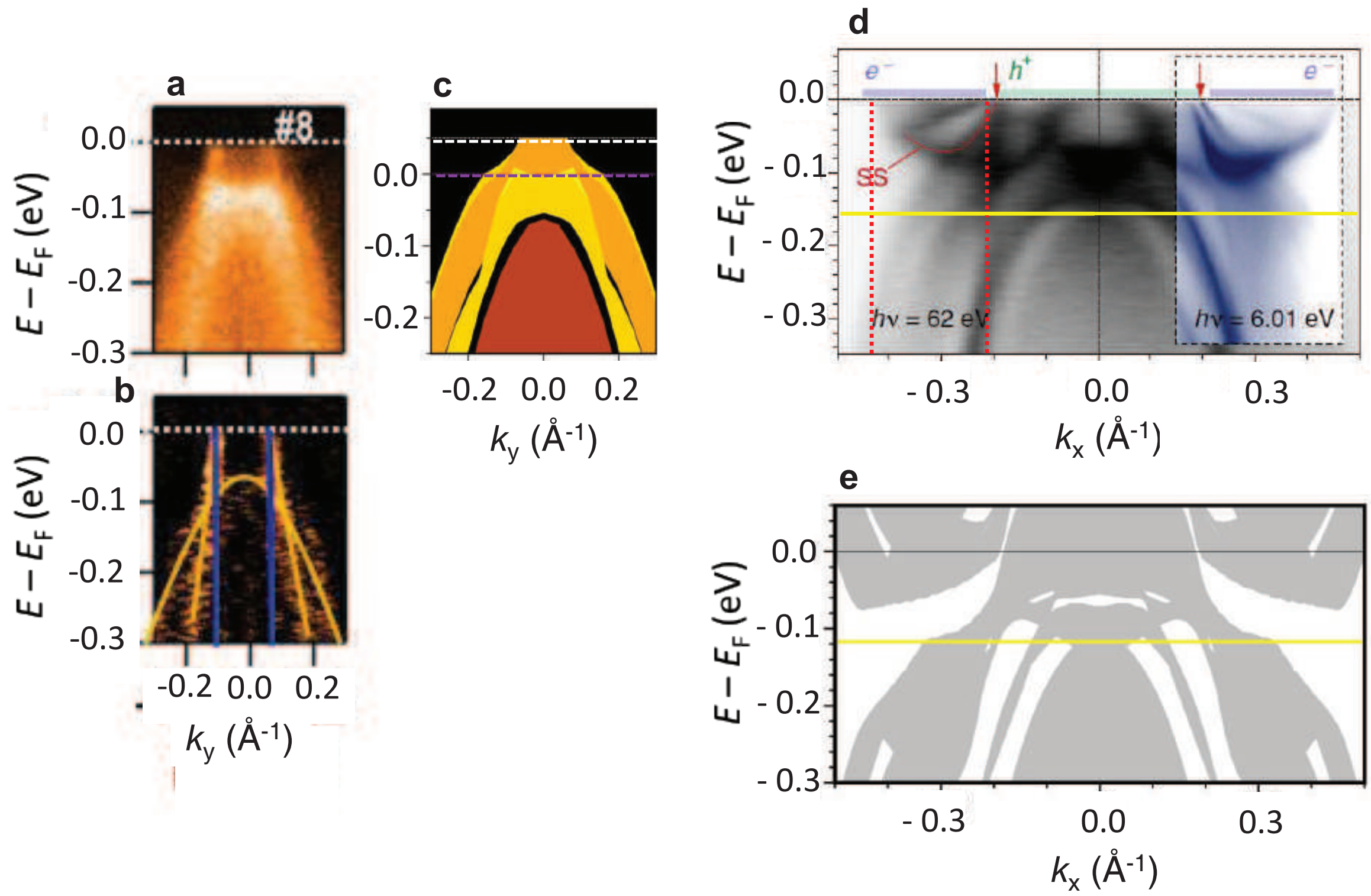

Figure 3.5: (a) ARPES energy distribution map along $k_y$ for $k_x = 0$ from Ref. [83]. (b) Derivative of the energy distribution map. Vertical blue lines indicate the diameter of the observed hole-pocket, with its area corresponding to a frequency $F \sim 0.33$ kT. (c) Ribbons obtained from our DFT calculations showing the $k_z$-projection for all bulk bands which are plotted along the same direction as the ARPES energy distribution map in (a). Different colors are chosen to indicate distinct bands. Purple dotted line corresponds to the original position of the Fermi level $\epsilon_F$ according to DFT, while the white one corresponds to the position of $\epsilon_F$ according to ARPES. Notice that the ARPES bands are shifted by $\sim -50$ meV with respect to the DFT ones. (d) ARPES energy distribution map corresponding to a cut along the $k_x-$direction with $k_y = 0$ , from Ref. [81] . Vertical red dotted lines indicate the diameter of the observed electron pockets or $\sim 0.2$ Å$^{-1}$. SS stands for "surface-state". (e) DFT Ribbons showing the $k_z$-projection for all bulk bands plotted along the same direction as the ARPES EDM in (d). In both panels black lines depict the original position of $\epsilon_F$ while the yellow lines are guides to the eyes illustrating the difference in energy between the top of the deepest DFT calculated hole-band and its equivalent according to ARPES.

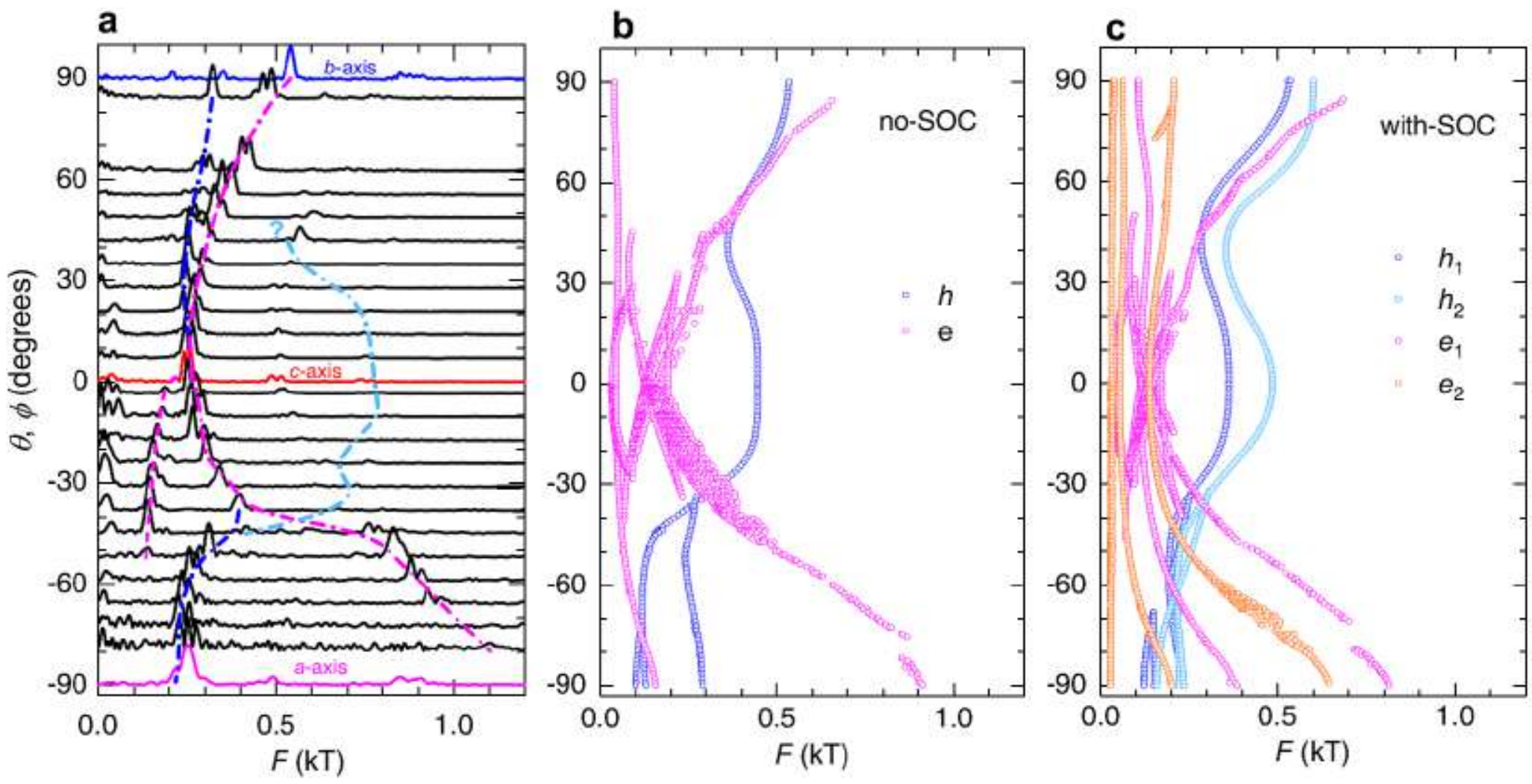

Figure 3.6: (a) Experimentally observed dHvA spectra as a function of $F$ for several angles $\theta$ and $-\phi$, where the magenta and the blue lines act as guides to the eyes and as identifiers of respectively, electron- and hole-like orbits according to the shifted band structure. Clear blue line depicts a possible hole-orbit associated with very small peaks in the FFT spectra. (b) Angular dependence of the dHvA orbits, or frequencies on the FS resulting from the shifted bands in absence of spin-orbit coupling, where magenta and blue markers depict electron- and hole-like orbits on the FS, respectively. Notice the qualitative and near quantitative agreement between the calculations and the experimental observations. (d) Angular dependence of the dHvA frequencies for the shifted electron and hole-bands in the presence of spin-orbit coupling. Here, electron-orbits are depicted by magenta and orange markers while the hole ones are indicated by blue and clear blue markers. The experimental data are better described by the non spin-orbit split bands.

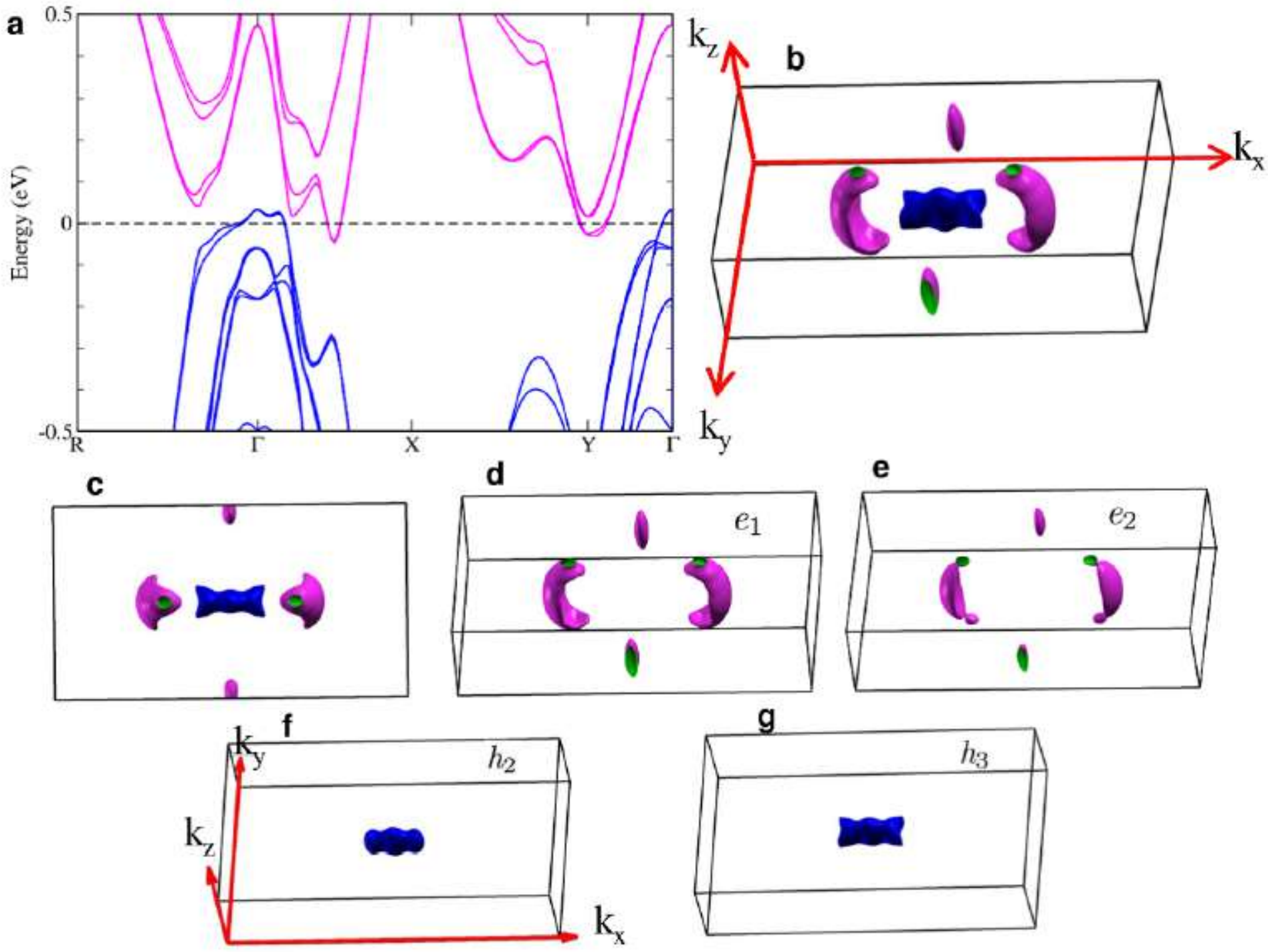

Figure 3.7: (a) Electronic band structure calculated with the inclusion of SOC after the hole-like bands have been shifted by −50 meV and the electron ones by +35 meV with goal of reproducing the observed dHvA frequencies and their angular dependence. Notice that these shifts suppress the crossings between the electron- and hole-bands and therefore the Weyl type-II points of the original band structure. (b) Fermi surface resulting from these band-shifts. (c) FS top view. (d) and (e) Electron-like FS sheets. (f) and (g) Hole-like sheets.

# CHAPTER 4

# TRANSPORT PROPERTIES OF PTTE$_2$

We investigate the transport properties of Dirac semimetal candidate $PtTe_2$ through systematic Hall effect measurements and detailed de Haas-van Alphen (dHvA) effect and Shubnikov de Haas (SdH) effect. The cross-sectional areas of Ferm surfaces are mapped through the detailed angular-dependent dHvA and SdH effect FFT spectra. We compared our DFT calculations including spin-orbit split with our experimentally observed Fermi surface cross-sectional areas.

## 4.1 Single Crystals of $PtTe_2$

Semi-metallic transition metal dichalcogenides (TMDC) $PtTe_2$ is a layered material and could be mechanically exfoliated as displayed in Figure 4.1. It crystallizes in the trigonal $1T$ structure with $P\bar{3}m1$ space group. The crystal structure is composed of edge-shared $PtTe_6$ octahedra with $PtTe_2$ layers tiling the $ab$ plane, as shown in Figure 1.10 . Based on X-ray diffraction on a $PtTe_2$ single crystal with size of $0.12 \times 0.12 \times 0.01 mm^3$ (see Figure 4.2), the extracted lattice constants from precession images have excellent agreement with Ref. [60], which confirmed that our $PtTe_2$ single crystal belongs to $P\bar{3}m1$ space group.

We have found that a rich spectrum of transport properties exist even among crystals extracted from the same batch. The residual resistivity remains at the same order even the residual resistivity ratio varies by a factor of 10 (Figure4.3(a)). We then turn to the large and non-saturating MR observed in all samples up to 9 T as shown in Figure4.3(b). In a transverse field ($H$ // $c$-axis), several crystals with relatively high residual resistivity ratio (i.e. black and red curves) show a striking $H$-linear MR, while a few other crystals with high $RRR$ exhibit quadratic behaviors. This could possibly be attributed to the sample-dependent mobility in crystals similar with $Cd_3As_2$ [51].

The magnetoresistance is still non-satuarting even up to 61 T at pulsed magnetic field. The MR% increases up to 60000% as shown in Figure4.4.

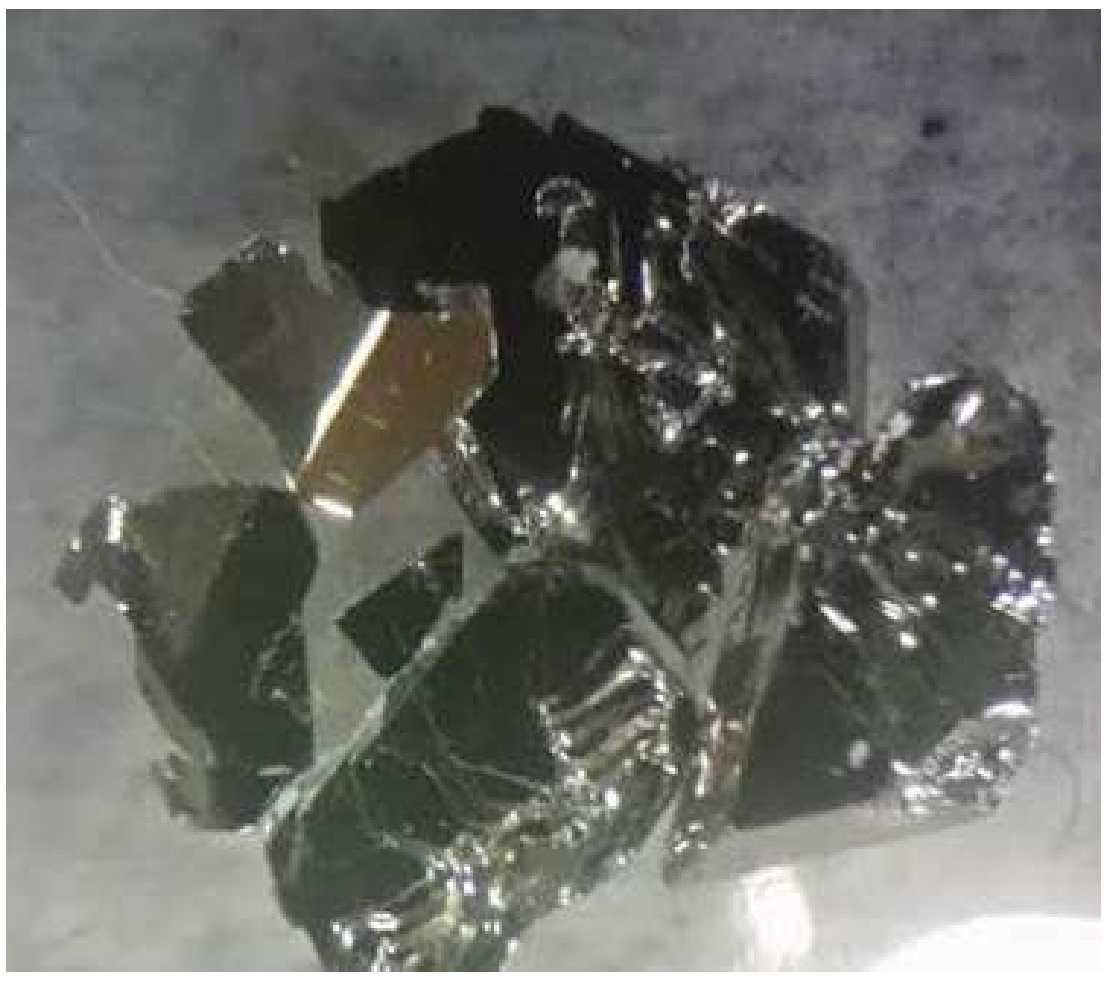

Figure 4.1: $PtTe_2$ single crystals grown by the Te flux method have a typical lateral size of ~ $3-8$ mm, rectangular shaped and have a metallic appearance. The layers in the $PtTe_2$ are stacked together via van der Waals interactions and can be exfoliated into thin two-dimensional layers.

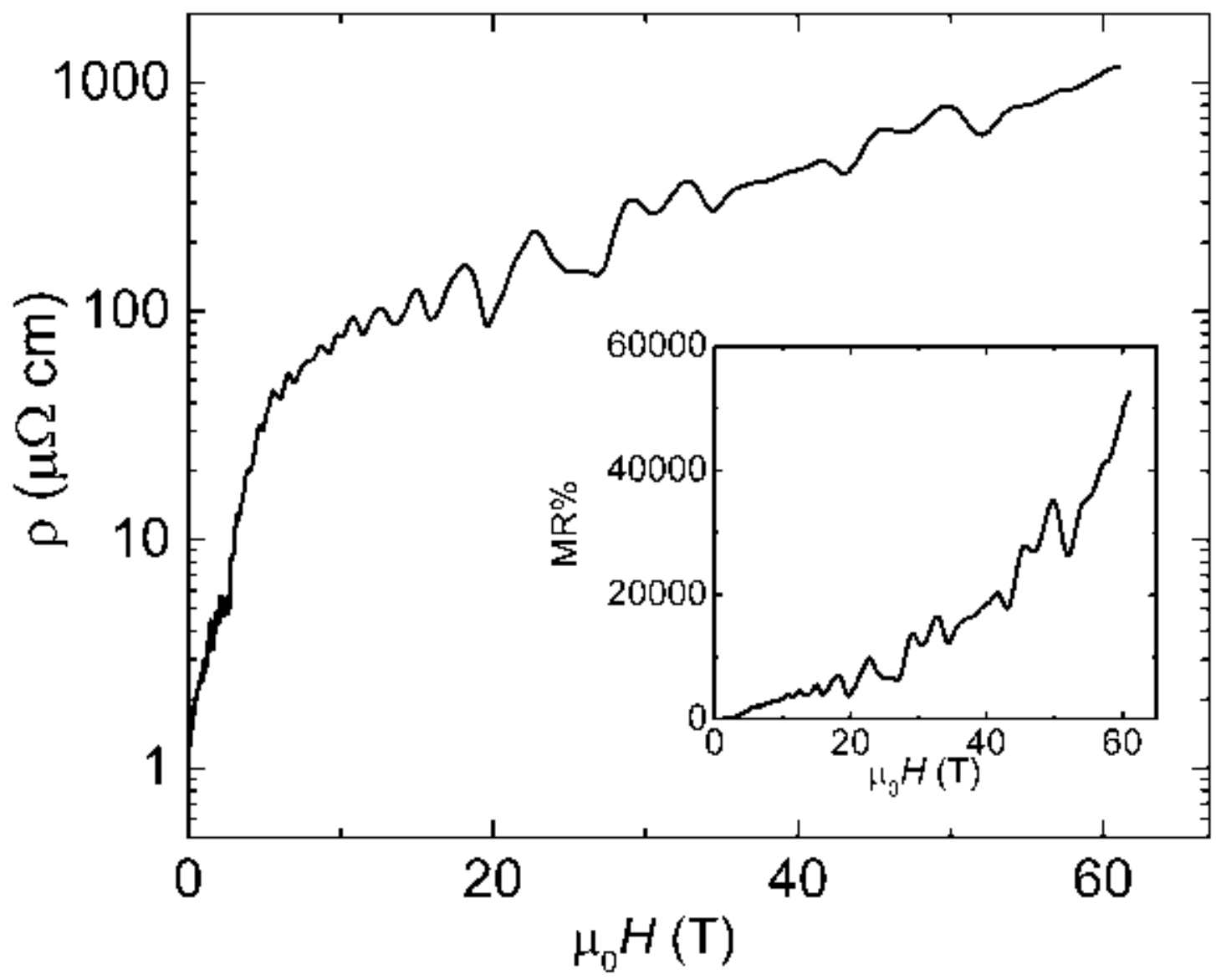

Figure 4.4: MR in pulsed field up to 61 T when the field is aligned around 45° with respect to the $c$ axis. The inset shows the magnetoresistance ratio $MR\% = 100\% \times (\rho(B) - \rho_a)/\rho_a$ .

The angular-dependent magnetoresistivity for the best sample with $RRR = 220$ when the $H$ is tilted 360° from $c$-axis (0°) then back to $c$-axis (360°). Apparently the magnetoresistance maximizes when $H//\ c$ axis and minimizes when $H \perp c$ axis, as shown in Figure 4.5 .

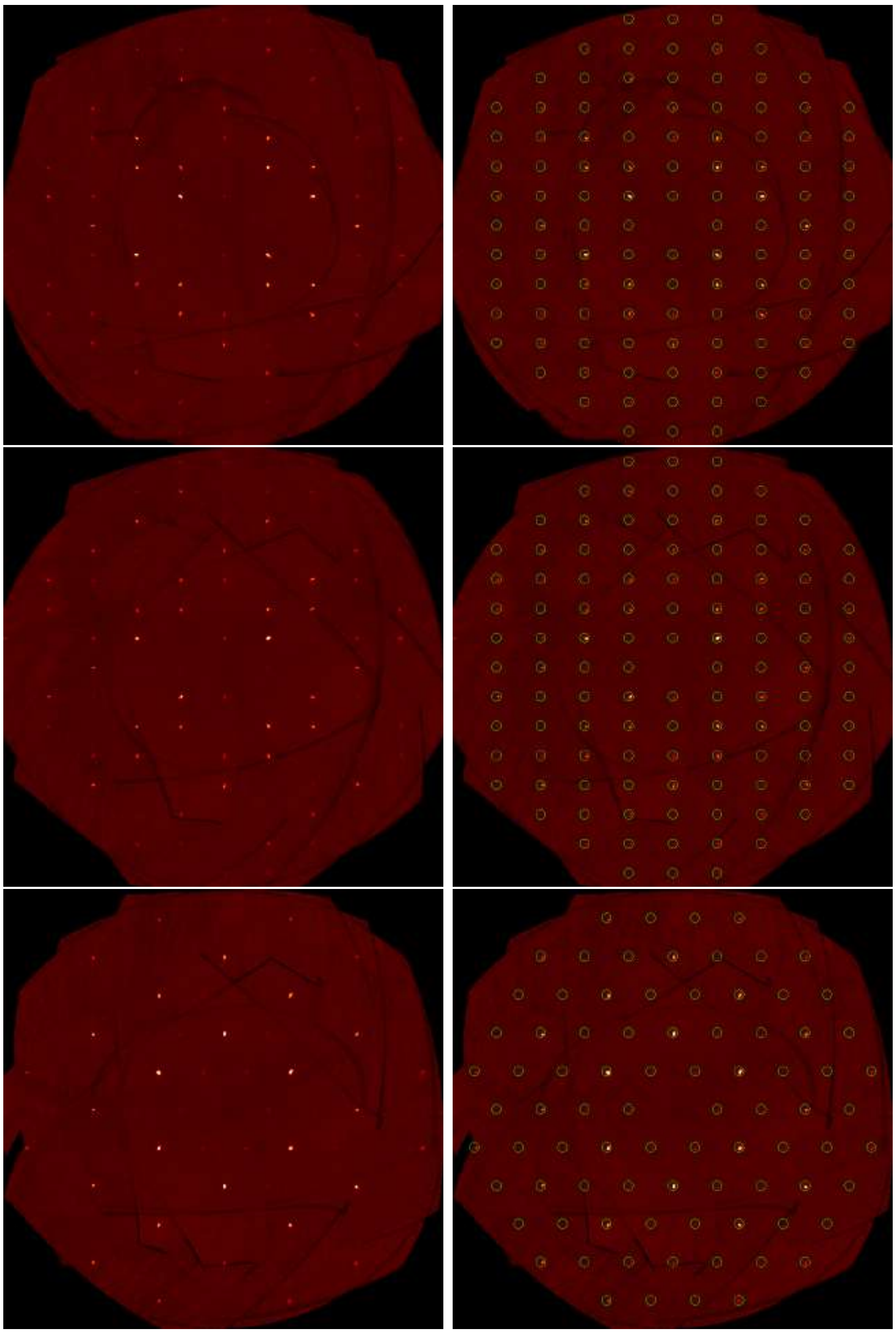

Figure 4.2: Precession images (0kl, h0l, and hk0 with and without green indexing circles) from lattice X-ray diffraction. The size of the $PtTe_2$ single crystal is 0.12 mm × 0.12 mm × 0.01 mm. The unit cell parameters are the following: a = 4.0257Å , c = 5.2235Å; Volume = 73.31$\text{Å}^3$.

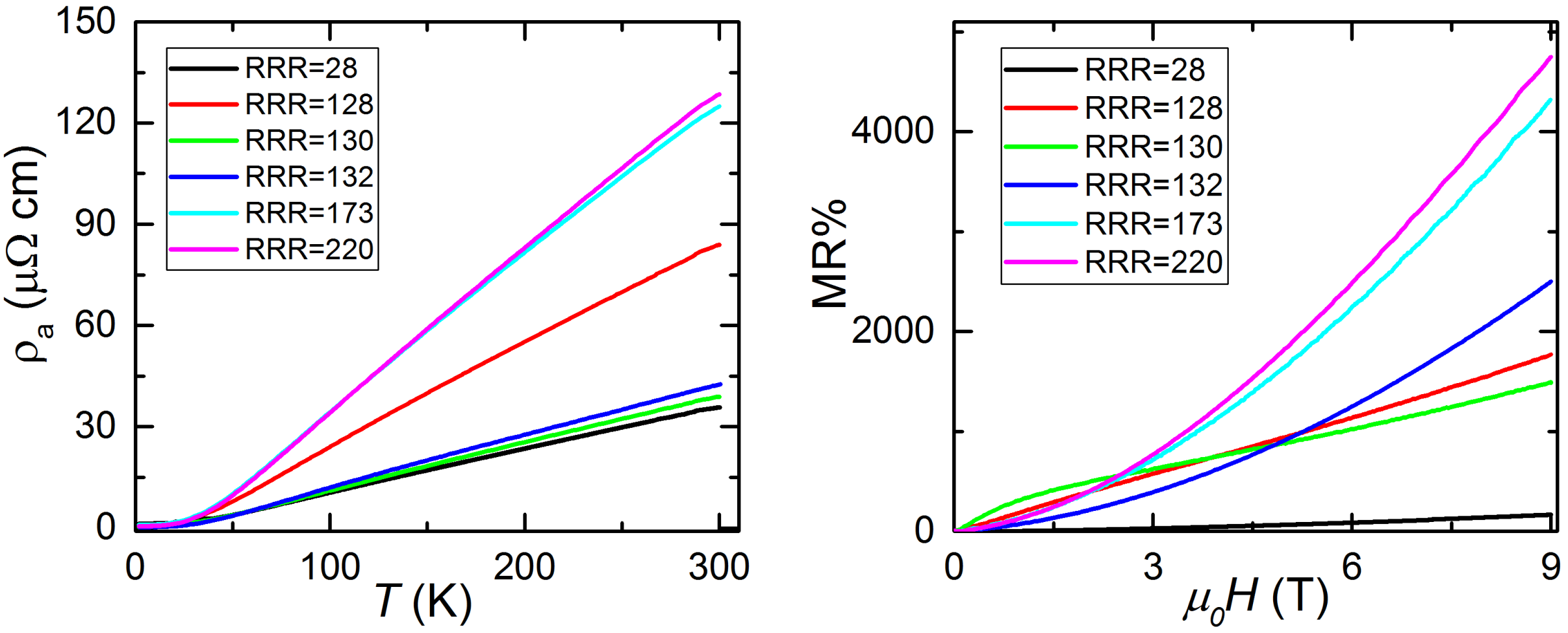

Figure 4.3: Transport measurements in a series of $PtTe_2$ samples. (a) Curves of the resistivity $\rho_a$ as a function of temperature $T$ measured along the needle axis from 2 K to 300 K. (b) Magnetoresistivity ratio and SdH oscillations in magnetic field parallel to the $c$-axis.

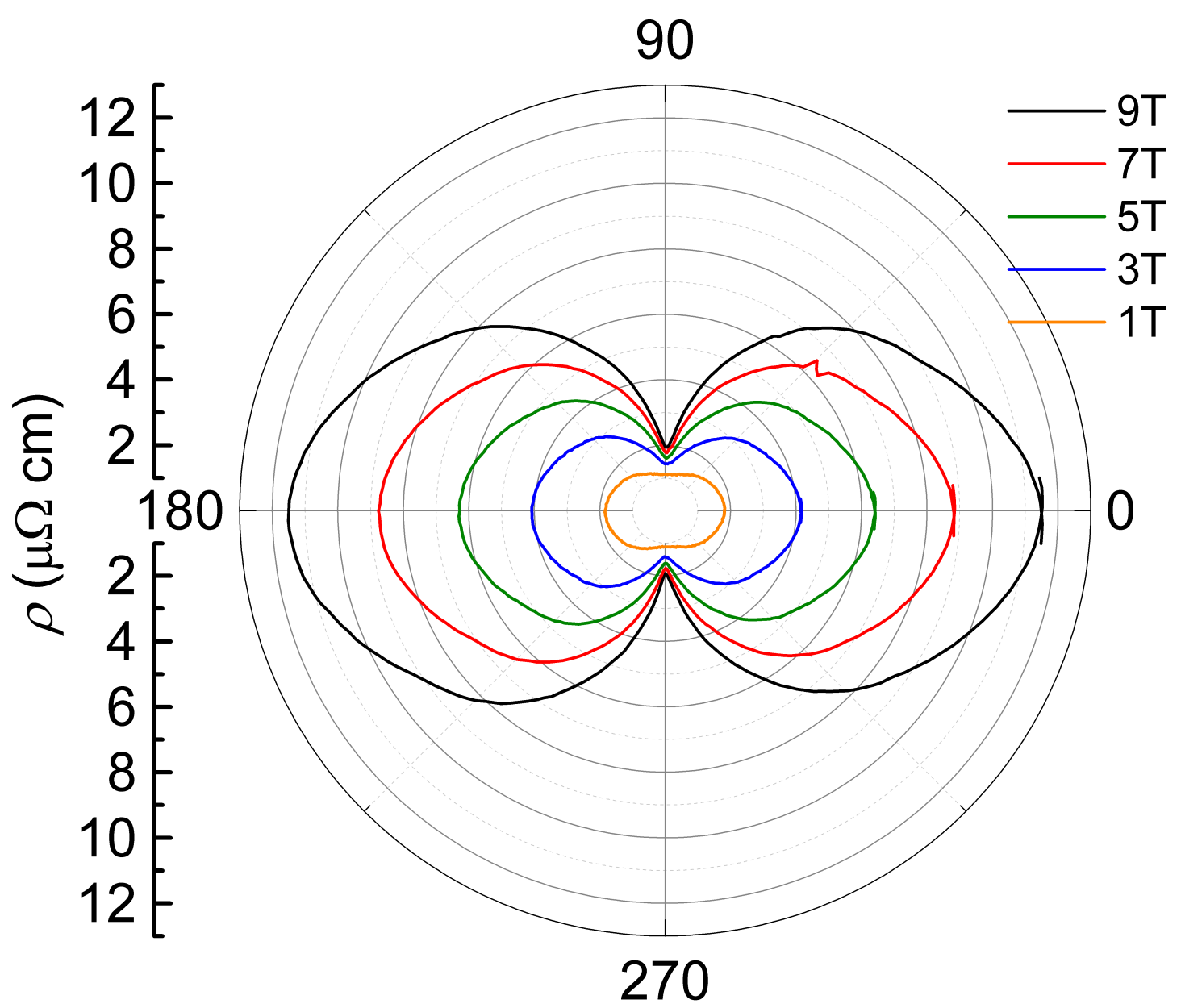

Figure 4.5: Magnetorestivity in tilted $H$ in $PtTe_2$ at 2K. The magnetoresistance reaches maximum as $H$ // $c$ axis while minimum as $H \perp c$ axis.

## 4.2 Hall Effect Analysis of $PtTe_2$

Given the large, non-saturating magnetoresistivity displayed by $PtTe_2$ , which is attributed to nearly perfect electron-hole compensation like $WTe_2$ and $MoTe_2$ [65, 43, 59, 93, 58]. The Hall resistivity as a function of field at several temperatures is displayed in Figure 4.6. The Hall resistivity shows non-linear behavior, particularly at low fields, which is a clear indication for electrical conduction by both types of carriers. As displayed in Figure 4.6(a), at low temperature $T < 30$ K, the Hall resistivity $\rho_{yx}$ switches from electron dominating to hole dominating at $\sim 5$ T. At low field area ($B < 3$ T) the Hall resistivity remains steady at several temperatures below 30 K . At high temperatures $T > 30$ K, the Hall resistivity $\rho_{yx}$ remains electron dominating . One interesting observation is that when the temperature is above 80 K, the Hall resistivity magnitude decreases as the temperature increases as depicted in Figure 4.6(b).

Assuming only a single band in the $PtTe_2$, the Hall mobility is determined by $\mu_H(T) = |R_H(T)|/\rho_{xx}(T)$ while the Hall coefficient $R_H = \rho_{xy}/B$. The temperature dependence of the Hall mobility $\mu_H$ and of Hall coefficient $R_{\rm H}$ are depicted in Figure 4.7. $R_{\rm H}(0$ T) has been determined from the initial slope of $\rho_{xy}$(B). It is worthy pointing out that we obtained a relatively high Hall mobility at 2 K, i.e. $\mu_H \sim 1.3 \times 10^4 \text{cm}^2\ \text{V}^{-1}\text{s}^{-1}$. The $\mu_H$(T) high mobilities decrease drastically downtrend as the temperatures increases above 5 K.

We use isotropic two-carrier model [65, 93, 58] to analyze the carrier density and their mobility. The fits of the Hall resistivity $\rho_{yx}$ in Figure 4.8(a) and the longitudinal resistivity $\rho_{xx}$ in 4.8(b) align very well with the experimental data. In Figure 4.9 we plot the density of electrons and the density of holes as determined from the Hall-effect measurements [65, 93, 58]. The temperature-dependent carrier densities and mobilities extracted from the isotropic two carrier model are shown in Figure 4.9. The electron and hole densities are close in value or of the same order of magnitude, over a broad range of temperatures especially at low temperatures as illustrated in Figure 4.9 (a) and 4.9 (b). Notice that the drastic increase in $n_e$ and $n_h$ as $T$ is reduced below $\sim 60$ K. Similarly, one observes a drastic increment in $\mu_e$ and $\mu_h$ as $T$ is reduced below $\sim 60$ K. The densities of electrons and holes are comparable with the density of holes slightly higher. At the same time that the mobility of electrons is slightly higher than those of the holes. The fitting results at high temperatures above 200 K are not quite robust perhaps due to lack of validity of the two-carrier model at high temperatures.

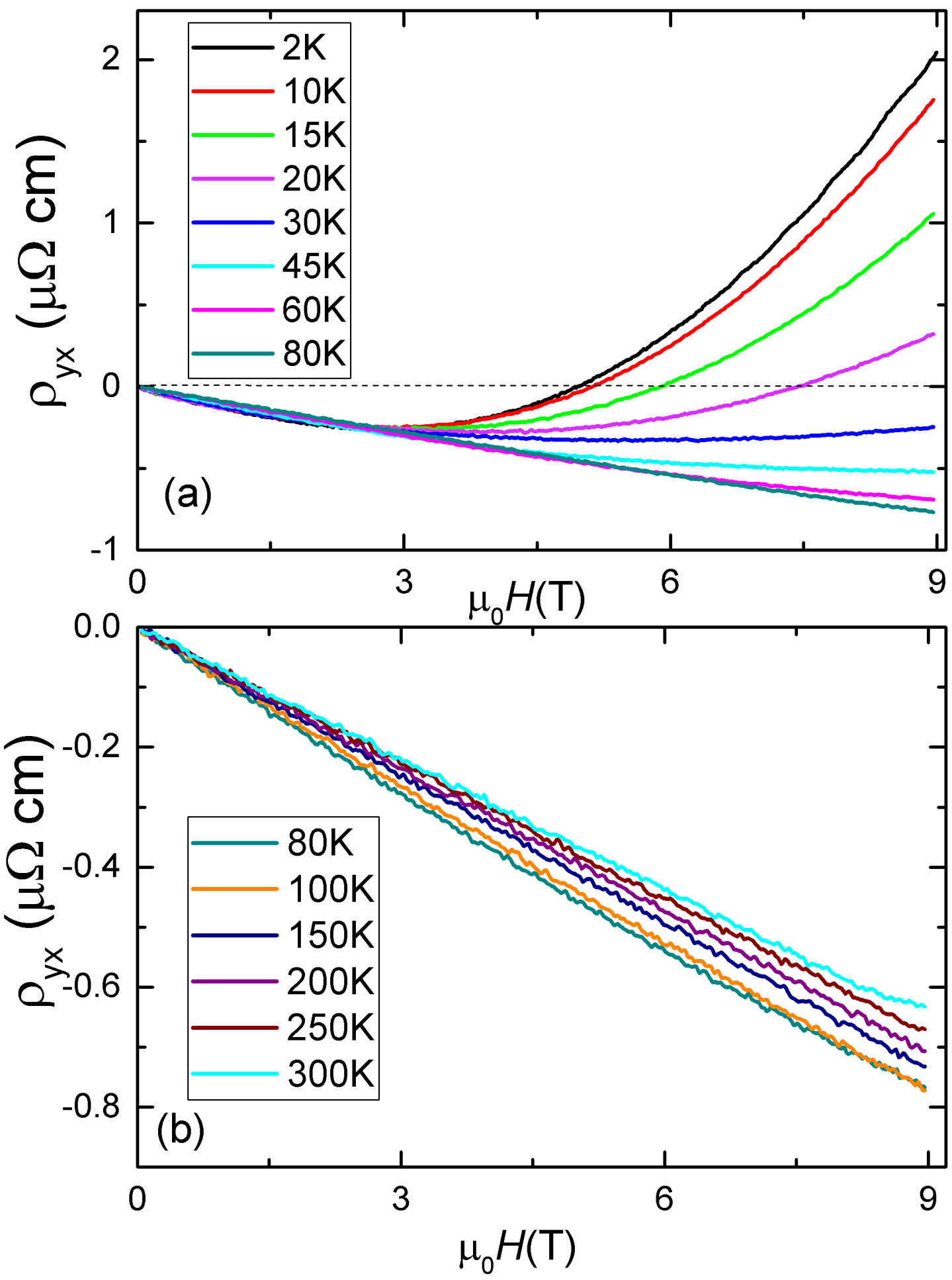

Figure 4.6: Hall resistivity $\rho_{yx}(B)$ at a series of temperatures ranging from 2 K to 300 K. (a) Magnetic field dependent Hall resistivity curves at temperature below 80 K. (b) Field dependent Hall resistivity curves at temperature above 80 K.

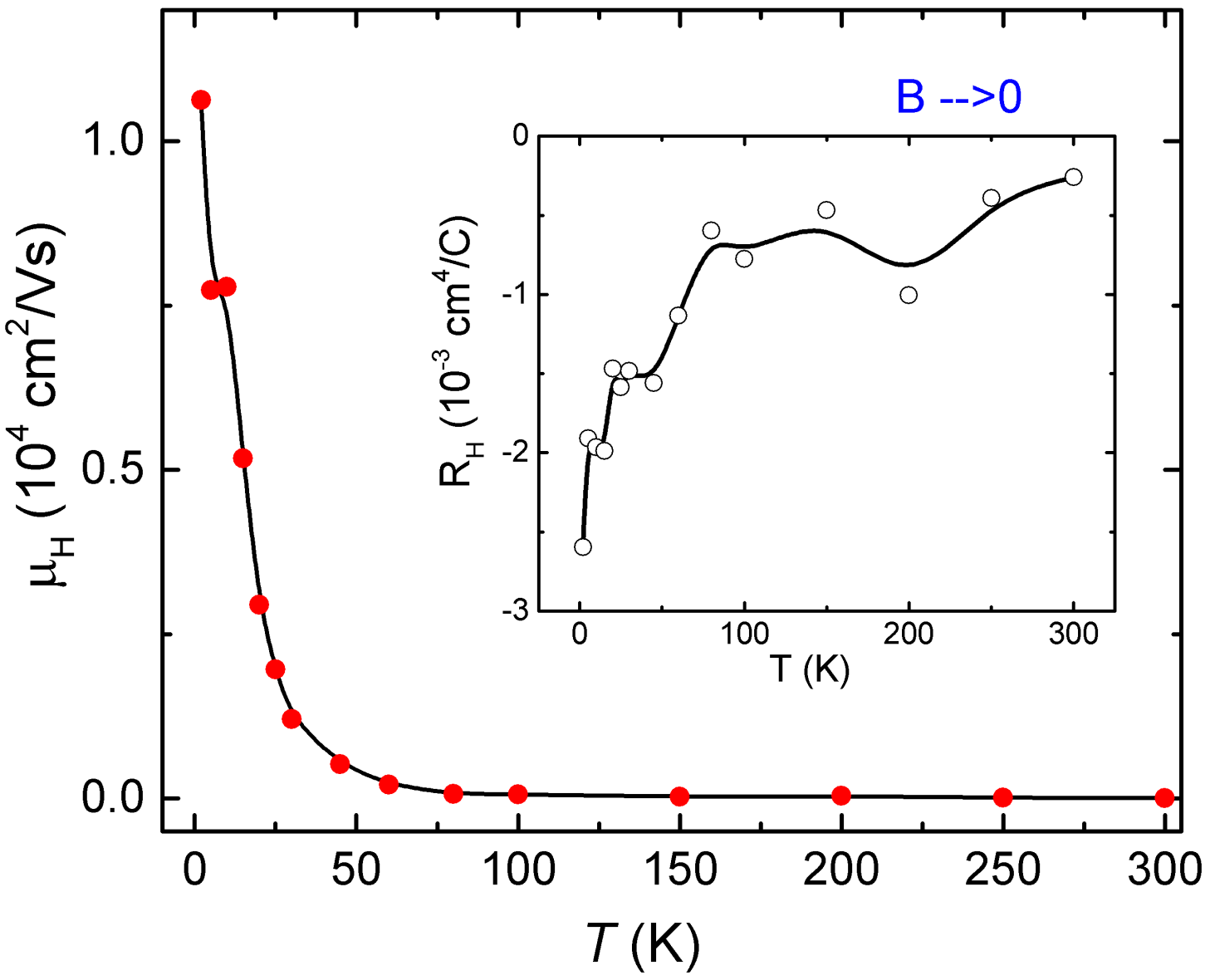

Figure 4.7: Hall mobility $\mu_H$ as a function of temperature at the range from 2 K to 300 K. Inset: Hall coefficient $R_H$ as a function of $T$. Open black symbols in the inset depict the value of Hall coefficient $R_{\rm H}(T)$ which is determined from the initial slope of the Hall resistivity $\rho_{xy}(B)$ as $B \to 0$.

Figure 4.10(a) shows longitudinal resistivity $\rho_{xx}$ as a function of $T$ under several magnetic fields applied perpendicularly to the electrical current. $\rho_{xx}$ shows a minimum at temperature between 20 K and 30 K, then the resistivity shoots up at high temperature. The observed behavior is similar to the one displayed by $WTe_2$ and $MoTe_2$. Namely, under a magnetic field $\rho_{xx}$ (T) essentially follows the zero-field curve until it reaches a minimum at a field dependent temperature $T^*$.

Figure 4.10(b) shows the Hall resistivity $\rho_{xy}$ as a function of $T$ under magnetic fields ranging from $\mu_0 H = 1$ up to 9 T. At high fields $B \geq 5$T, below a field-dependent temperature $T^{neg}$, the Hall resistivity $\rho_{xy}$ shows a pronounced increment towards positive values, while at lower fields below 5 T, the Hall resistivity remain negative even at higher temperatures. It turns out that below $T^{neg}$, one observes a sharp increase in the mobility of both carriers and in the ratio of densities $n_h/n_e$.

## 4.3 De Haas-Van Alphen Quantum Oscillation and Analysis of Berry Phase of $PtTe_2$

The de Haas-van Alphen quantum oscillations in the magnetization were measured with a SQUID magnetometer with the magnetic field both parallel and perpendicular to the $c$-axis. Figure 4.11 displays the oscillatory signal superimposed onto the magnetic torque of $PtTe_2$ as a function of field. One interesting observation is that $PtT_2$ is paramagnetic when $H$ // $c$ while it is diamagnetic when $H//ab$. Both orientations shows beautiful quantum oscillations at several temperatures.

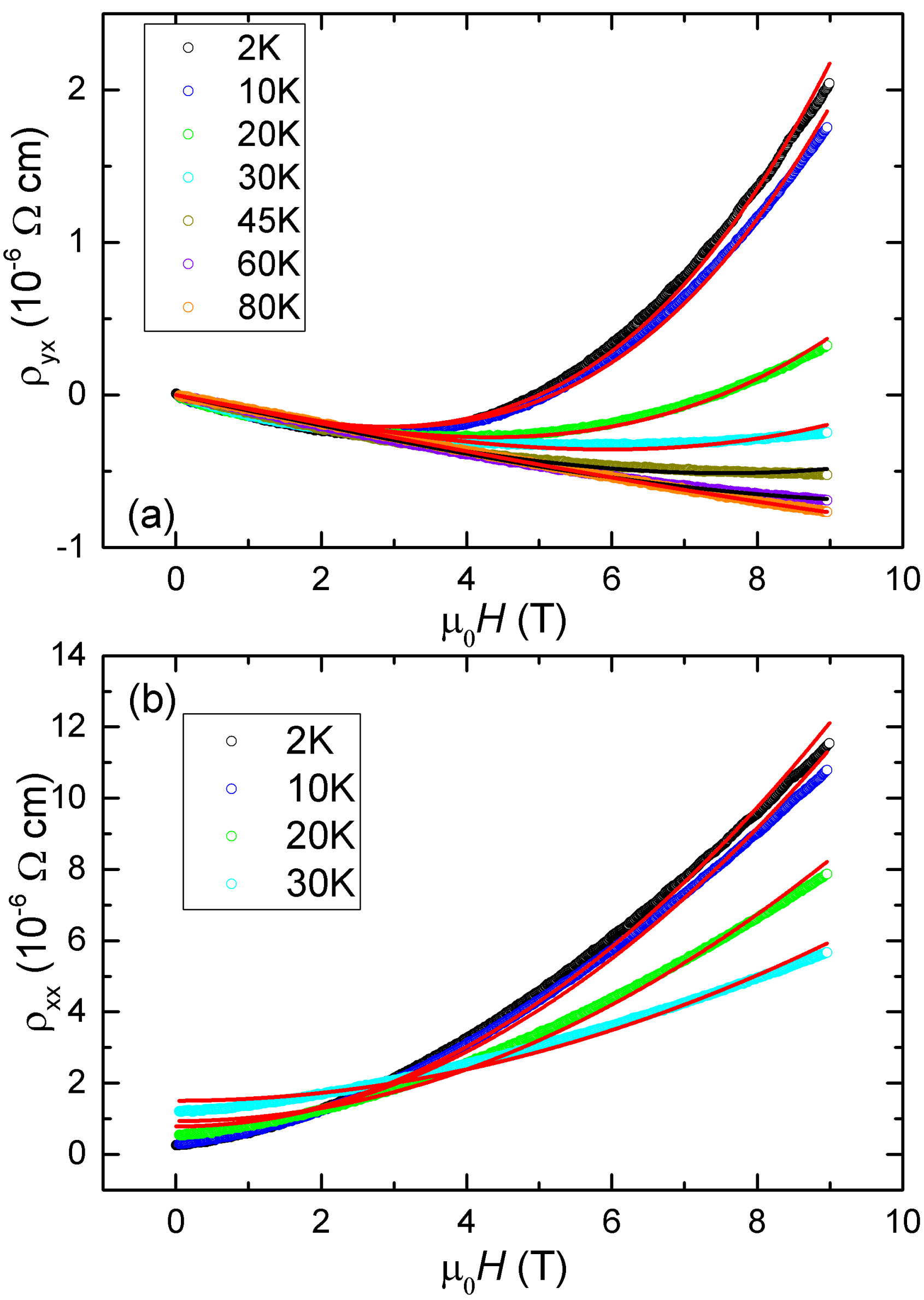

Figure 4.8: Fits to the two-band model of the experimental data for $PtTe_2$.

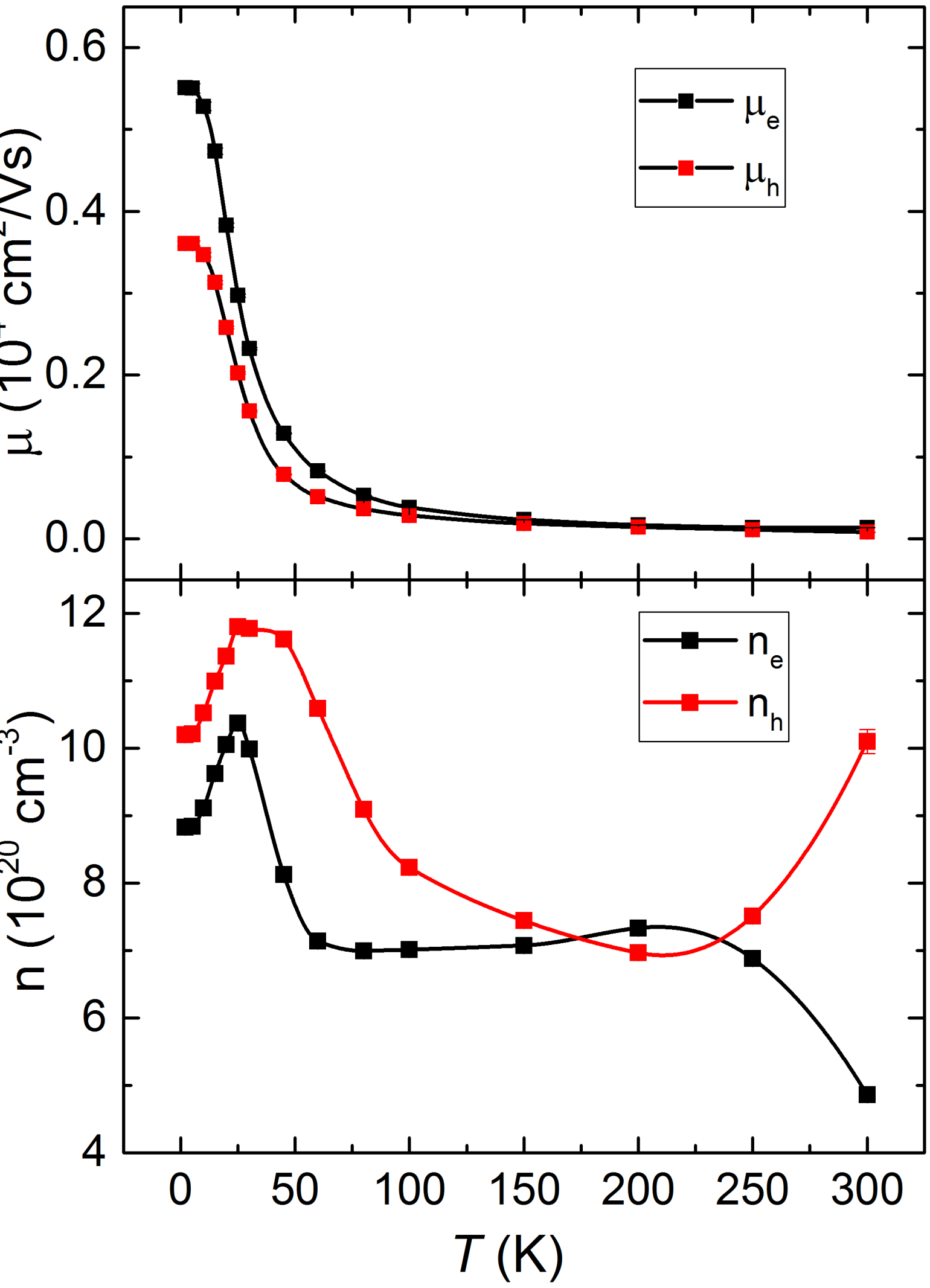

Figure 4.9: Fitting parameters from the two-carrier model applied to the Hall effect for $PtTe_2$. Top figure depicts the density of electrons $n_e$ and density of holes $n_h$ from the two-carrier model analysis of resistivity. The figure at the bottom depicts the carrier mobility $\mu_e$ and $\mu_h$ as a function of temperature extracted from $\rho_{yx}$ and $\rho_{xx}$.

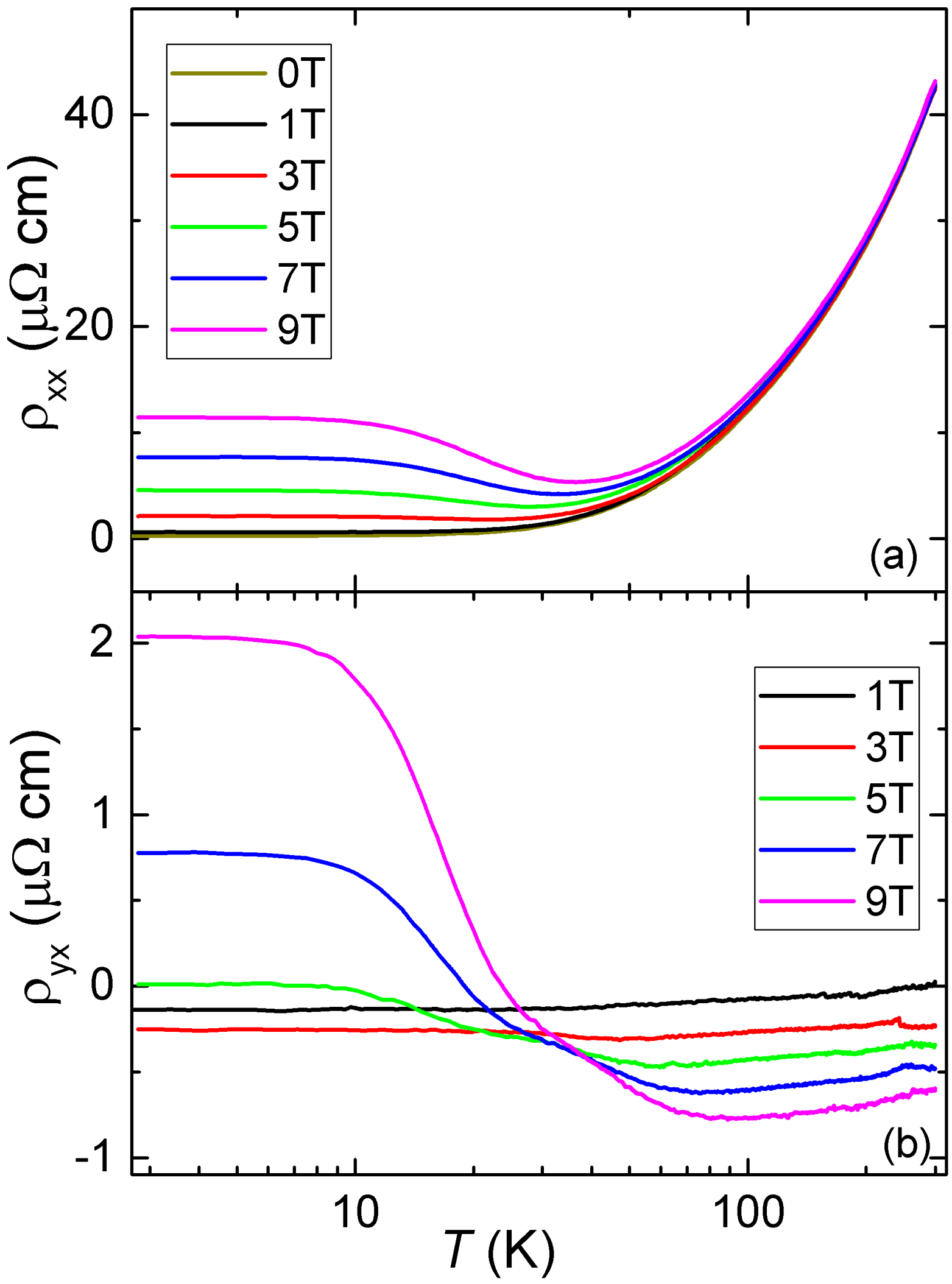

Figure 4.10: (a) $\rho_{xx}$ as a function of $T$ under various magnetic fields applied perpendicularly to the electrical current. $\rho_{xx}$ shows a minimum at temperature between 20 K and 30 K. (b) Hall resistivity as a function of $T$ under various fields.

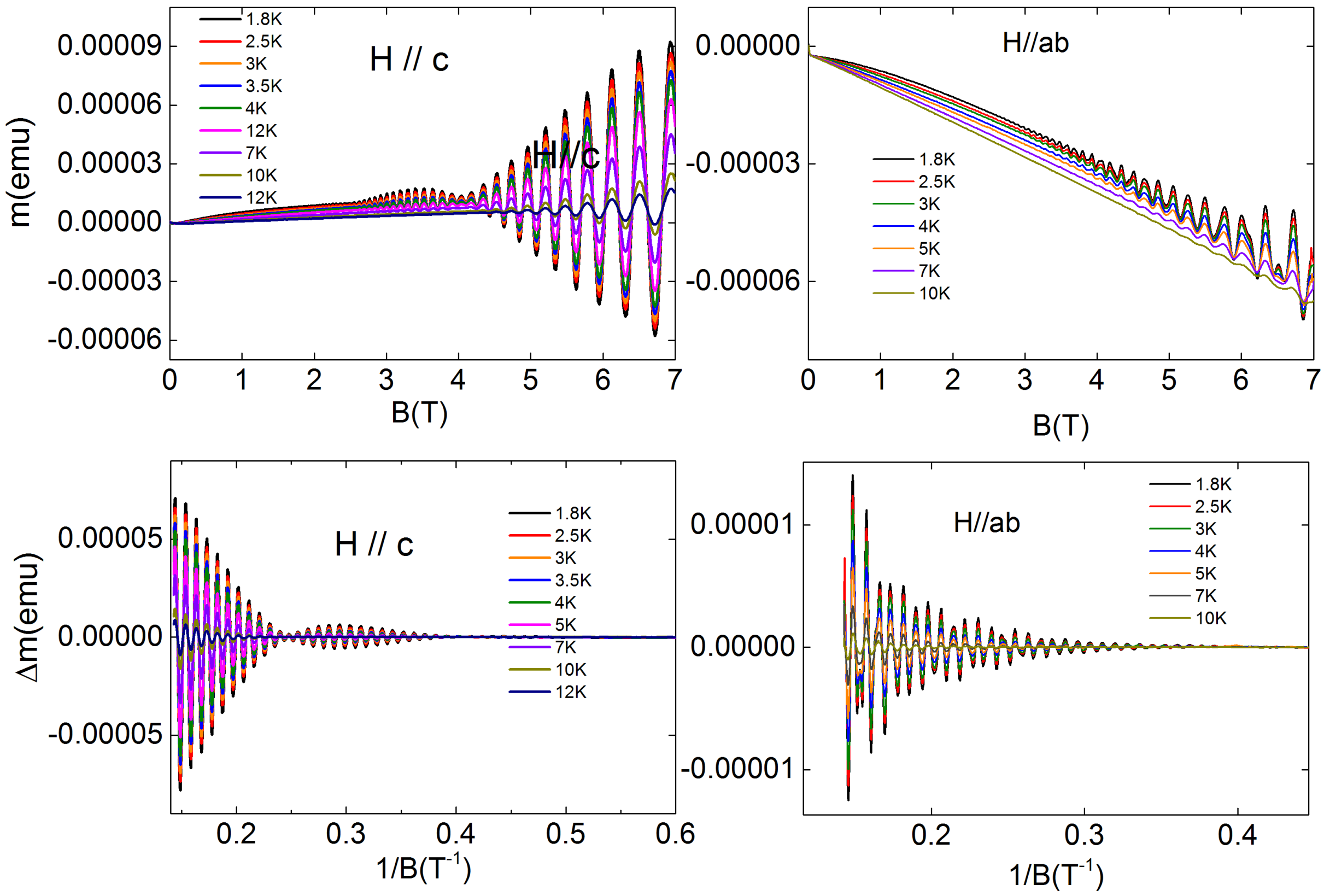

Figure 4.11: De Haas-van Alven (dHvA) effect from $PtTe_2$ at a few temperatures. The raw magnetization curves are in the top left panel and in the top right panel. While the dHvA signal as a function of the inverse field $H^{-1}$ for both $H//c$ axis and $H//ab$ plane are at the bottom left and bottom right, respectively.

Using the standard Lifshitz-Kosevich theory we fitted the temperature-dependent dHvA oscillation amplitude for both field orientations. As shown in the two left figures in Figure 4.12, The FFT of the oscillatory signal for fields along the $c$-axis displays 3 main frequencies around 100 T, 108 T and 243 T, while the FFT for fields along the $ab$-plane shows 3 main frequencies around 123 T, 205 T and 223 T, each is associated with an external cross-sectional area $A$ of the Fermi surfaces through the Onsager relation $F = A(\hbar/2\pi e)$ where $\hbar$ is the Plank constant and $e$ is the electron charge. In the right two figures of Figure 4.12, we plot the amplitude of the main peaks observed in the FFT spectra for fields along the $c$-axis and along the $ab$-plane as a function of the temperature. Red lines are fits to the Lifshitz-Kosevich (LK) temperature damping factor, i.e. $x/\sinh x$ with $x = 14.69\mu T/H$ and $\mu$ being the effecrtive mass, from which we extracted the mass associated with each frequency. We get effective mass around $0.11m_e$ for both of the lower frequencies of and 0.25 $m_e$ for the higher frequency.

Figure 4.13 displays the oscillatory signal superimposed onto the derivative of the magnetic torque (black trace) as a function of inverse field along with fit (red trace) to Lifshitz-Kosevich oscillatory components performed over the range of field from 2.5 T to 7 T. Both traces are nearly in-phase. The relatively small deviation in their respective phases is due to the difference in angle between the magnetic field and $c$-axis of the crystal given that torque measurement requires an angle. The object of this fit is to evaluate the Berry-phase and therefore the topological character of $PtTe_2$. One observes that the Dingle temperatures for the two main frequencies are relatively high, i.e. $T_D$ for the $\alpha-$orbit is 12.5 K while its value for $\beta-$orbit is 11.2 K. We found possible a non-trivial Berry phase $\sim \pi$ for the $\beta$ orbit (108 T), which corresponds to the minimum cross-sectional area in the Fermi surface. However, according to our DFT calculation, when the field is aligned parallel to $c$-axis, the corresponding Fermi surface cross-sectional area would be maximum instead of minimum as displayed in the bottom panel of Figure 4.13. We are not convinced of the non-trivial Berry phase given the contradiction between our data and DFT calculation.

## 4.4 Mapping the Fermi Surface of $PtTe_2$ Through SdH and DHvA Quantum Oscillations

To investigate the electronic structure and Fermi surfaces, we performed detailed temperature and angular-dependent study of the frequencies extracted from the SdH and dHvA effects in $PtTe_2$ .

Figure 4.14 display the FFT from the SdH signals up to 35 T measured in several temperatures from 0.3 K to 5 K for $H//$ $c$-axis. Besides the low frequencies as we discussed above, 3 high frequencies 1701 T, 1971 T, and 6068 T were observed. We plot the amplitude of the main high-frequency peaks observed in the FFT spectra for fields along the $c-$axis as a function of the temperature. Red lines are fits to the Lifshitz-Kosevich (LK) temperature damping factor, from which we extract the masses associated with each high frequency. As seen, for $H//$ $c-$axis one obtains $\mu_{\alpha_1} = 1.6\mathrm{m}_e$ , $\mu_{\beta_1} = 1.6\mathrm{m}_e$ and $\mu_{\gamma_1} = 3.6\mathrm{m}_e$ .

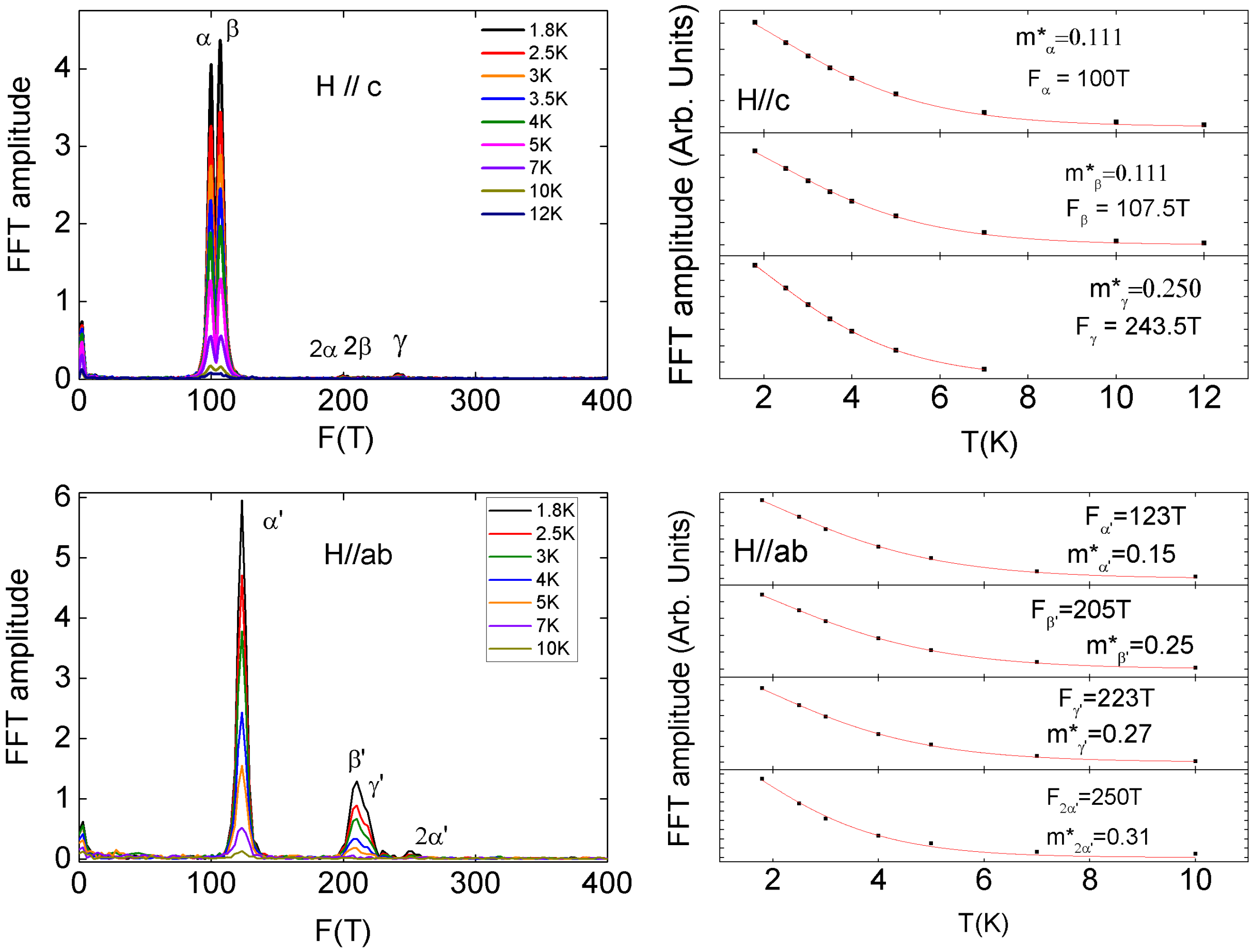

Figure 4.12: (Left panel) Fast Fourier Transform (FFT) of the dHvA signals collected at temperatures ranging from 1.8 K to 12 K for H aligned nearly along the *c*-axis and *ab*-plane. (Right panel) Amplitude if the peaks observed in the FFT spectra for $H//c$ axis and for $H//ab$ as a function of $T$ including the corresponding fits to the LK formula from which the effective masses are extracted.

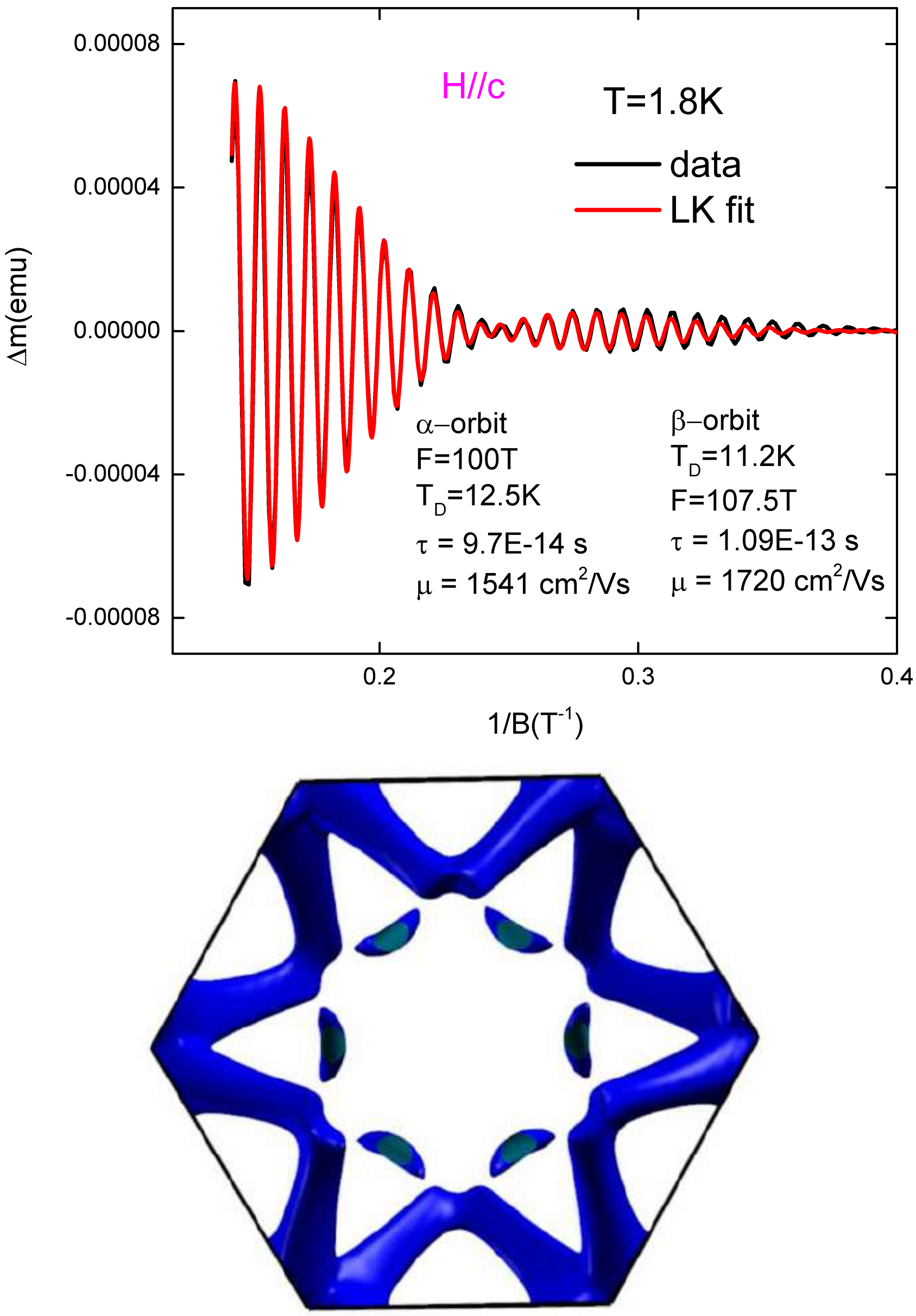

Figure 4.13: From top to bottom panel: fit (red line) of the dHvA signal superimposed onto the magnetic susceptibility (as extracted from the derivative of the magnetic torque $\tau$ with respect to $\mu_0 H$) to the standard Lifshitz Kosevich formula. The bottom one shows electron pockets (band 2) of Fermi surface of $PtTe_2$ via Density Functional Theory (DFT) .

However, as the field is aligned along the *ab*-plane, the FFT spectra reveals a high frequency of 4048 T, with field-dependent effective mass , as demonstrated in Figure 4.15. The effective mass is $\sim 5.1\text{m}_e$ at $10-35$ T window while it is $\sim 6.1\text{m}_e$ at the $10-35$ T window.

Angular-dependent magnetoresistivity at high field up to 35 T is displayed in Figure 4.16 . The magnetoresistivity decreases as the field is aligned from *c*-axis to *ab*-plane which agrees with Figure 4.5.

Figure 4.17 depicts the magnetic torque as the field is rotated from the *c*-axis to the *ab*-plane as temperature of 1.3 K under pulsed field up to 69 T, and the de Haas-van Alphen effect superimposed onto the magnetic torque as a function of the inverse field $H^{-1}$ for several angles between the *c*-axis and the *ab* plane. Figure 4.17 also displays the field dependent magnetic torque for the fields aligned along the *c*-axis at several temperatures.

The FFT spectra extracted from the angular-dependent dHvA effect under pulsed fields and the corresponding frequencies associated to the Fermi surface cross-sectional areas are displayed in Figure 4.18. Only the low frequencies below 300 T are observed due to the signal to noise ratio of the pulsed field. The black markers in both figures depict the frequencies associated to the Fermi surface cross-sectional areas based on our DFT calculations including spin-orbit split. The calculation is able to capture several of the main frequencies between 90 T and 250 T, although most of the frequencies as the angle is between 30 and 80 are not revealed by the DFT.

The comparison between the measured SdH cross-sectional areas and those resulting from the shifted DFT calculations are shown in Figure 4.19, which displays the Fast Fourier Transform spectra from the SdH signals with the markers acting as identifiers of the angular-dependence of the SdH frequencies on the Fermi surfaces according to the DFT calculations. The frequencies resulting from shifted electron (red and green traces) and hole bands (blue traces) as functions of field orientation when rotating from from the *c*-axis to the *ab*-plane for including the spin-orbit-coupling are also shown in 4.19.

The experimental data is better described by the spin-orbit split bands when the field is aligned at an angle of 30° plane between the *a*-axis and the *b*-axis. The Fermi surfaces from the experimental data is rather complicated while the DFT calculation captures several of the main pockets. Still the calculation shows a significant discrepancy between the theoretical calculation and the experimental data, which is highlighted by the absence of a frequency around $\sim 6$ kT associated with a two-dimensional pocket, as shown in Figure 4.20 which displays the angular-dependence of this Fermi surface cross-sectional area which follows a $1/\cos\theta$ dependence.

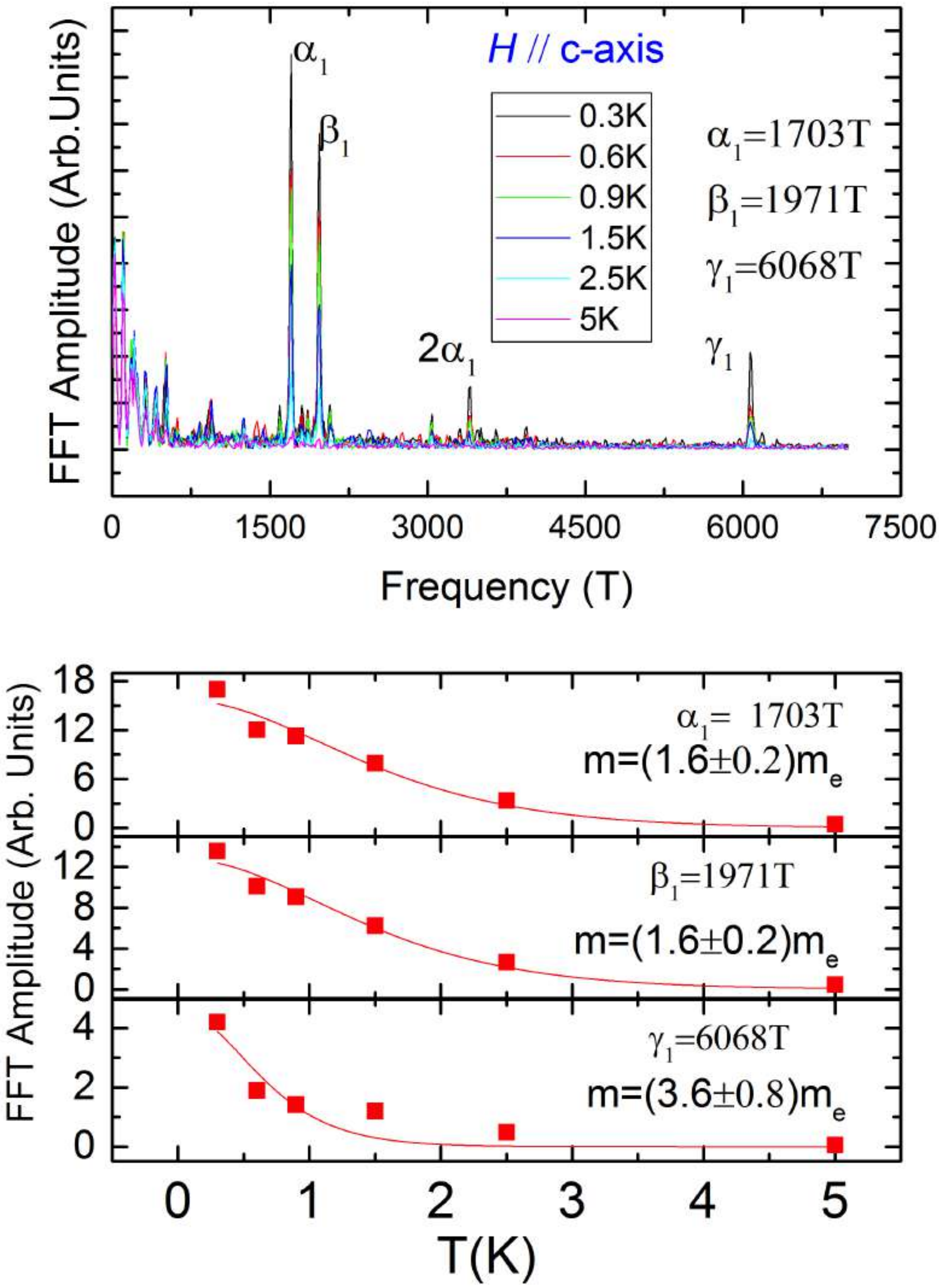

Figure 4.14: (Top panel) FFT spectra for Shubnikov-de Haas signals superimposed on the magnetoresistivity for fields aligned along the *c*-axis of $PtTe_2$ single crystals at several temperatures. (Bottom panel) Amplitude of the peaks observed in the FFT spectra for $H$ // $c$-axis and the corresponsing fits to the LK formula to extract the effective masses.

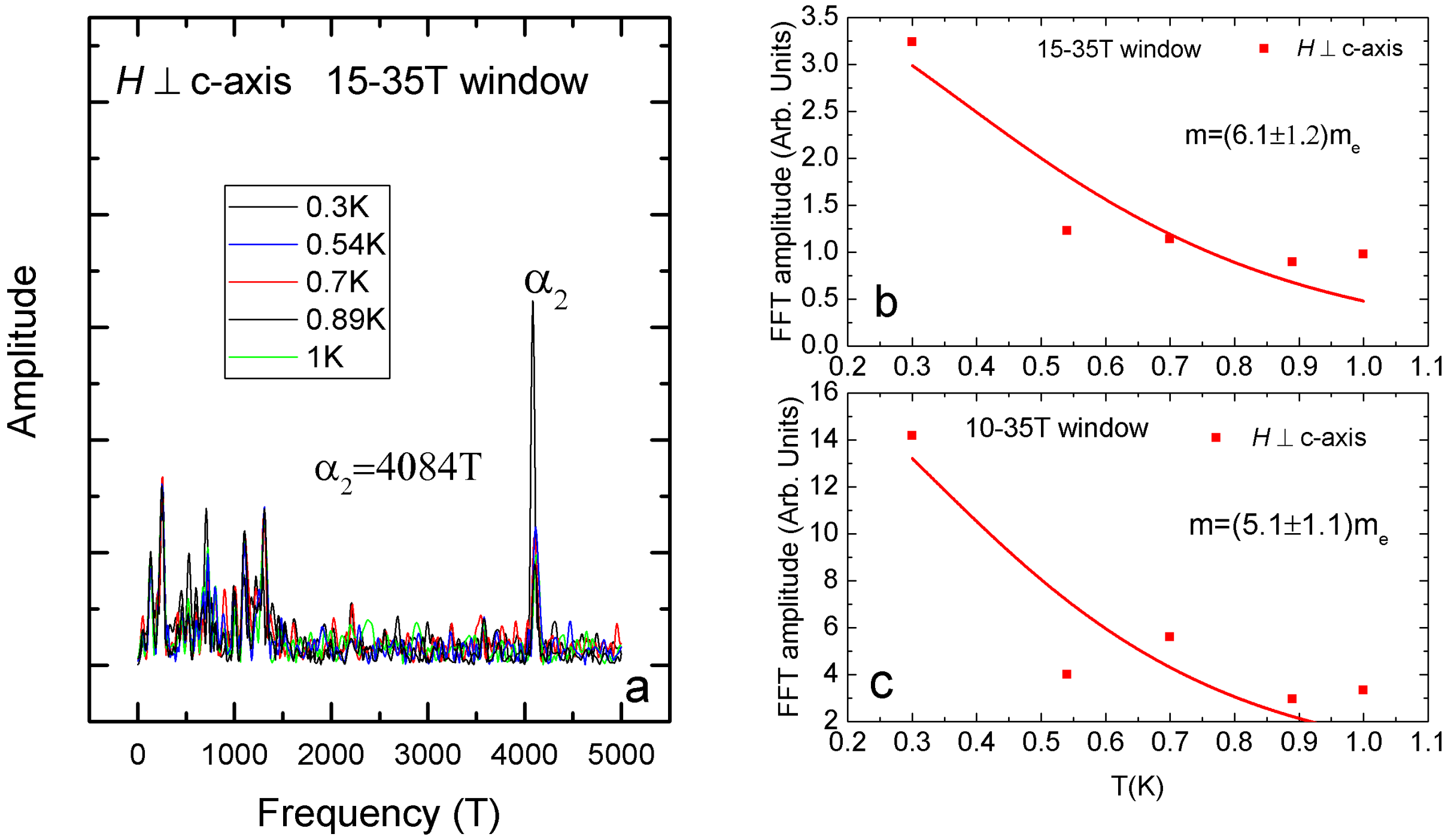

Figure 4.15: (Top panel) FFT spectra of the Shubnikov-de Haas signals superimposed on the magnetoresistivity for fields aligned along the *ab*-plane of $PtTe_2$ single crystal at several temperatures. (Bottom panel) Amplitude of the peaks observed in the FFT spectra for $H$ // *ab*-plane and the corresponding fits to the LK formula to extract the effective masses.

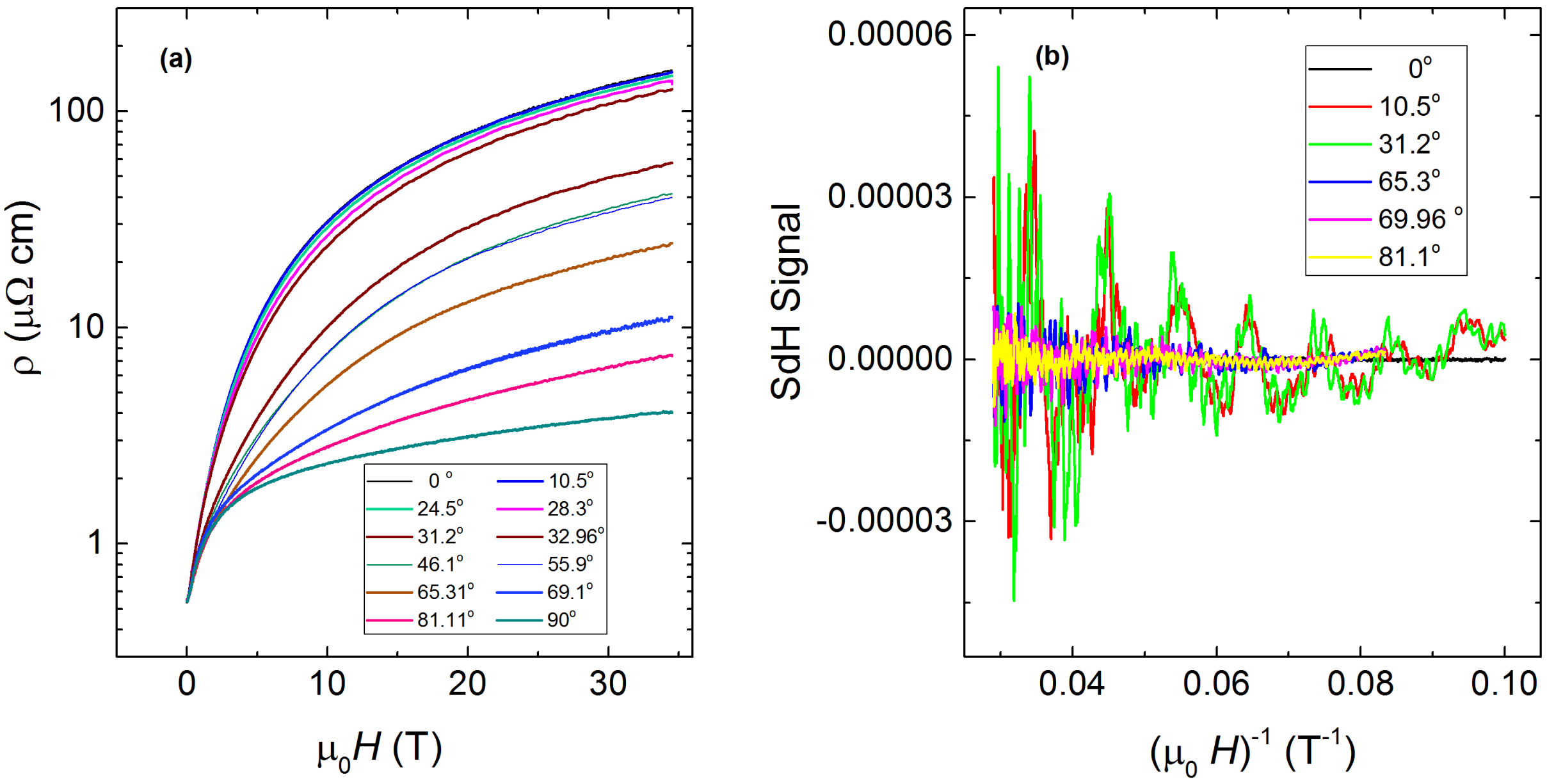

Figure 4.16: (Left panel) Experimental observed angular-dependent magnetoresistivity for $PtTe_2$. (Right panel) SdH spectra from $H//c$-axis to $H \perp c$-axis.

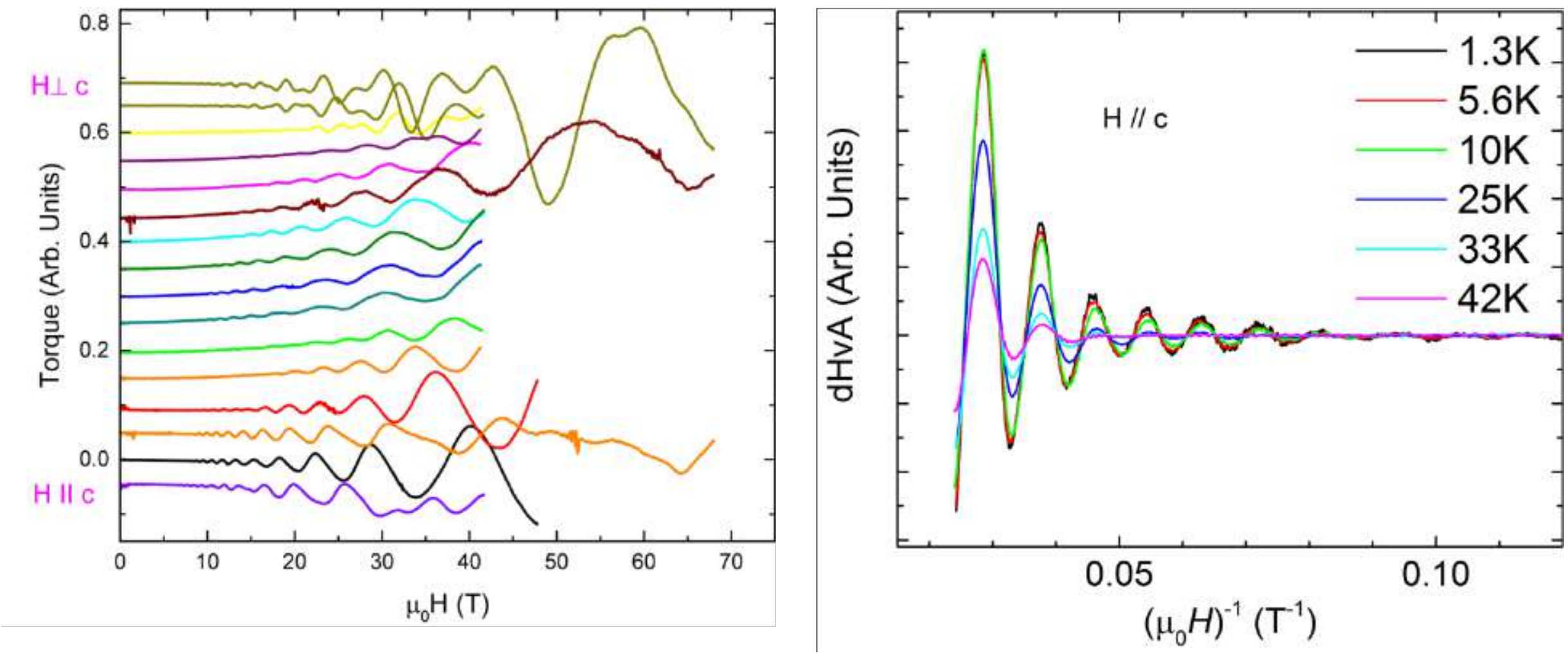

Figure 4.17: (Left panel) Field dependent magnetic torque for fields rotated from the $c$-axis to the $ab$-plane. (Right panel) Oscillatory component , or the dHvA effect, superimposed onto the magnetic torque as a function of inverse field and for several temperatures between 1.3 K and 42 K when $H$ // $c$-axis.

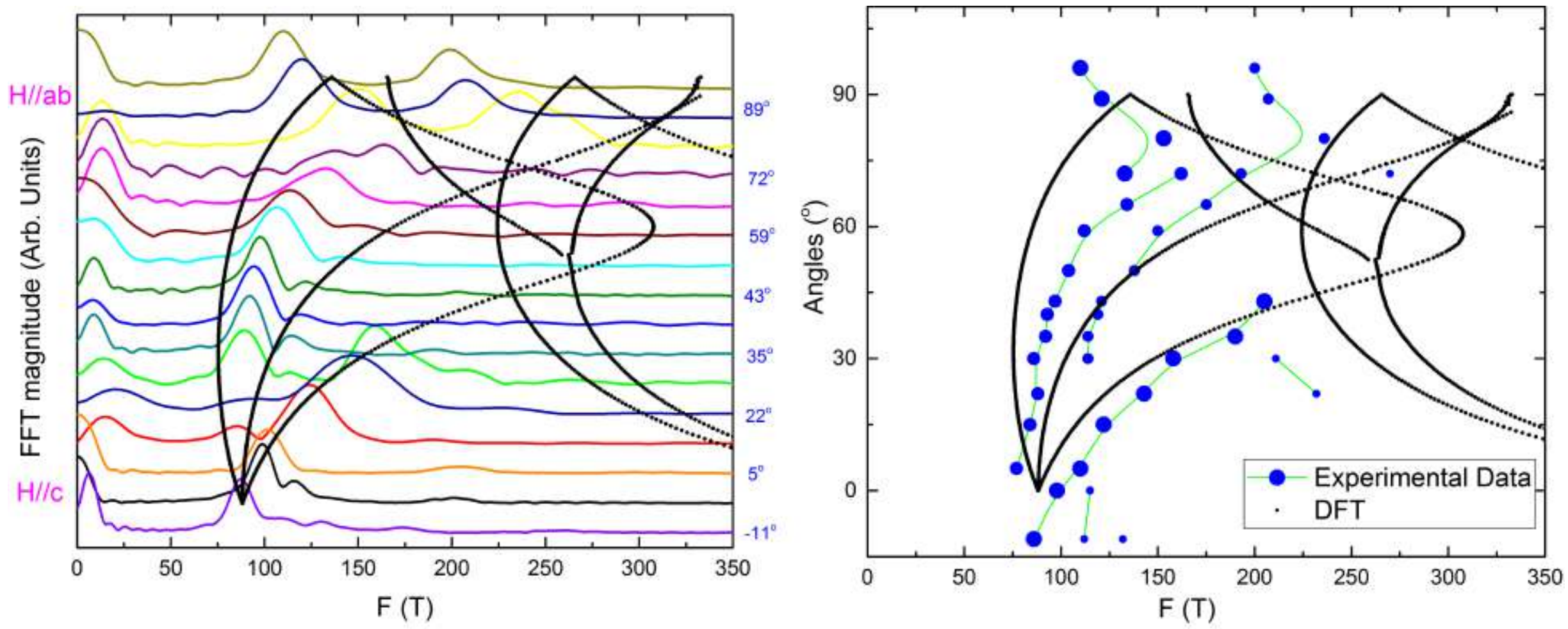

Figure 4.18: (a) Experimental observed dHvA spectra as a function of the frequency $F$ for several angles between $c$- and the $ab$-plane. (b) Angular dependence of the de Haas-van Alphen frequencies , through the Onsager relation, the Fermi surface cross-sectional areas as with respect to the main crystallographic axes, with the size of blue circles are proportional to the amplitude of FFT peaks.

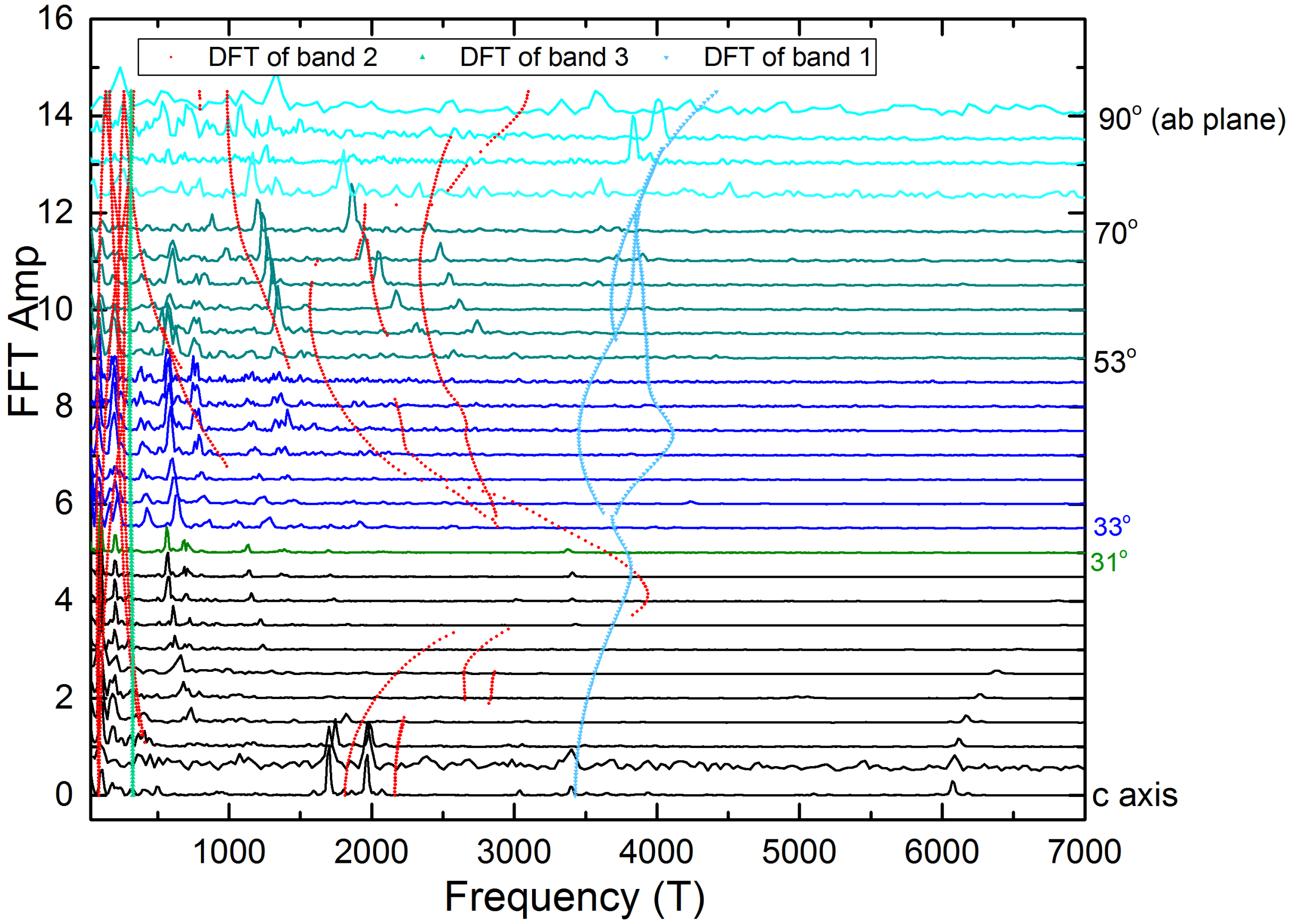

Figure 4.19: Experimentally observed SdH spectra as a function of $F$ for several angles from the $c$-axis to the $ab$-plane for $PtTe_2$. Red, green and blue lines act as identifiers of the angular dependence of the SdH effect, or frequencies on the Fermi surfaces resulting from the shifted electron and hole-bands in the presence of spin-orbit coupling according to the DFT calculation. Here electron-orbits are depicted by red and green markers ( bands 2 and 3) while the hole ones (band 1) is indicated by blue markers.

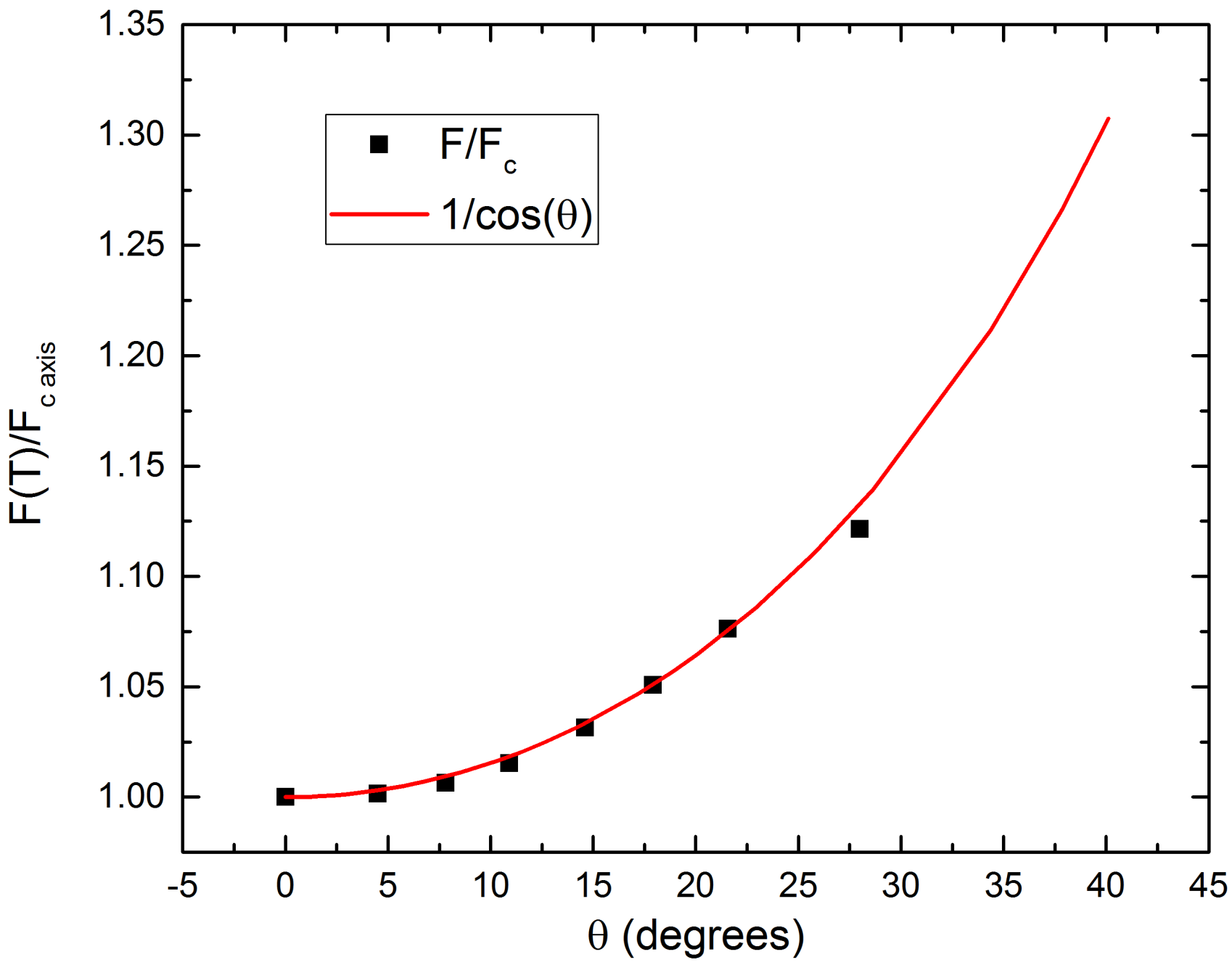

Figure 4.20: Ratio between the angle dependent frequency and the $c$-axis frequency as a function of $\theta$, the angle between the field and the $c$-axis of the crystal, for the ∼6 kT band observed in the experimental data. Red trace is a fit to $1/\cos\theta$.

# CHAPTER 5

# CONCLUSION AND OUTLOOK

We observed non-saturating magnetoresistance which is ascribed to nearly perfect electron-hole compensation in $\gamma$-$MoTe_2$. Anomalies observed in the carrier densities, carrier mobilities, Hall coefficient and in the heat capacity as a function of the temperature suggests a cross over or a electronic transition between 30 and 70 K in $\gamma$-$MoTe_2$. We speculated that the crystallographic structure evolve upon lattice cooling below 100 K, however the results from synchrotron X-ray don't support the hypothesis. Poor agreement between quantum oscillations and DFT calculations including spin-orbit coupling suggests the role of spin-orbit coupling is overestimated in $\gamma$-$MoTe_2$.

For the predicted type-II Dirac semimetal $PtTe_2$, two-carrier model might not be able to explain the non-saturating magnetoresistance given that we have far higher value of Hall mobility than the estimated mobility of carriers according to two-carrier model. We found non-trivial Berry phase $\pi$ for $\beta$ orbit but it is not convincing due to the contradiction with the DFT calculations. The Fermi surface extracted from the dHvA effect and SdH effect is relatively complicated and small pockets have light effective masses. There are several large pockets are not able to be captured by the DFT calculations including spin-orbit split in $PtTe_2$.

To further probe the band structure of $PtTe_2$ to see if it is possible to describe the angular-dependence of the observed dHvA and SdH Fermi surface cross-sectional areas, more comprehensive work on the DFT calculations including tuning the Fermi level or even shift the bands guided by the most recent ARPES results of $PtTe_2$ could be needed. Secondly, it will be interesting to investigate further on the sample-dependent magnetoresistivity in $PtTe_2$.

# APPENDIX A

# ANALYSIS OF BULK FERMI SURFACES FOR $\gamma$-MOTE$_2$

## A.1 Anisotropy in Superconducting Upper Critical Fields

As displayed in Figure A.1(a), superconducting upper critical fields Hc2 for a $\gamma$-MoTe$_2$ single-crystal (sample #2 or blue trace in Fig. 3.1(b)) as a function of the temperature $T$ and for fields applied along all three main crystallographic axes. Solid lines are fits to a Ginzburg-Landau expression $H^i_{c2}(T) = \phi_0/(2\pi\xi_j(0)\xi_k(0))[(1-t^2)/(1+t^2)]$, $\phi_0$ is the quantum of magnetic flux, $\xi_i(0)$ is the zero temperature superconducting coherence length along a certain crystallographic axis, and $t$ is the reduced temperature $T/T_C$. Figure A.1(b) shows anisotropy in upper critical fields $\gamma = H^i_{c2}/H^j_{c2}$ , for fields applied along the $a$-and the $c$-axis (violet markers) and for fields along the a- and the b-axis (pink markers), respectively. Notice how $\gamma$ decreases slightly upon decreasing $T$, suggesting multi-band superconductivity.

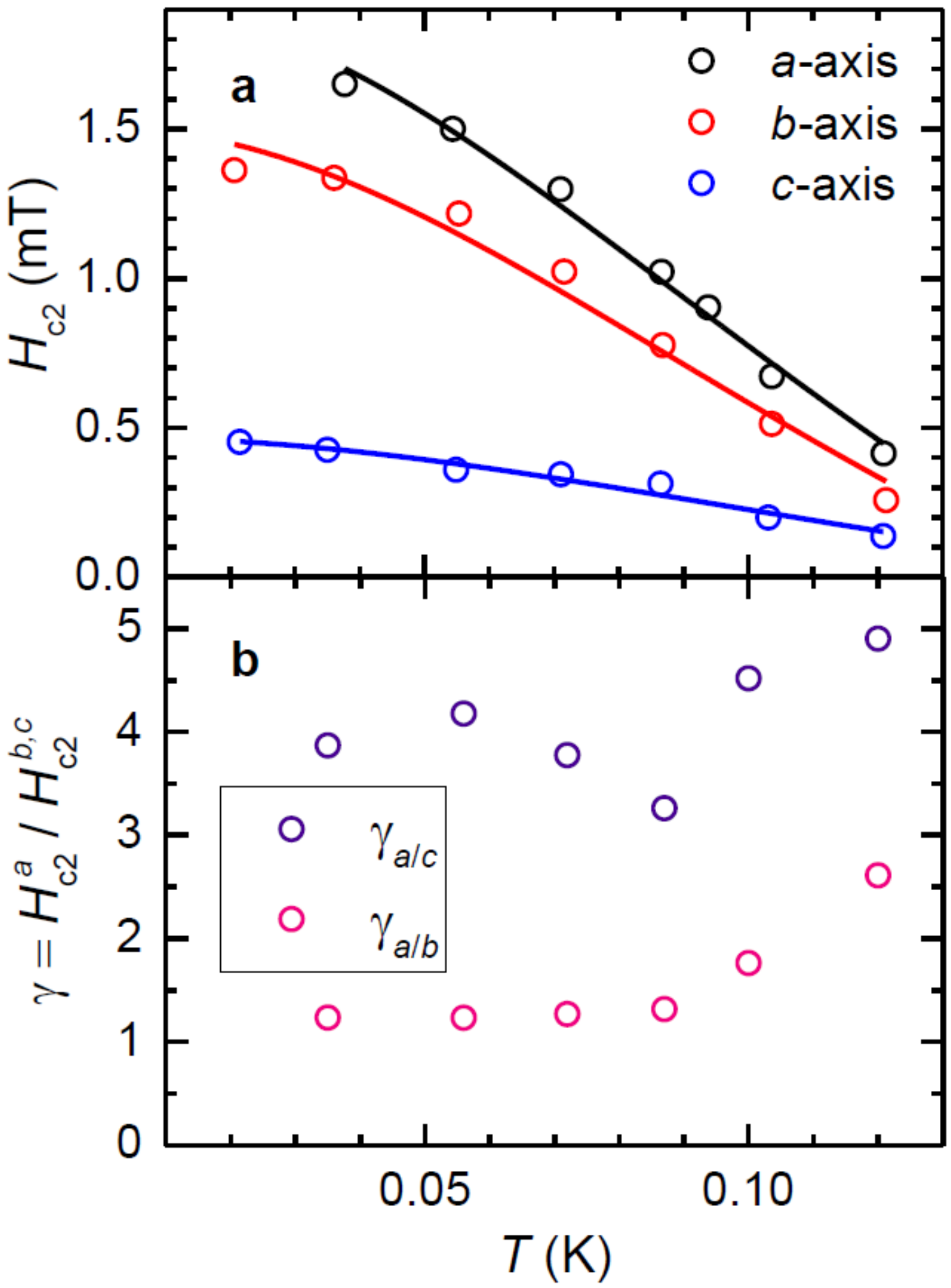

Figure A.1: Anisotropy in superconducting upper critical fields for a $\gamma$-$MoTe_2$ crystal.

## A.2 Hall-Effect in $\gamma$-$MoTe_2$: Possible Lifshitz Transition

Given the enormous, non-saturating magnetoresistance displayed by $\gamma$-$MoTe_2$, which is ascribed to nearly perfect electron-hole compensation [65, 43, 59], as claimed to be the case for $WTe_2$, we plot its density of electron and holes as determined from Hall-effect measurements [58]. Our detailed, two-band analysis of the Hall response of $\gamma$-$MoTe_2$ can be found in Ref. [58].

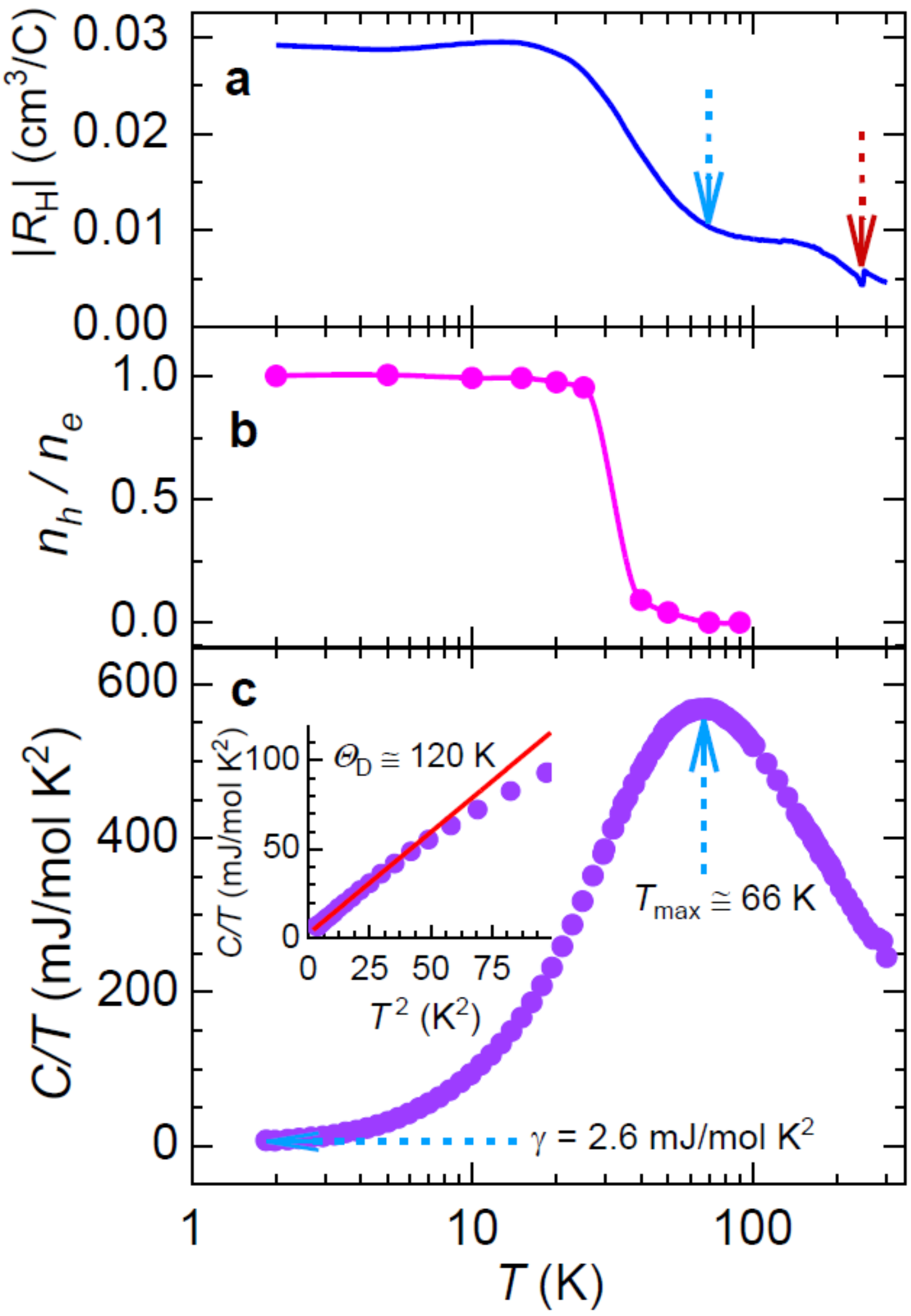

Figure A.2: (a) Absolute value of the Hall constant $|R_{\rm H}|$ measured under a field $H = 9$ T as a function of the temperature. Brown arrow indicates the temperature $T \sim 240$ K where the monoclinic to orthorhombic transition occurs. Blue arrow indicates the temperature below which a sharp increase is observed in the Hall response, suggesting a crossover or a possible electronic phase transition. (b) Ratio between the density of holes $n_h$ and the density of electrons $n_e$ as a function of $T$ and as extracted from a two-band analysis of the Hall-effect. Notice how $n_h$ increases very quickly below 40 K reaching parity with $n_e$ below $\sim 15$ K. Hence, the Hall-effect indicates that this compound becomes nearly perfectly compensated at low temperatures. (c) Heat capacity $C$ normalized by the temperature $T$ and as a function of $T$ for a $\gamma$-MoTe$_2$ single-crystal. A broad anomaly is observed at $T_{max} \cong 66$ K around which $\mu_{\rm H}$ is seen to increase abruptly. At low $T$s, $C/T$ saturates at $\gamma \cong 2.6$ mJ/molK$^2$ which corresponds to the electronic contribution to C/T. Inset: $C/T$ as a function of $T^2$. Red line is a linear fit from which one extracts the phonon coefficient $\theta$ in $C/T \propto \theta T^2$ which yields a Debye temperature $\Theta_D \cong 120$ K.

As seen in Figure A.2 (a), the Hall-effect yields two important observations: two anomalies in its temperature dependence (measured under $\mu_{\mathrm{H}}$ = ± 9 T), the first associated with the monoclinic to orthorhombic transition (indicated by the brown arrow) and the second yielding a sharp increase in the Hall constant below ~ 80 K (blue arrow) which leads to a very sharp increase in the density holes making the system almost perfectly compensated below 30 K (see Figure A.2 (b)). In fact, at low temperatures the Hall resistivity is basically linear all the way up to $\mu_{\mathrm{H}} = 9$ T indicating that $\gamma$-$MoTe_2$ is better compensated than $WTe_2$, see e.g. Ref. [93]. A sharp increase in the Hall constant at lower Ts is suggestive of an electronic phase-transition akin to the temperature-dependent Lifshitz-transition [106] reported for $WTe_2$. To evaluate this hypothesis, we performed heat capacity measurements in a $\gamma$-$MoTe_2$ single-crystal. Figure A.2 (c) displays the specific heat $C$, normalized by the temperature and as a function of $T$ in a logarithmic scale. $C/T$ displays a broad anomaly around $T \cong 66$ K roughly at the same $T$ where one observes a sharp increase in the Hall-constant. Extrapolation to $T = 0$ K yields a sizeable electronic contribution term $\gamma_e \cong 2.6$ mJ/mol$K^2$, a surprising value for a semi-metal characterized by a relatively small density of carriers. For instance, metallic copper which contains a ~ $10^2$ larger density of carriers yields $\gamma_e \cong 0.69$ mJ/mol$K^2$. Therefore, the large $\gamma_e$ suggests a relatively high density of states at the Fermi level which is consistent with the observation of superconductivity. The broad anomaly cannot be reconciled with either a first- or a second-order phase-transition. Instead, it suggests a crossover between two regimes, and since it occurs at temperatures where $C/T$ is dominated by phonons, we speculated that the crystallographic structure would evolve upon cooling below $T = 100$ K. The Hall-effect, on the other hand, indicates that the electronic structure would be affected by the evolution of the lattice upon cooling. However, our detailed low temperature synchrotron based structural analysis does not support this scenario, see Figure A.3 below.

## A.3 Crystallographic Structure Characterization From Synchrotron X-Ray Diffraction

Three single-crystals were measured at the CHESS-A2 beam line. Crystal 1 was measured with 19.9 keV incident photons (with an energy bandwidth of $\sim 2$ eV). This crystal, had dimensions of $\sim 100 \times 30 \times 10$ $\mu m^3$, and was fully bathed by a nitrogen gas jet when subjected to the $200 \times 200$ $\mu m^2$ incident beam. For Sample 2, the incident beam was reduced to $80 \times 80$ $\mu m^2$ in to illuminate fewer grains from a larger crystal (few mm long) characterized by significant structural disorder. The temperature of this sample was controlled using a closed-cycle cryostat. The diffracted X-rays were recorded using a Pilatus3 300 K area detector mounted on the $2\theta$ arm of a 4-circle diffractometer.

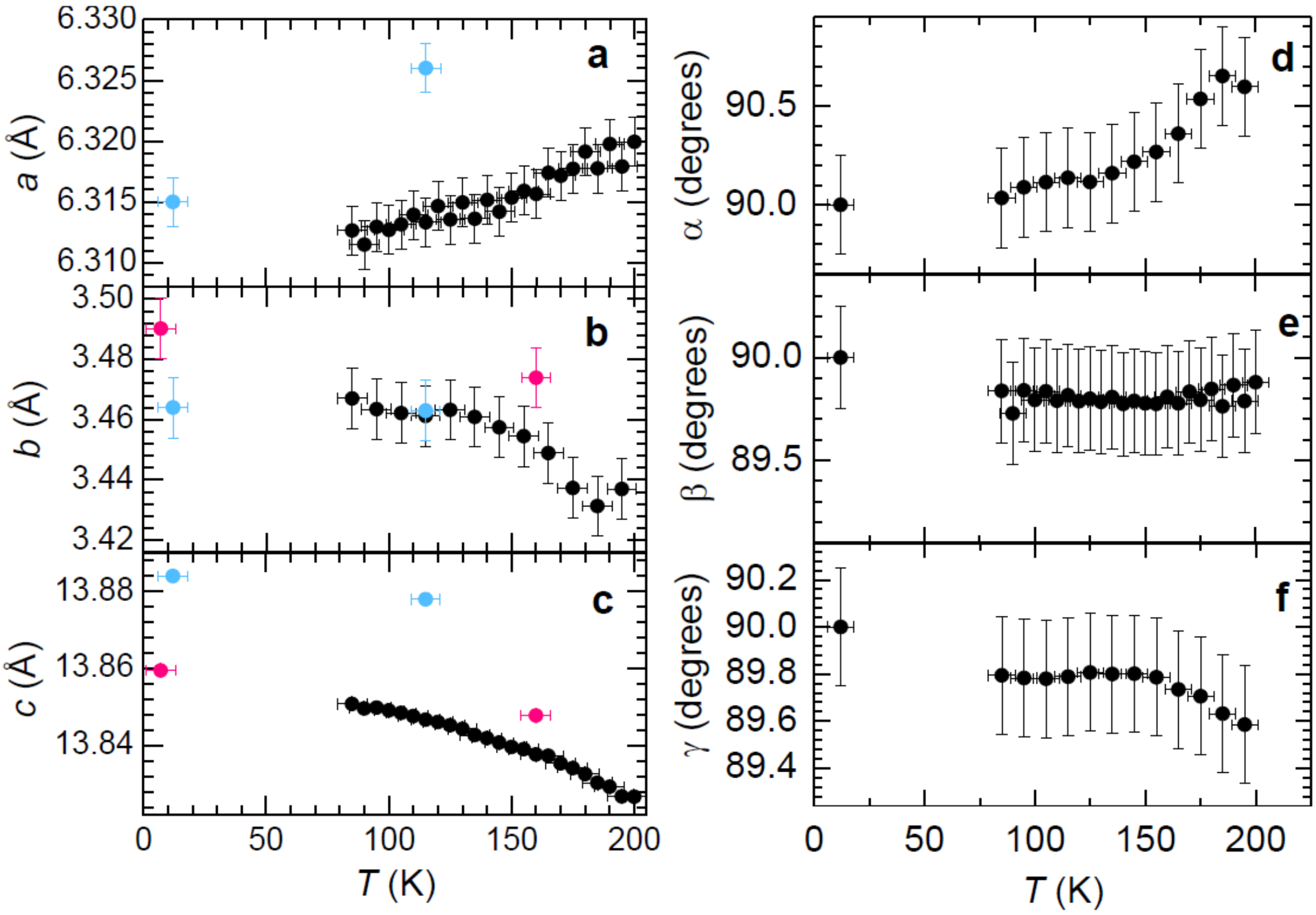

Figure A.3: (a), (b), (c) $\gamma$-$MoTe_2$ lattice constants a, b and c of as functions of the temperature $T$. (d), (e), (f) Angles $\alpha$, $\beta$ and $\gamma$ between the lattice vectors as functions of $T$. Black markers depict Sample 1, colored markers were collected from Samples 2 (in pink) and 3 (in blue).

A third crystal of dimensions $\sim 80 \times 20 \times 5$ $\mu m^3$ was measured with 55 keV photons and a 6 Mega-pixel photon counting area detector. The single-crystal was continuously rotated over 370° during shutterless data collection, with 0.1° of rotation integrated into each frame. Three complete datasets were collected by varying the position of the detector relative to the incident beam. Reflections from the reduced datasets were scaled and merged together for structure solution and refinement. Diffraction data at $T = 115$ K was collected with a nitrogen cryostream system to bathe the crystal in a stream of cold gas. Data at $T = 12$ K was collected in a similar fashion using a helium cryostream. For both setups the temperatures of the cold stream were confirmed with a silicon diode sensor immediately prior to the measurements.

Table A.1: Low temperature structural parameters of $\gamma$-$MoTe_2$ resulting from synchrotron X-ray scattering. Orthorhombic space group #31 or $Pmn2_1$ with $\alpha = \beta = \gamma = 90°$ . $T = 115$ K: $a = 3.463$Å, $b = 6.326$Å, $c = 13.878$Å.

| Atom | X | Y | Z | site |
|---|---|---|---|---|
| Te1 | 0.00000 | 0.85960 | 1.02325 | 2a |
| Te2 | 0.00000 | 0.21033 | 0.77018 | 2a |
| Te3 | 0.50000 | 0.36298 | 1.06678 | 2a |
| Te4 | 0.50000 | 0.71658 | 0.81359 | 2a |
| Mo5 | 0.00000 | 0.47030 | 0.92545 | 2a |
| Mo5 | 0.50000 | 0.10570 | 0.91135 | 2a |

We investigated the temperature evolution of the crystalline-structure by performing single-crystal synchrotron X-ray diffraction measurements at low temperatures; Figure A.3 displays the lattice constants and the angles between lattice vectors as functions of the temperature for 7 K $\leq T \leq$ 200 K as collected from three single crystals sample 1 (black dots) and samples 2 and 3 (colored markers). Unit cell parameters were co-refined by collecting three-dimensional center-of-mass measurements in reciprocal space of a set of 4 Bragg peaks, measured in temperature intervals of 5 K upon cooling. Within the sphere of confusion of the instrument all angles between lattice vectors approach 90° as $T$ is lowered below 150 K and as expected for an orthorhombic structure. The low temperature structural parameters of $\gamma$-$MoTe_2$ resulting from synchrotron X-ray scattering are displayed in Table A.1 and Table A.2 .

To summarize, synchrotron X-ray diffraction confirms that the crystallographic structure of $\gamma$-$MoTe_2$ is temperature dependent down to$\sim$ 125 K, with this compound displaying a remarkable negative thermal expansion coefficient along the b- and the c-axes. Perhaps more important, little dependence on temperature is observed below ~ 100 K. The pronounced deviations with respect to orthorhombicity observed in the range of temperatures 150 K $\leq T \leq$ 250 K, are probably a manifestation of the hysteresis associated with the structural transition at ~ 250 K.

Table A.2: Low temperature structural parameters of $\gamma$-$MoTe_2$ resulting from synchrotron X-ray scattering. Orthorhombic space group #31 or Pmn21 with $\alpha = \beta = \gamma = 90°$. $T = 12$ K: $a = 3.464$Å, $b = 6.315$Å, $c = 13.884$Å.

| Atom | X | Y | Z | site |
|---|---|---|---|---|
| Te1 | 0.00000 | 0.86030 | 1.02332 | 2a |
| Te2 | 0.00000 | 0.21070 | 0.77028 | 2a |
| Te3 | 0.50000 | 0.36360 | 1.06682 | 2a |
| Te4 | 0.50000 | 0.71710 | 0.81347 | 2a |
| Mo5 | 0.00000 | 0.47060 | 0.92537 | 2a |
| Mo5 | 0.50000 | 0.10560 | 0.91139 | 2a |

## A.4 Berry Phase Analysis and Angular Dependence of the De Haas-Van Alphen Signal

The correct phase is determined via a comparison with the magnetic susceptibility, $\chi_{zz} = \partial M_z/\partial(\mu_0 H)$, as measured with a SQUID magnetometer (blue trace, low fields). As seen, both traces are nearly in-phase. The relatively small deviation in their respective phases is due to the difference in angle between the magnetic field and $c$-axis of the crystal for either measurement: torque, which requires an angle, was measured at $\theta \sim 3°$, while susceptibility $\theta \sim 0°$.

Figure A.4 displays the oscillatory signal superimposed onto the magnetic torque (black trace) of $\gamma$-$MoTe_2$ for fields nearly along its $c$-axis as a function of the inverse field. To this trace we have superimposed the oscillatory signal measured in a SQUID magnetometer for fields also aligned nearly along the $c$-axis. The goal of this figure is to demonstrate that the oscillations from both measurements are nearly in phase, given that the sign of the torque signal is angle dependent, and therefore that we can extract the correct values of the Berry–phase of $\gamma$-$MoTe_2$ through fits of the oscillatory signal to Lifshitz-Kosevich formalism. Figure A.5 displays the oscillatory signal superimposed onto the derivative of the magnetic torque (black trace) as a function of inverse field along with fits (red traces) to four Lifshitz-Kosevich oscillatory components performed over an increasing range in $(\mu_0 H)^{-1}$. Two of these components describe the main fundamental frequencies observed for this orientation while the other two depict their first harmonics, respectively. The objective of these fits is to evaluate the Berry-phase and therefore the topological character of $\gamma$-$MoTe_2$. As seen through the four panels of Figure A.5, one obtains $F_\alpha = 240.4$ T and $F_\beta = 259$ T for the fundamental frequencies and for fields below 10 T, but they are observed to continuously

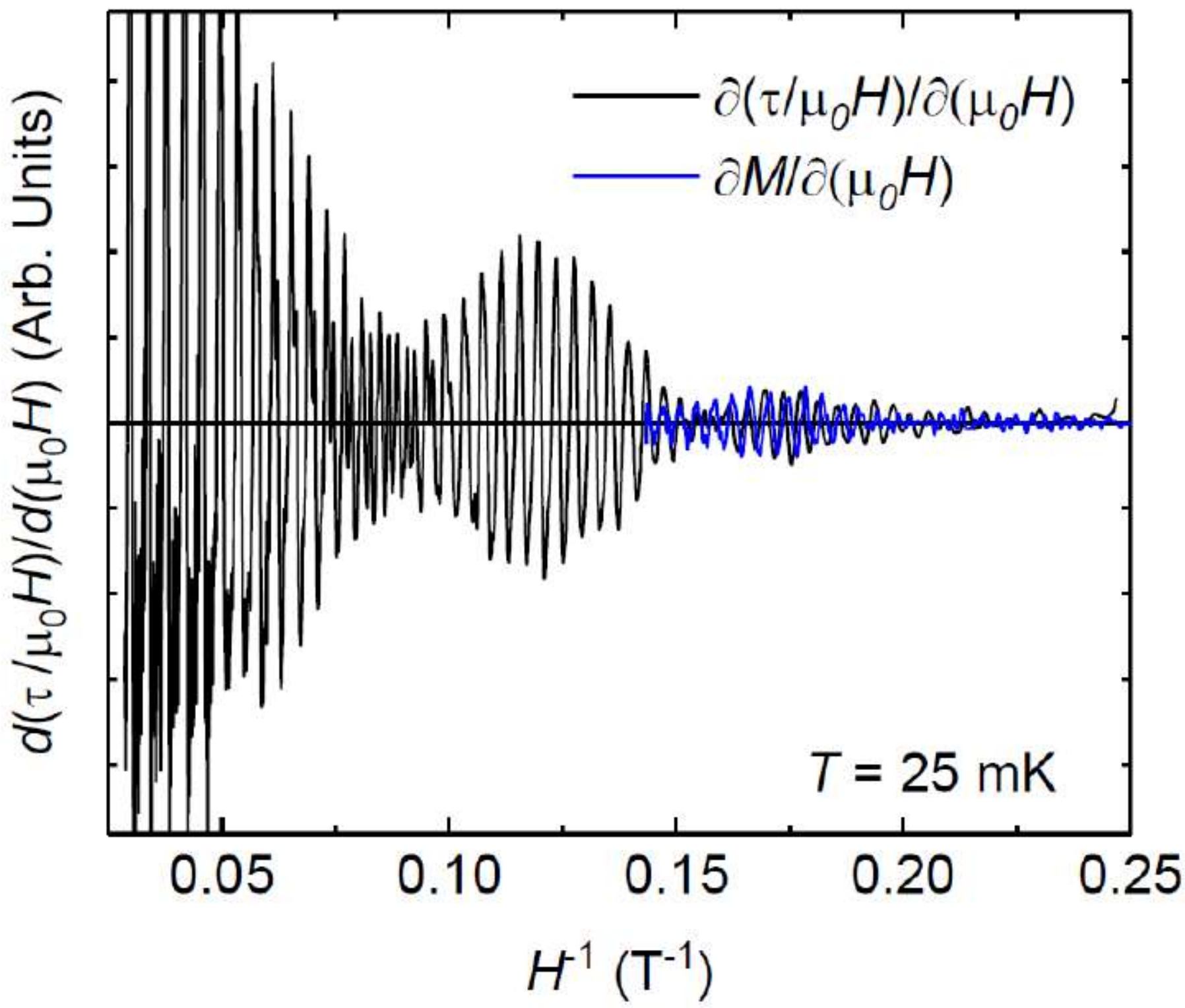

Figure A.4: Given that the magnetic torque for a layered system is given by $\tau = (\chi_{xx} - \chi_{zz})H^2/2\sin 2\theta$, where $\chi_{xx}$ is the component of the magnetic susceptibility for fields applied along a planar direction and $\chi_{zz}$ is the component of the susceptibility for fields applied along the inter-planar direction with $\theta$ being the angle between the magnetic field and the inter-planar direction, the sign of $\tau$ is actually $\theta$-dependent. Therefore, the phase of the oscillatory signal (observed at high fields) superimposed onto magnetic torque $\tau$ (black trace) is also $\theta$-dependent.

evolve as the field interval is increased, i.e. $F_\alpha$ increases to 244.3 T while $F_\beta$ decreases slightly to 256.6 T when one fits the entire fit range. Meanwhile, and assuming constant values for the effective masses (shown below and in the main text), one observes the Dingle temperatures to decrease as the fit progressively includes the whole field range. For example, $T_D^\alpha$ decreases from 3.3 K to 1.9 K.

Similarly, the phase-factors from which one extracts the Berry phases $\phi_B^{\alpha,\beta}$ are also found to be strongly field-dependent. The values of $\phi_B^{\alpha,\beta}$ indicated in the figure were calculated assuming minima and maxima extremal-cross-sectional areas, respectively, and a three-dimensional FS. In summary, the geometry of the Fermi surface of $\gamma$-$MoTe_2$ evolves as a function of the field due to the Zeeman-effect, which precludes the extraction of its Berry phase. However, even at the lowest fields the values of $\phi_B^{\alpha,\beta}$ differ from $\pi$ indicating that it is not a topological compound.

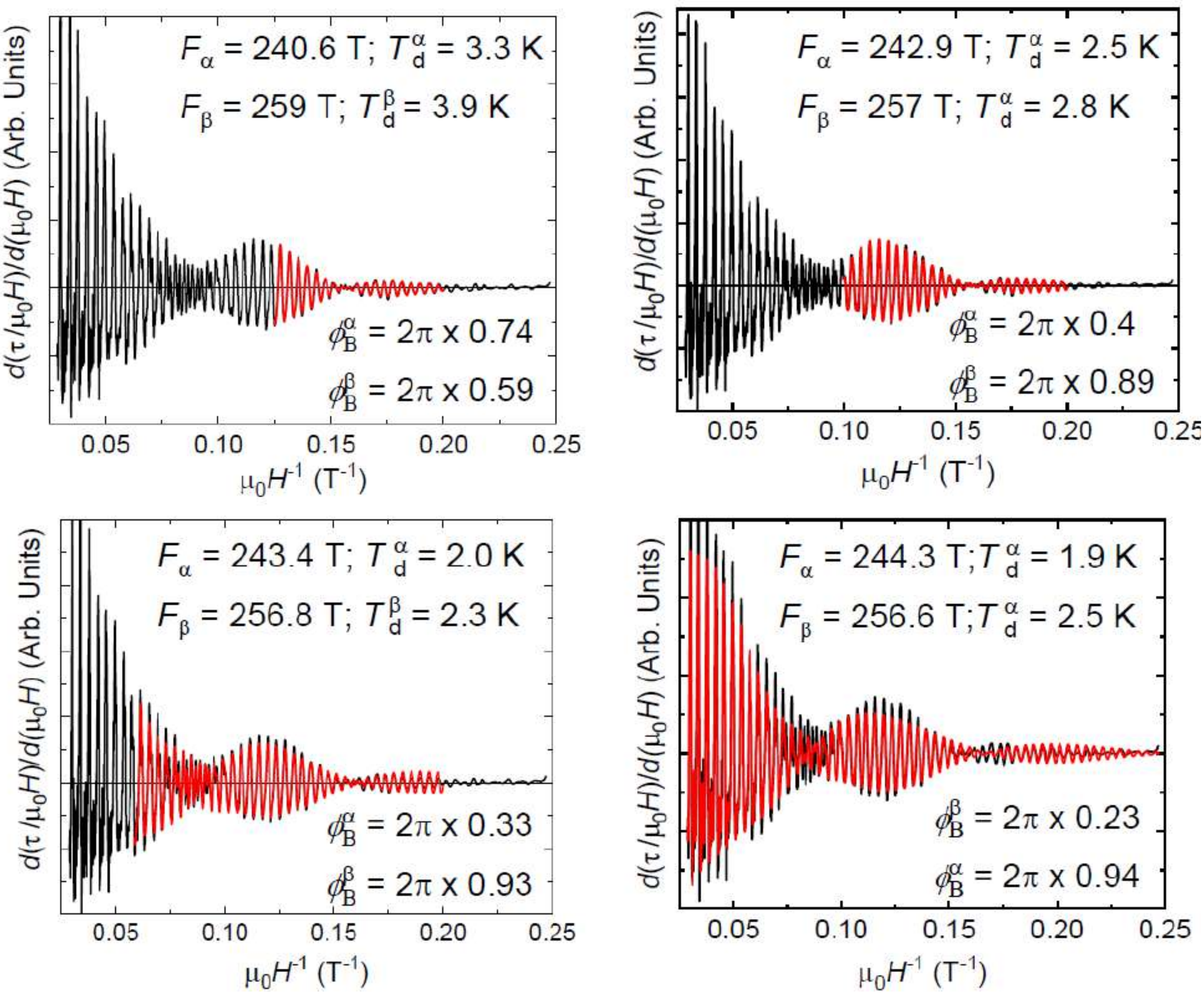

Figure A.5: From left to right and top to bottom: fits (red lines) of the de Haas van Alphen signal (dHvA) superimposed onto the magnetic susceptibility (as extracted from the derivative of the magnetic torque $\tau$ with respect to $\mu_0 H$) to four Lifshitz-Kosevich oscillatory components. The four components describe the two fundamental frequencies $F_\alpha$ and $F_\beta$ observed for fields applied along the $c$-axis along with their first harmonics. Notice that the fundamental frequencies shift as the interval in $(\mu_0 H)^{-1}$ increases. The obtained Berry phases $\phi_B^{\alpha,\beta}$ and extracted Dingle temperatures $T_D^\alpha$ are also field-dependent. Hence, the electronic structure of $\gamma$-$MoTe_2$ is affected by the Zeeman-effect which prevents the extraction of its Berry phase. Notice however, that the Berry-phases extracted at the lowest fields are distinct from $\pi$ and therefore at odds with the prediction of a Weyl type-II semi-metallic state in $\gamma$-$MoTe_2$.

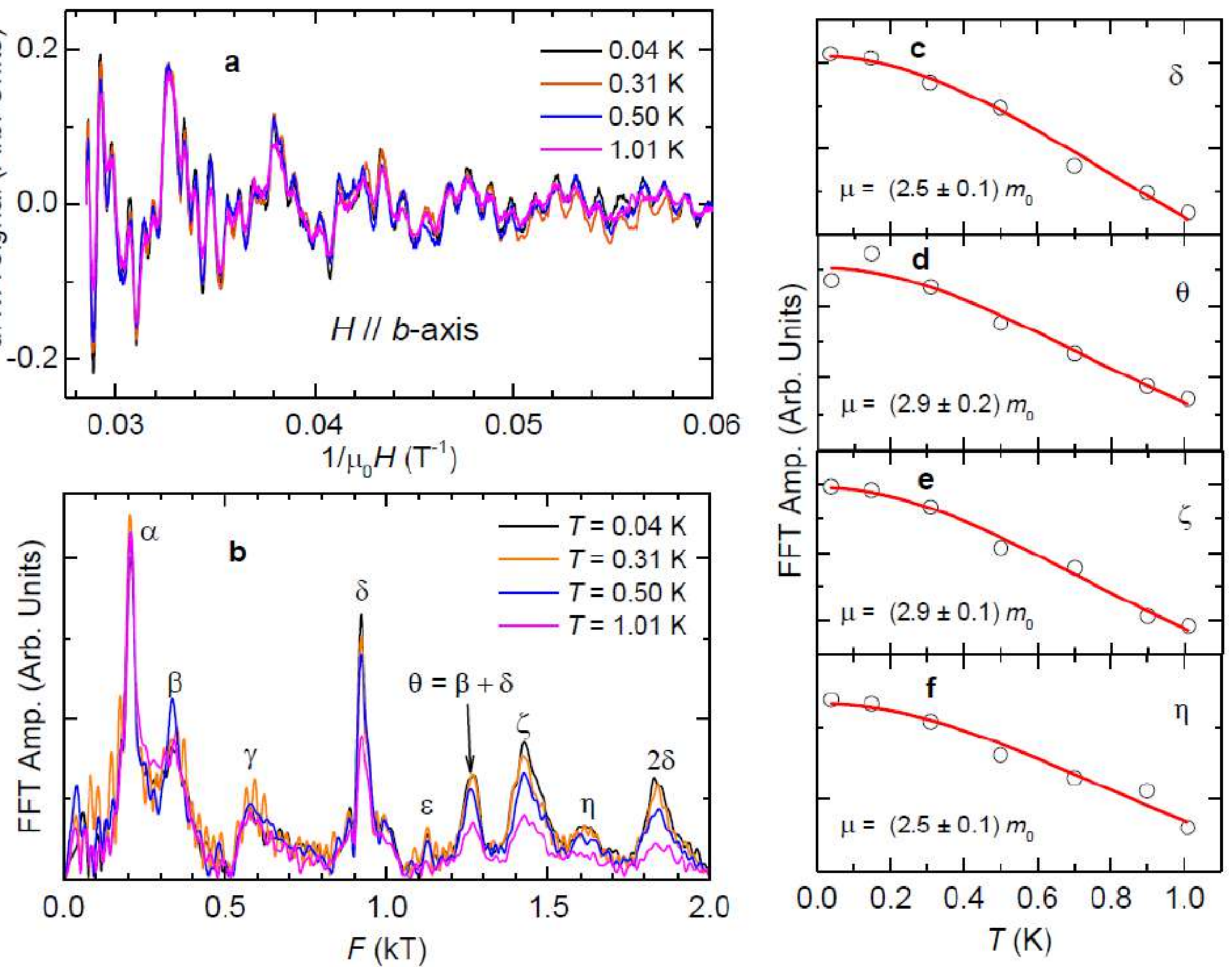

Figure A.6: Oscillatory signals and carrier effective masses for fields along the *b*-axis. (a) Oscillatory component of the Haas-van Alphen effect, superimposed onto the magnetic torque as a function of the inverse field $H^{-1}$ and for several values of $T$. Here, $H$ was aligned nearly along the *b*-axis. (b) FFT spectra taken from a window in $H$ ranging from 20 to 35 T. (c), (d), (e), and (f), Amplitude of selected peaks observed in the FFT spectrum as a function of $T$. Red lines are fits to the Lifshitz-Kosevich expression from which we extract the effective masses indicted in each panel.

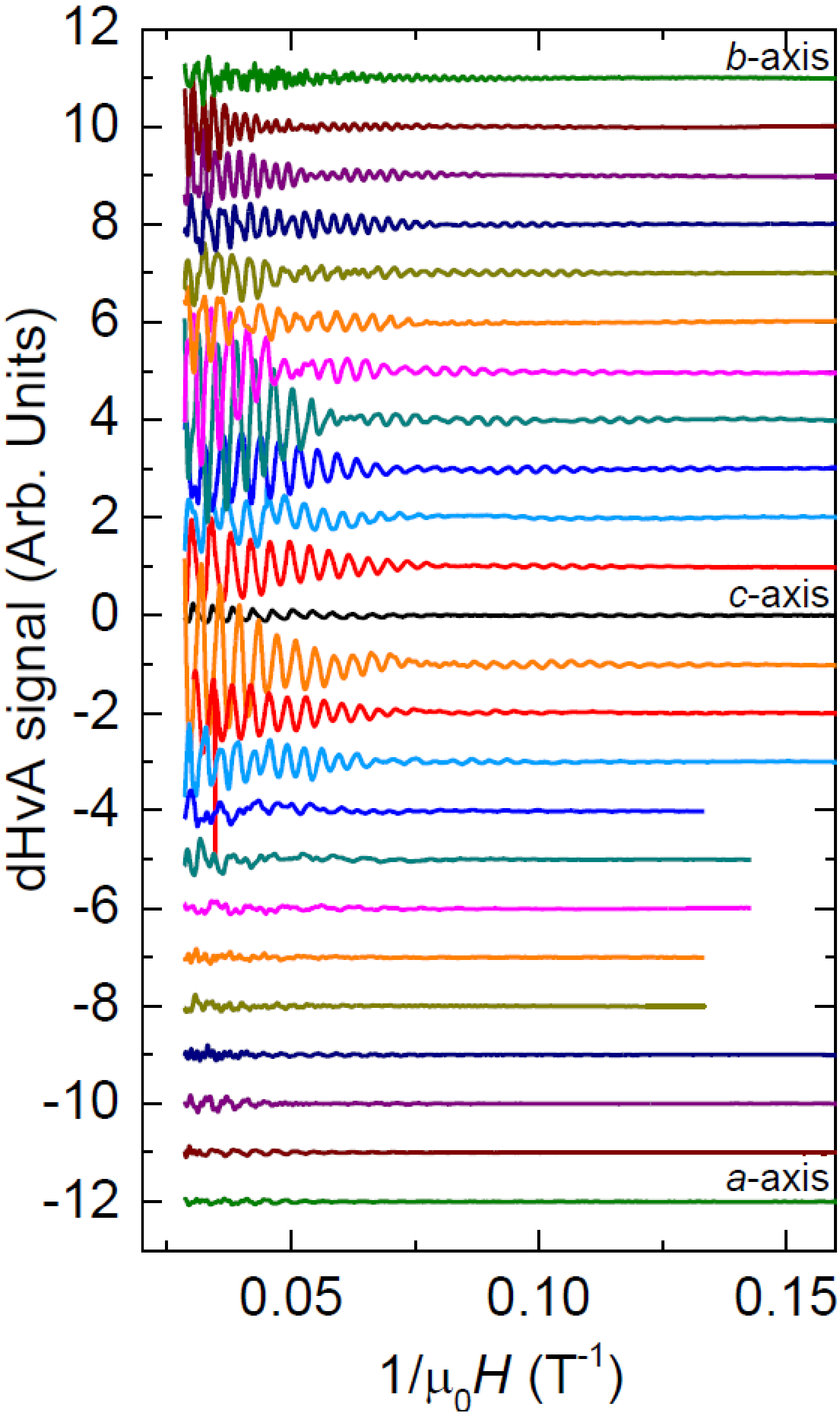

Figure A.7: Oscillatory component, or the dHvA effect, superimposed onto the magnetic torque as a function of the inverse field $H^{-1}$ and for several angles between the $c$-axis and the other two crystallographic axes. This data was collected at 35 mK and was used to extract the fast Fourier-transform spectra as a function of the angle displayed in Figs. 3.4(b) and 3.6(a) within Chapter 3.

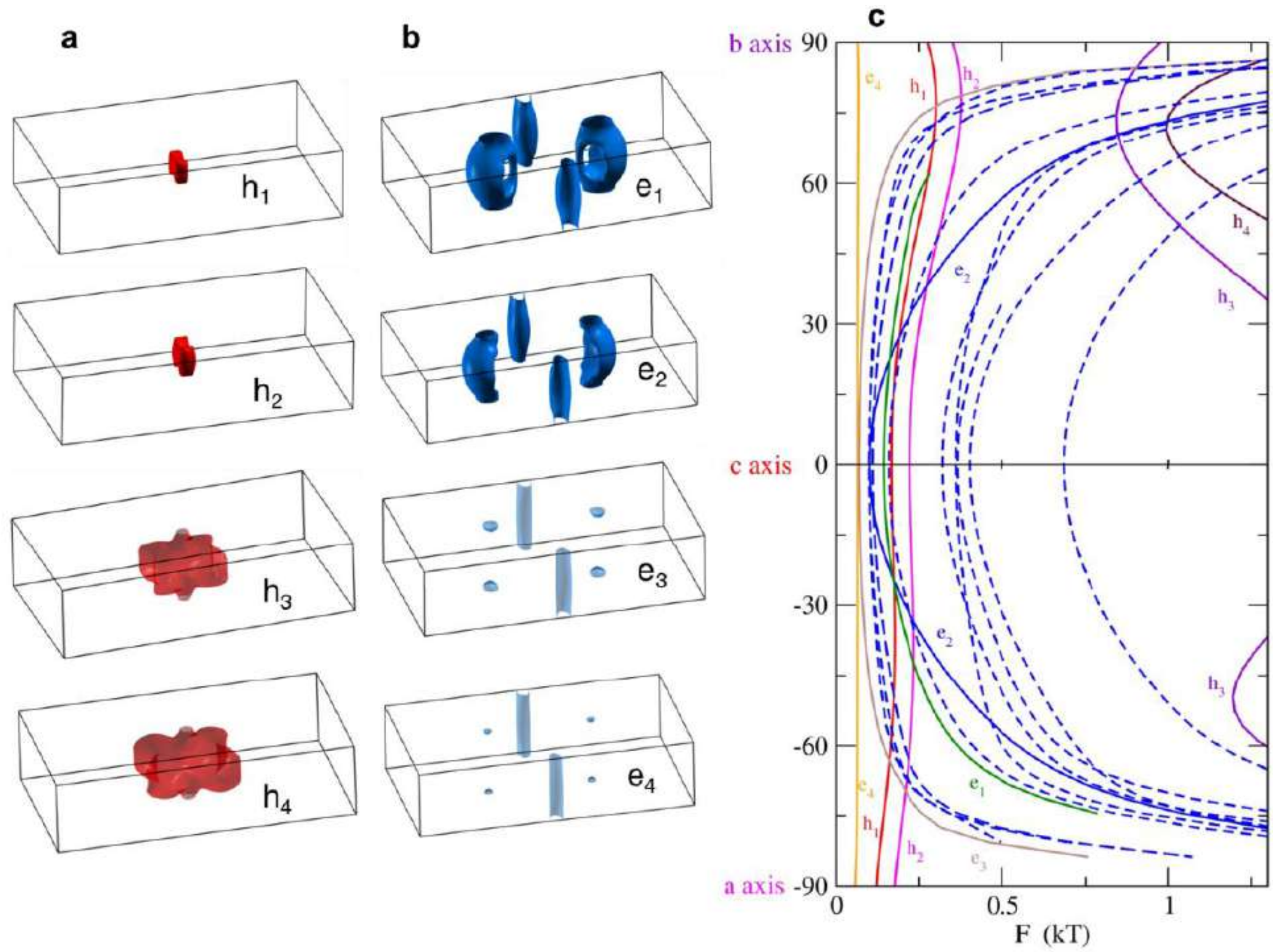

Figure A.8: Fermi surface of $\gamma$-$MoTe_2$ according to the VASP implementation with the inclusion of spin-orbit coupling. (a) Hole-like Fermi surfaces around the $\Gamma$-point. In contrast to the Wien2K implementation, VASP predicts small hole-like Fermi surfaces at $\Gamma$. In addition, the larger four-fold symmetric hole-like surfaces would touch the boundary of the first Brillouin zone. This is attributable to a different position of the Fermi energy relative to the Wien2K implementation. (b) Electron-like Fermi surfaces. In contrast to Wien2K, VASP predicts small electron ellipsoids in addition to three sets of cylindrical electron-like Fermi surfaces at the $X$-point. (c) Fermi surface cross-sectional areas as functions of the angle of rotation of the external magnetic field relative to the crystallographic axes.